\documentclass{PoS}
\usepackage{amssymb,amsmath}
\usepackage{mwe}
\usepackage{subfig,url}
\usepackage[absolute]{textpos}
\usepackage[inline]{enumitem}
\usepackage[export]{adjustbox}
\usepackage{amsmath}
  \newcommand{\ttbar}{\ensuremath{\rm{t\bar{t}}}\xspace}

  \newcommand{\pt}{\ensuremath{p_T}\xspace}

\title{Results on TOP physics from CMS}

\ShortTitle{}

\author{\speaker{Georgios Konstantinos Krintiras} on behalf of the CMS Collaboration\\
        Universit\'e catholique de Louvain, Louvain-la-Neuve, Belgium\\
        E-mail: \email{gkrintir@cern.ch}}

\abstract{
After the discovery of the top quark more than 20 years ago, 
top quark production cross sections have been meticulously studied.
The rich variety of results from the LHC experiments are complemented with increasingly accurate
theoretical predictions of heavy quark production and decay.
Measurements of the top quark production provide a
benchmark test of perturbative quantum chromodynamics and the standard model (SM), 
constraining at the same time the background in Higgs boson searches as well as extensions 
beyond the SM.
Recent top quark measurements from  CMS  are  reviewed,  illustrating  
past and current experimental methods along with their attained precision.
A perspective of top quark physics at the High-Luminosity LHC and at future colliders is briefly given.
}

\FullConference{
  International Workshop on Future Linear Colliders, LCWS2017\\
  23-27 October 2017\\
  Strasbourg, France}

\begin{document}

\tableofcontents{}

\clearpage
\section{Setting the scene}
\label{sec:intro}

\begin{figure}[!ht]
  \centering
  \includegraphics[scale=0.25]{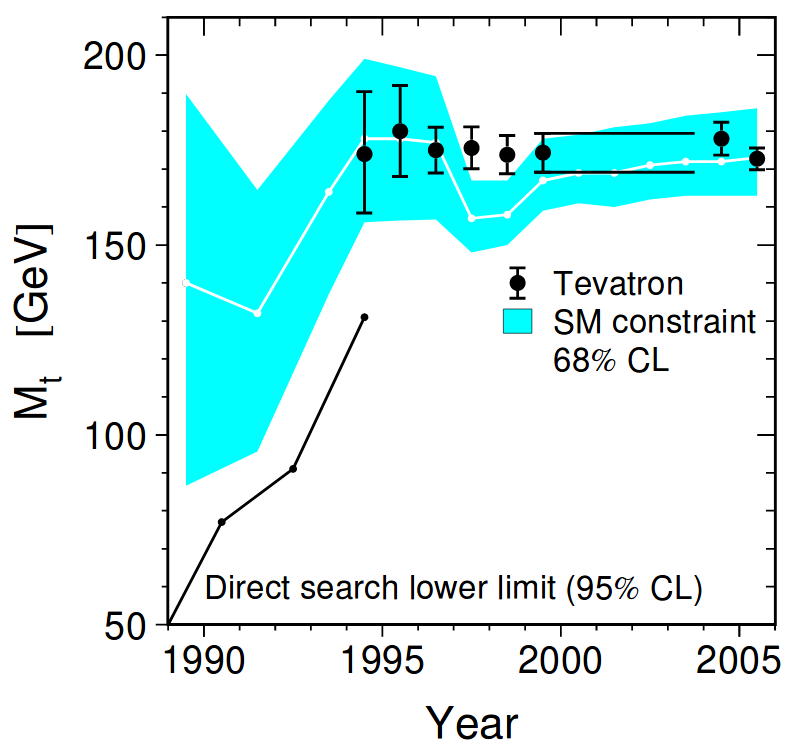}
    \caption{
      Comparison of direct (at Tevatron) and indirect determinations of the top quark mass as a function of time~\cite{aleph}.
      The good agreement between the top quark mass measured directly and the predicted mass determined 
      indirectly on the basis of measurements at the Z-pole is a convincing illustration of the validity of radiative 
      corrections. Also shown is the 95\% confidence level lower limit from the direct searches before the discovery of the top quark~\cite{cdf_obs,d0_obs}.
}
    \label{fig:topmass_time}
\end{figure}

%The top quark is the heaviest elementary particle with its mass close to a gold atom, yet its size is less than $10^{-3}$ fm.  
With a mass close to the one of a gold atom, yet a size less than $10^{-3}$ fm, the top quark is the heaviest elementary particle known.
When the top quark was independently discovered by the CDF~\cite{cdf_obs} and D{\O}~\cite{d0_obs} Collaborations in 1995 at the Tevatron proton-antiproton ($\rm{p\bar{p}}$) collider,
a  decades-long search for one of the last missing pieces of the standard model (SM) had come to an end.
%In hadron-hadron collisions at high center-of-mass energies ($\sqrt{s}$), 
%the actual interaction occurs between internal quarks (qq) and  gluons ($gg$). 
%the relative importance of qq- as opposed to gg-initiated processes is simply dictated by $\sqrt{s}$.%
%Most of the times, each  quark  or  gluon  
%carries only a modest fraction of the total energy of the incoming hadron, meaning  the  collision  must  be  
%energetic enough  to  generate  top  quarks.  
%Such collisions  are  rare,  and  the  higher the top mass 
%($m_{\rm{top}}$\footnote{Alternative notations, like $M_{\rm{top}}$, $M_{\rm{t}}$, or $m_{\rm{t}}$, exist and are typically used interchangeably.}), 
%the rarer they are.
Despite an early blunder, that had resulted in claiming a ``clear signal'' compatible with a W boson decaying into a 40 GeV top quark in 1984~\cite{ua1_blunder},
indirect hints of a large top quark mass ($ \gg 50$ GeV) later came from significant flavor oscillation frequencies in $B_d^0$ meson mass eigenstates
and electroweak (EW) precision data. In Fig.~\ref{fig:topmass_time}, the limits on the top mass, $m_{\rm{top}}$ \footnote{Alternative notations, like $M_{\rm{top}}$, $M_{\rm{t}}$, or $m_{\rm{t}}$, exist and are typically used interchangeably.},
as a function of time are compared to direct measurements at the Tevatron.
The indirect determination within the framework of the standard model (SM) adheres to remarkably good agreement with the direct measurements.
%meaning the top quark is predicted and measured extremely heavy mass and hence short lifetime, 
It is thus intriguing to test whether the top quark with such extremely heavy mass 
readily influences predictions regarding the stability of the Higgs field 
and its effects on the evolution of the Universe, 
as well as the sensitivity of SM consistency checks. 
%top quark therefore offers a unique possibility to study the properties of an elemenetary particle in
%a quasi-free-quark state.

\subsection{A special role in the EWSB mechanism? }
\label{sec:ewsb}

\begin{figure}
\begin{minipage}{.5\linewidth}
\centering
\subfloat[]{\label{fits:a}\includegraphics[scale=0.22]{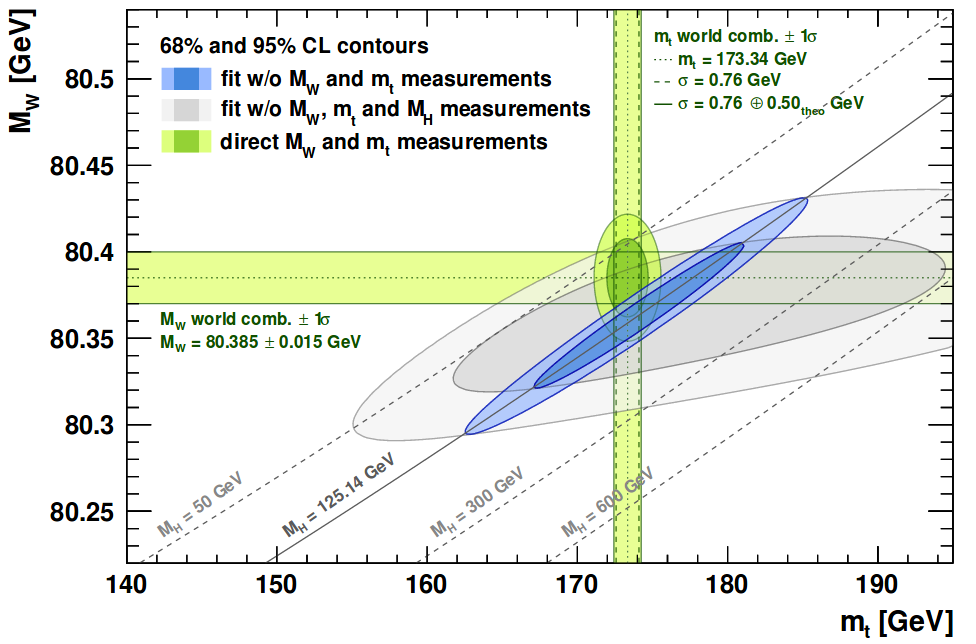}}
\end{minipage}%
\begin{minipage}{.5\linewidth}
\centering
\subfloat[]{\label{fits:b}\includegraphics[scale=0.22]{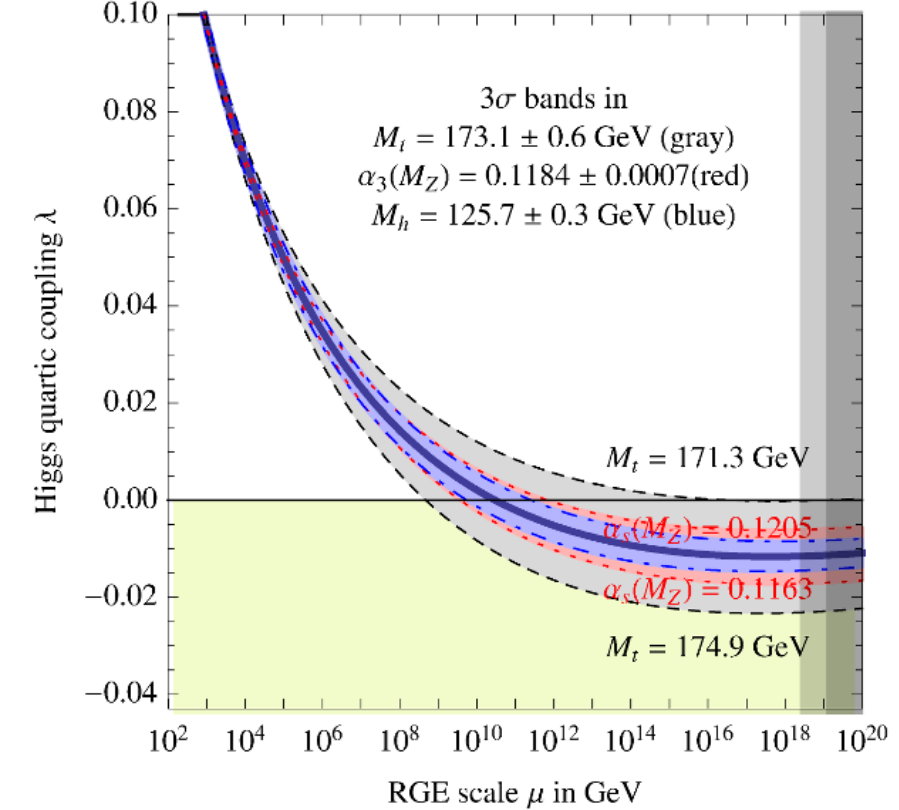}}
\end{minipage}\par\medskip

\caption{
(a) Contours at 68\% and 95\% confidence level 
obtained from scans of fits~\cite{global_EW_fit} with fixed variable pairs of the W boson against the top quark mass. 
The corresponding direct measurements are excluded from the fit.
The narrower blue and larger grey allowed regions are the results of the fit including and excluding the Higgs boson mass 
measurements~\cite{atlas_h_mass,cms_h_mass}, respectively. The horizontal bands indicate the $\pm 1\sigma$ regions of the world averages used for the W boson~\cite{pdg_2017} and top quark~\cite{topmass_world}.
A theoretical uncertainty of 0.5 GeV--non-perturbative theoretical uncertainty of $\mathcal{O}(\Lambda_{\rm{QCD}})$--is added to the direct top mass measurement.
(b)
Renormalisation-group evolution of the Higgs boson self-coupling varying the top mass, strong $\rm{SU(3)_{c}}$ coupling and Higgs boson mass within a $\pm 3\sigma$ range~\cite{vacuum_stability_1}.
Residual theoretical uncertainties~\cite{vacuum_stability_2} are safely smaller than the experimental uncertainty, dominated by the uncertainty on the top quark mass.
}
\label{fig:fits}
\end{figure}

%Whereas the W and Z boson masses can be calculated from the SM, %
In the SM, the quark and lepton masses have to be
``inserted'' by means of adjustable parameters  that  describe  how  feebly or strongly  each type  interacts with the Higgs boson.
%For an electron, the interaction strength is $3 \times 10^{-6} $, while% 
For a top quark the Yukawa coupling is almost exactly unity, unlike for any other fermion.
This relatively strong coupling with the Higgs boson, and to some extent the mystique associated  with  a  value  of  unity, suggests  
that  the  top  quark  may  play  a special role in the mechanism of electroweak symmetry breaking (EWSB).
Quadratically divergent radiative corrections appear in the computation of the Higgs boson mass and
are dominated by top quark loop contributions. 
Repeating the exercise of indirect determination of $m_{\rm{top}}$ from global SM fits to EW precision data 
is thus essential for testing the overall SM consistency.
A scan~\cite{global_EW_fit} of the confidence level profile of W boson mass versus $m_{\rm{top}}$ is shown in Fig.~\ref{fits:a} for
the scenarios where the direct Higgs boson measurements~\cite{atlas_h_mass,cms_h_mass} 
are included (blue) or not (grey) in the fit.  Both contours agree with the direct measurements (green bands and ellipse).
%This induces a fine-tuning in the renormalisation
%procedure, i.e., an unnatural cancellation among the quantum corrections to the Higgs boson mass, 
%often referred to as the naturalness problem.

In addition, owing to its large value, top quark has a direct impact on extrapolations of the EWSB- to high-energy scales.
Radiative corrections can drive the Higgs boson self-coupling ($\lambda$) towards zero or even negative values, potentially
leading to an unstable vacuum. 
The determination of such energy scale ($\mu$), either requiring or not beyond the SM (BSM) physics at lower energies like the EWSB scale,
is strongly influenced by the precision of the measurements in the top quark sector and by their unambiguous interpretation
in a clear theoretical framework (in Fig.~\ref{fits:b} $\mu$ varies by several orders of magnitude under a seemingly insignificant variation of $\pm 2$ GeV in $m_{\rm{top}}$).
Last but not least, the indication of a flat Higgs potential ($\lambda=0$) at very high $\mu$ values could imply 
the possibility the Higgs boson to play the role of the scalar field in the early universe, as an appealing phenomenological consequence (e.g.~\cite{higgs_inflation}).

\subsection{ Theory predates experiment; experiment keeps a tight rein on theory }
\label{sec:theo}

\begin{figure}[!ht]
  \centering
  \includegraphics[scale=0.65]{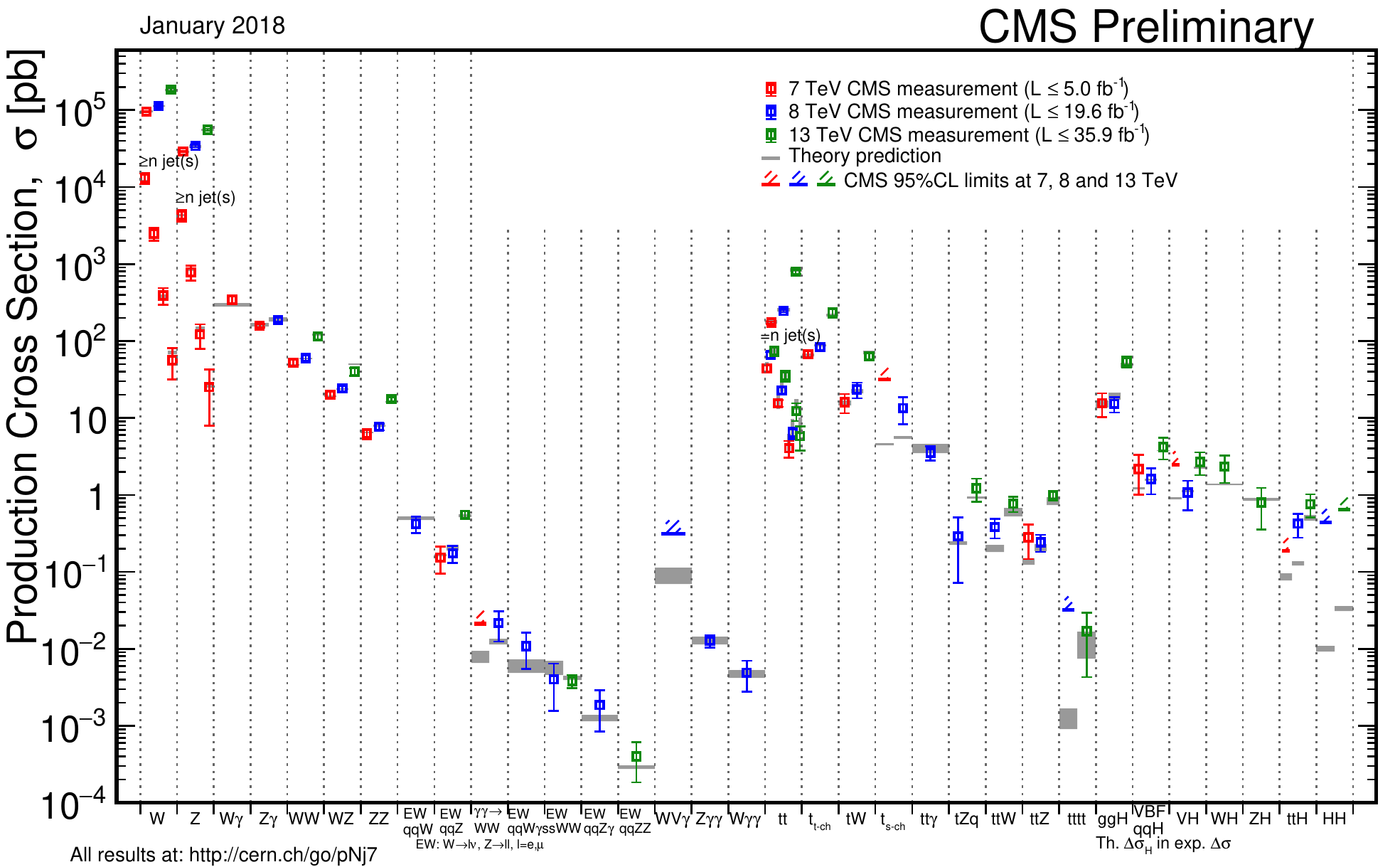}
    \caption{
	Compilation of CMS measurements and SM predictions in proton-proton (pp) collisions at $\sqrt{s}=7$, 8, and 13 TeV for a variety of 
	processes as indicated on the horizontal axis~\cite{cms_summary_SM}.
}
    \label{fig:cms_summary_SM}
\end{figure}

The basic tool for top quark physics is a high energy particle collider. For instance, the dominant production
of top quark pairs ($\rm{t\bar{t}}$),
which is a quantum chromodynamics (QCD) process,
is accessible both at lepton and at hadron
colliders, provided the $\sqrt{s}$ value of  the  collisions  is  above  the production threshold of twice $m_{\rm{top}}$.
The SM predicts the EW production of single top quarks in addition to the dominant $\rm{t\bar{t}}$ process,
while the associated production of top quarks with the EW gauge bosons or the Higgs boson is also possible (Fig. \ref{fig:cms_summary_SM}).
%Based on the best theoretical calculations, it is expected that pp collisions produce top quarks
%one thousand times less copiously than W bosons (Fig. \ref{fig:cms_summary_SM}).
A natural question can be then addressed: how theory and experiment have coped with the broad range of questions in top quark
physics, from inclusive production cross sections and precise measurements of the top quark mass and of its couplings 
to a variety of searches for BSM physics with top quarks.

\begin{figure}
\begin{minipage}{.5\linewidth}
\centering
\subfloat[]{\label{pdfs:a}\includegraphics[scale=0.2]{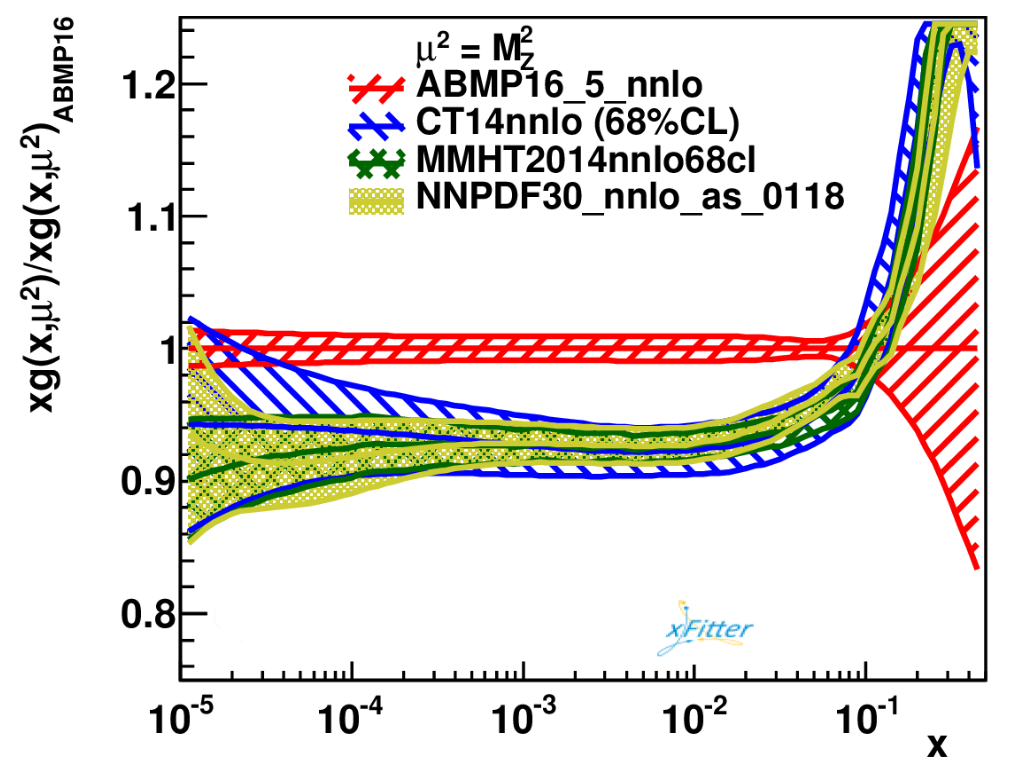}}
\end{minipage}%
\begin{minipage}{.5\linewidth}
\centering
\subfloat[]{\label{pdfs:b}\includegraphics[scale=0.215]{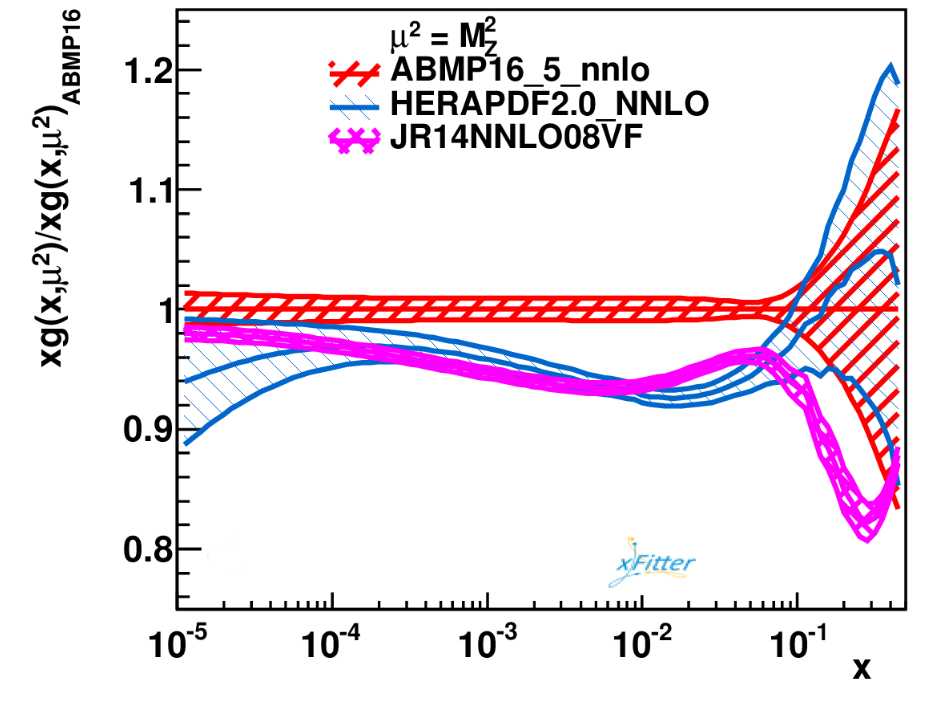}}
\end{minipage}\par\medskip

\caption{
The proton gluon PDF against the longitudinal momentum $x$ as determined from the (a) ABMP16~\cite{abmp16},
CT14~\cite{ct14}, MMHT14~\cite{mmht14}, NNPDF3.0~\cite{nnpdf30}, (b) HERAPDF2.0~\cite{hera20}, and JR14~\cite{jr14} sets
with their $\pm 1\sigma$ uncertainties at the factorization scale equals to the Z boson mass.
The distributions are normalized to the central values from ABMP16 for comparison~\cite{abmp16}.}
\label{fig:pdfs}
\end{figure}

The probability (cross section), $\sigma$, for any process can be expressed as a weighted product of a high-energy (``hard'') parton-parton scattering
multiplied with parton distribution functions (PDFs), integrated over all parton momenta and  summed  over  all  parton  
types (``phase space''):
\begin{align*}
\sigma=\sum_{jk}^{\rm{partons}}\int\limits_0^1 dx_j dx_k f_k(x_k,\mu_{\rm{F}}^2) \hat{\sigma} (sx_jx_k,\mu_{\rm{F}},\alpha_{\rm{s}}) \ .
\end{align*}
While the  hard  scattering  cross  section, $\hat{\sigma}$, is process specific and can be computed in perturbative QCD at different levels of accuracy,
PDFs, $f_i(x_i,\mu_{\rm{F}}^2)$, are deemed universal functions that describe the probability to find a parton $i$ with a given longitudinal 
momentum  fraction $x_i$, when  the hadron  is  probed  at  a  momentum  transfer scale of $\mu_{\rm{F}}$ (Fig.~\ref{fig:pdfs}). 
The PDF parameterization absorbs all long-distance effects in the  initial  state, and ultraviolet divergences in $\hat{\sigma}$ are treated with
a renormalization procedure of the strong coupling constant , introducing an additional energy scale ($\mu_{\rm{R}}$). 
The  ``default'' (but not unique) choice of energy scales are process specific, e.g., for $\rm{t\bar{t}}$ production a typical choice is $m_{\rm{top}}$.
Cross sections for $\rm{t\bar{t}}$ and single top quark production include higher-order corrections,
and are currently available up to  next-to-next-to-leading order (NNLO), including  soft-gluon  resummation  
at next-to-next-to-leading logarithm (NNLL), in perturbative QCD plus NLO EW and NNLO accuracy, respectively.

QCD color confinement requires the final state particles to be color-neutral, but the process of turning partons into hadrons 
cannot be treated perturbatively, and it thus relies on phenomenological models. 
To compare calculations of hadron-hadron collisions to the data, software tools based on the Monte Carlo (MC) method 
are employed. ``General-purpose'' MC event generators (e.g.~\cite{amc@nlo}-~\cite{sherpa}) can automatically compute  the  full  set  of
contributions to the hard process at NLO given the Feynman rules of the underlying theory (both SM and BSM).
For almost all processes, the normalization of the MC event generation is corrected to match the most precise
calculation, ignoring the effect that higher-order corrections may cause on the shape of kinematic observables.
The final process of turning partons into hadrons is separated into two steps; a probabilistic method to model 
the fragmentation of partons that effectively resums soft and collinear radiation (parton shower), and hadron formation (hadronization).
In order to avoid double-counting parton emissions due to higher order processes, specialized matching and merging techniques (e.g.~\cite{mlm,fxfx})
are available to consistently interface NLO MC event generators  to  the  parton  shower, typically to leading-logarithmic (LL) precision.  
Hadronization is described and validated (e.g.~\cite{cms_tuning_1,cms_tuning_2}) with phenomenological approaches, the most popular being the Lund string~\cite{pythia6,pythia8} 
and the cluster~\cite{herwig} models, 
albeit their unique implementation has been lately questioned in view of multiparticle production effects~\cite{alice}.

\clearpage
\subsection{ That could well be a top quark! }
\label{sec:exp}

\begin{figure}[!ht]
\begin{minipage}{.5\linewidth}
\centering
\subfloat[]{\label{lumi:a}\includegraphics[scale=0.35]{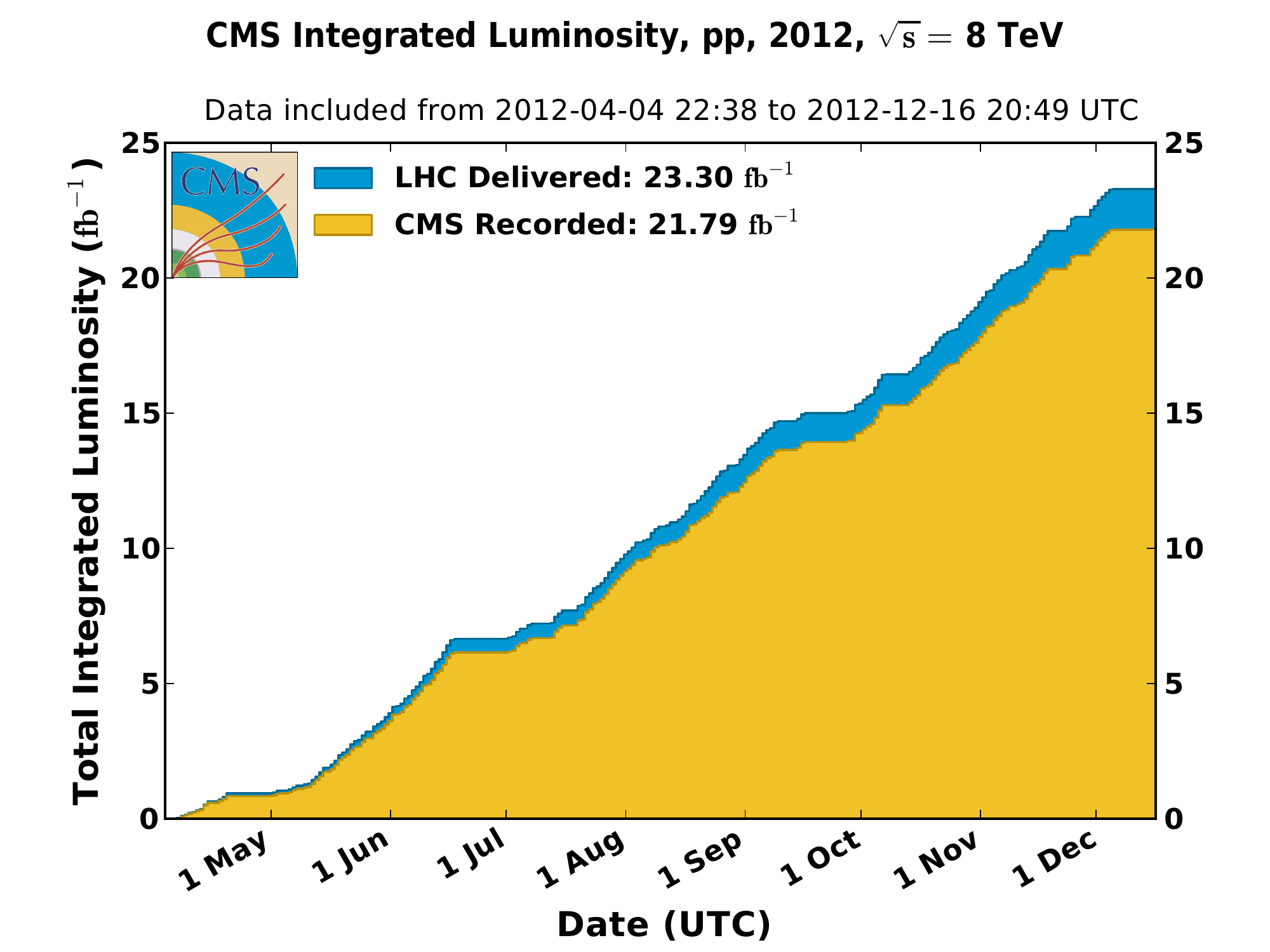}}
\end{minipage}%
\begin{minipage}{.5\linewidth}
\centering
\subfloat[]{\label{lumi:b}\includegraphics[scale=0.35]{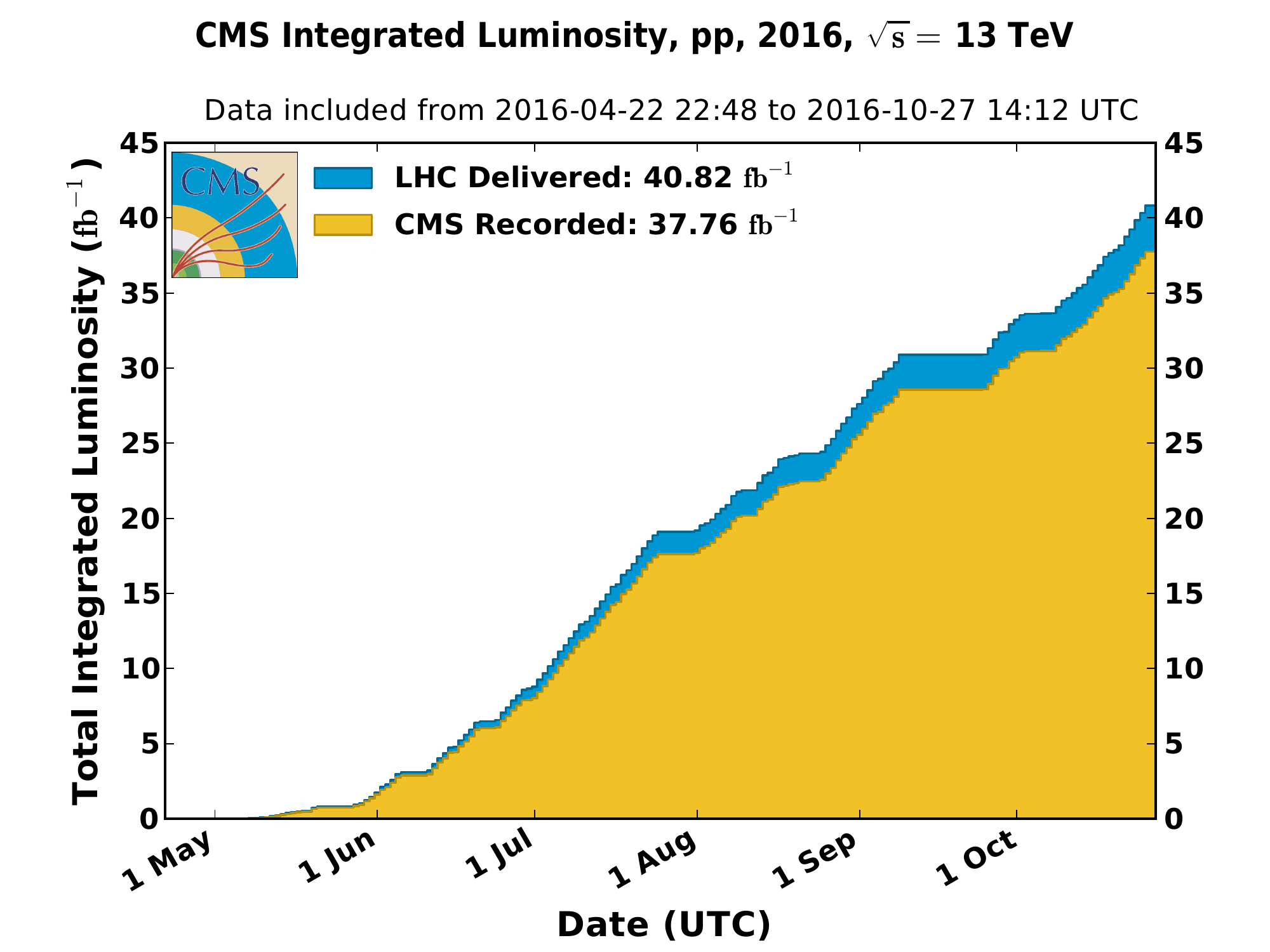}}
\end{minipage}\par\medskip

\caption{
Cumulative (``integrated'') delivered and measured luminosity versus time for pp collisions during LHC Run 1 (2012) and Run 2 (2016) period 
at $\sqrt{s}=8$ (a) and 13 (b) TeV, respectively. The measurements make use of the best available offline 
calibration conditions~\cite{lum13001,lum17001}.
With an availability for luminosity production of 50\% an integrated luminosity of almost 50 fb$^{-1}$ has been further 
delivered in 2017~\cite{evian_boyd}.  
}
\label{fig:lumi}
\end{figure}

Until the end of Tevatron data-taking in 2011, the collider had delivered a  total  of  $12$ fb$^{-1}$ of  integrated  luminosity.
The successful attempt from LHC at $\sqrt{s}=7$ TeV in 2010 (``Run 1'') has marked the beginning  of a new era  in  top quark  physics.   
Enhanced cross  sections for top quark production, on the one hand, and ever increasing data sets culminating at $25$ fb$^{-1}$ 
by the end of 2012 (Fig.~\ref{lumi:a}),
on the other hand, rendered LHC the first ``top quark factory''.
After  a  two-year  shutdown  for  maintenance  of  the  LHC  machine  and
experiments,  the  LHC  has restarted  in  early  2015 (``Run 2''). 
The  $\sqrt{s}$ value has further increased to 13 TeV, boosting typical top quark cross sections by about a factor of 
three compared to Run 1 (Fig.~\ref{lumi:b}).

Unlike  the  up and down quarks, which are stable, the top quark has a mean lifetime of only about $10^{-24}$ sec. 
In the SM, the top quark will decay almost instantly and nearly all the time into a  W boson and a bottom (b) quark. 
Neither the W boson nor the b quarks 
can be directly observed. When a quark emerges from a collision, it gets ``dressed up'' by a cloud of quarks and antiquarks. 
What is then experimentally observed is a jet, 
a directed spray of particles that in the vacuum have roughly the same energy and direction as the original quark.
The W boson decays with a probability of almost 70\%~\cite{pdg_2017} into a quark and an antiquark of the same generation (``hadronically''). In this case, the  quark and antiquark show up as two
jets correlated in phase space. But the W boson also decays, though with a decreased probability of around 10\%~\cite{pdg_2017}, ``leptonically'', i.e., into  a  charged  and  a  neutral lepton of the same 
generation, such as a muon and a neutrino.
The  neutrino traverses a detector completely unobserved. Its presence can be indirectly deduced due to a significant amount of missing momentum, 
assumed to have been carried away by the neutrino.

\begin{figure}
\begin{minipage}{.5\linewidth}
\centering
\subfloat[]{\label{btagging:a}\includegraphics[scale=0.35]{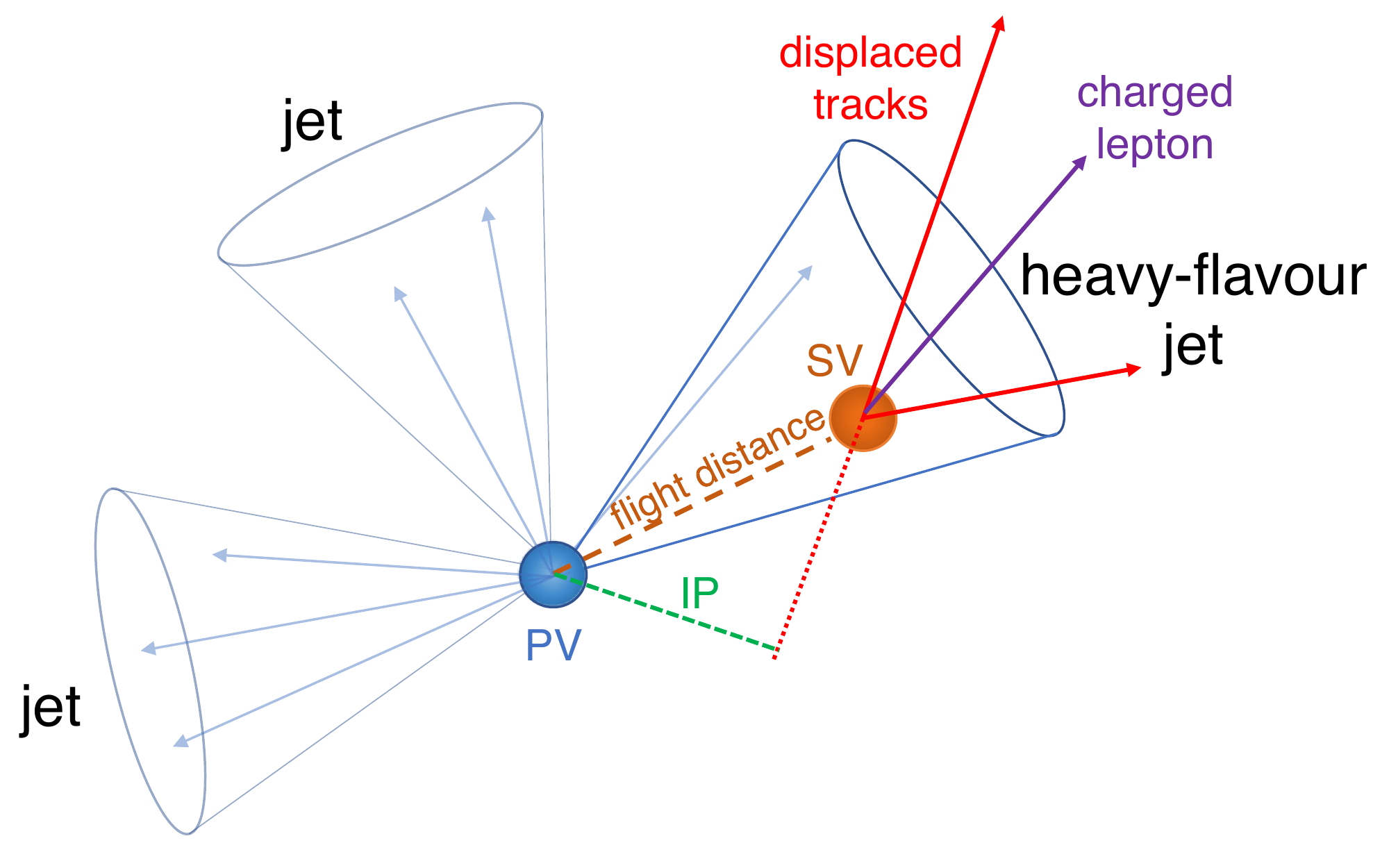}}
\end{minipage}%
\begin{minipage}{.5\linewidth}
\centering
\subfloat[]{\label{btagging:b}\includegraphics[scale=0.35]{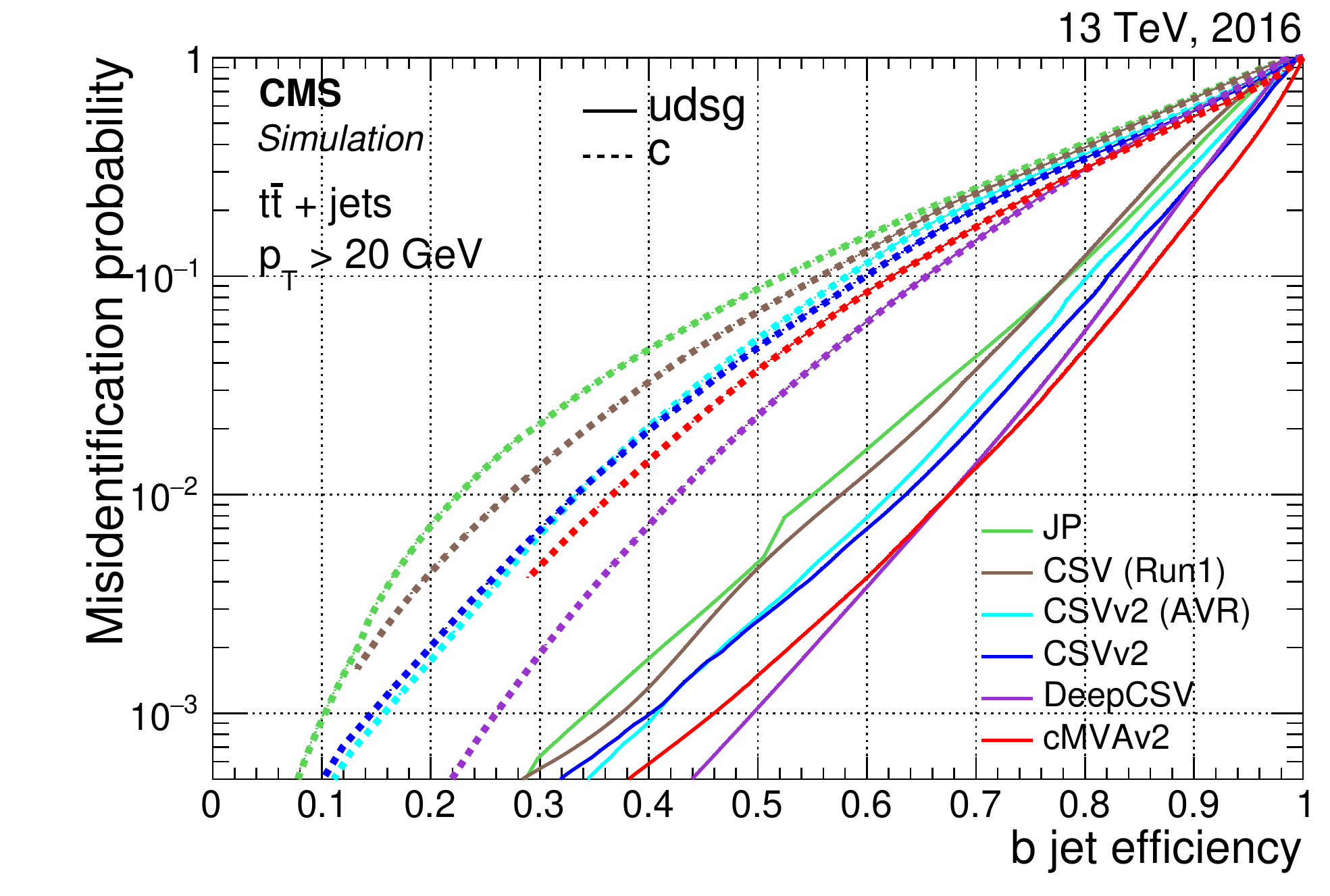}}
\end{minipage}\par\medskip

\caption{
(a)
Illustration~\cite{btagging_cms} of a heavy-flavor jet with a secondary vertex (SV) from the decay of a b or c hadron 
resulting in charged-particle tracks that are displaced relative to the primary interaction vertex (PV),
and hence with a large impact parameter (IP) value. 
(b)
Misidentification probability for c and light-flavor jets against b jet identification efficiency 
for various b tagging algorithms applied to jets (with a minimum transverse momentum requirement) in \ttbar events from a 
full simulation of the CMS detector response~\cite{geant4} based on POWHEG~\cite{powheg,hvq} (v2) at NLO along with the NNPDF3.0~\cite{nnpdf30} PDF set, 
and using \textsc{PYTHIA}~\cite{pythia8} (v8.205) for hadronization~\cite{btagging_cms}.  
}
\label{fig:btagging}
\end{figure}

Among the most striking features of a top quark signature are therefore the jets containing b quarks. 
General-purpose detectors, like CMS~\cite{cms}, put great emphasis on the ability to accurately track the paths of individual particles 
in a magnetic field, and in parallel, rely on extremely precise segmented calorimeters.
The b quark lies within a jet as part of a hadron, then  decays  roughly  half  a  millimeter from where it was created (Fig.~\ref{btagging:a}). 
The particles in jets can be very precisely tracked using a silicon vertex detector.
%, placed exactly on top of the region where the opposing beams collide.%
More specifically, the CMS  silicon  vertex detector can locate the path of a particle  to  within  $\pm 15\ \mu \rm{m}$.  
By identifying most of the tracks in a jet and extrapolating them backwards towards the primary interaction vertex,  
the point where the bottom quark decayed could thereby be attained and the jet identified (``tagged'') as a b jet (Fig.~\ref{btagging:b}).

\clearpage

\section{ A multifaceted quark }
\label{sec:measurements}

\subsection{ Most probably flies in pairs }
\label{sec:tt}

\begin{figure}
\begin{minipage}{.5\linewidth}
\centering
\subfloat[]{\label{tt_prod:a}\includegraphics[scale=0.25]{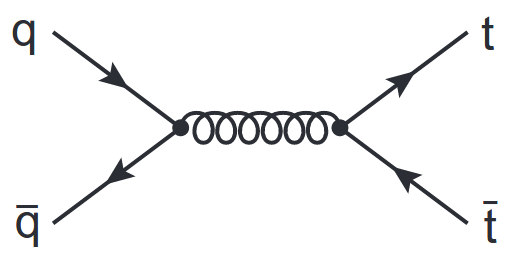}}
\end{minipage}%
\begin{minipage}{.5\linewidth}
\centering
\subfloat[]{\label{tt_prod:b}\includegraphics[scale=0.25]{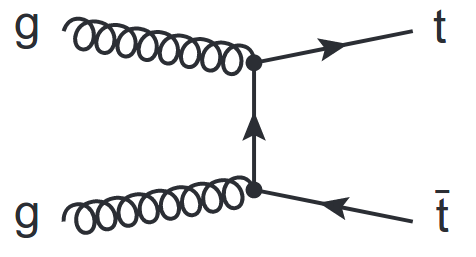}}
\end{minipage}\par\medskip
\begin{minipage}{.5\linewidth}
\centering
\subfloat[]{\label{tt_prod:c}\includegraphics[scale=0.25]{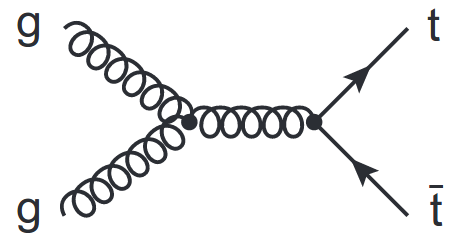}}
\end{minipage}%
\begin{minipage}{.5\linewidth}
\centering
\subfloat[]{\label{tt_prod:b}\includegraphics[scale=0.25]{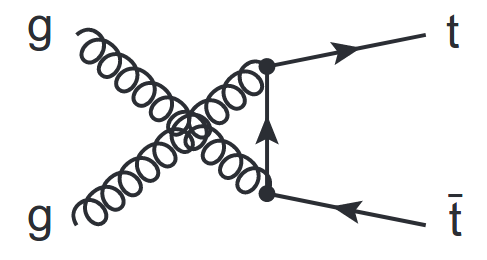}}
\end{minipage}\par\medskip

\caption{Representative diagrams~\cite{husemann} for \ttbar production at LO: $\rm{q\bar{q}}$ annihilation (a),
gg fusion in the \textit{t}- (b), \textit{s}- (c), and \textit{u}-channels (d).}
\label{fig:tt_prod}
\end{figure}

\begin{figure}
\begin{minipage}{.5\linewidth}
\centering
\subfloat[]{\label{tt_frac:a}\includegraphics[scale=0.35]{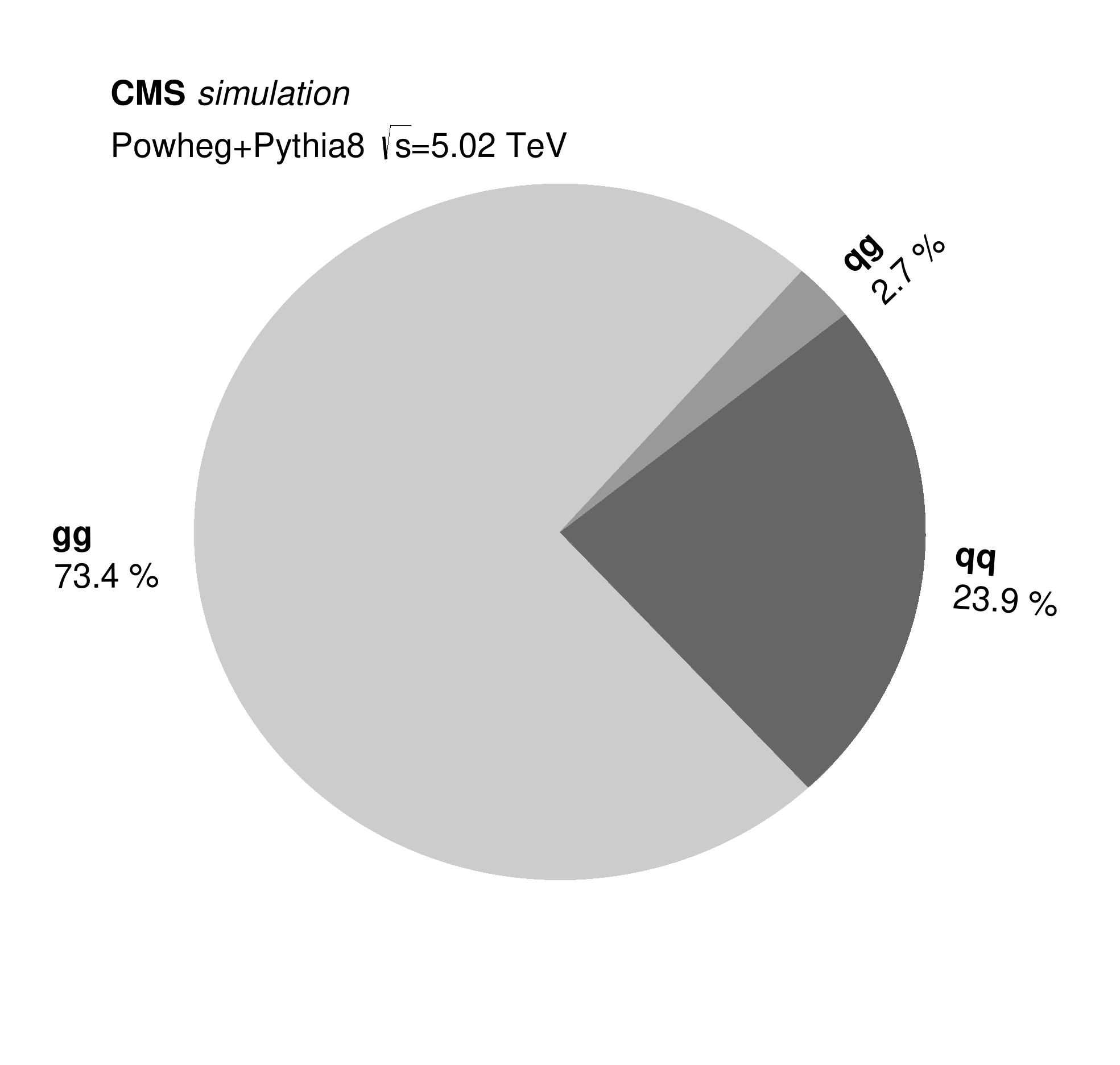}}
\end{minipage}%
\begin{minipage}{.5\linewidth}
\centering
\subfloat[]{\label{tt_frac:b}\includegraphics[scale=0.35]{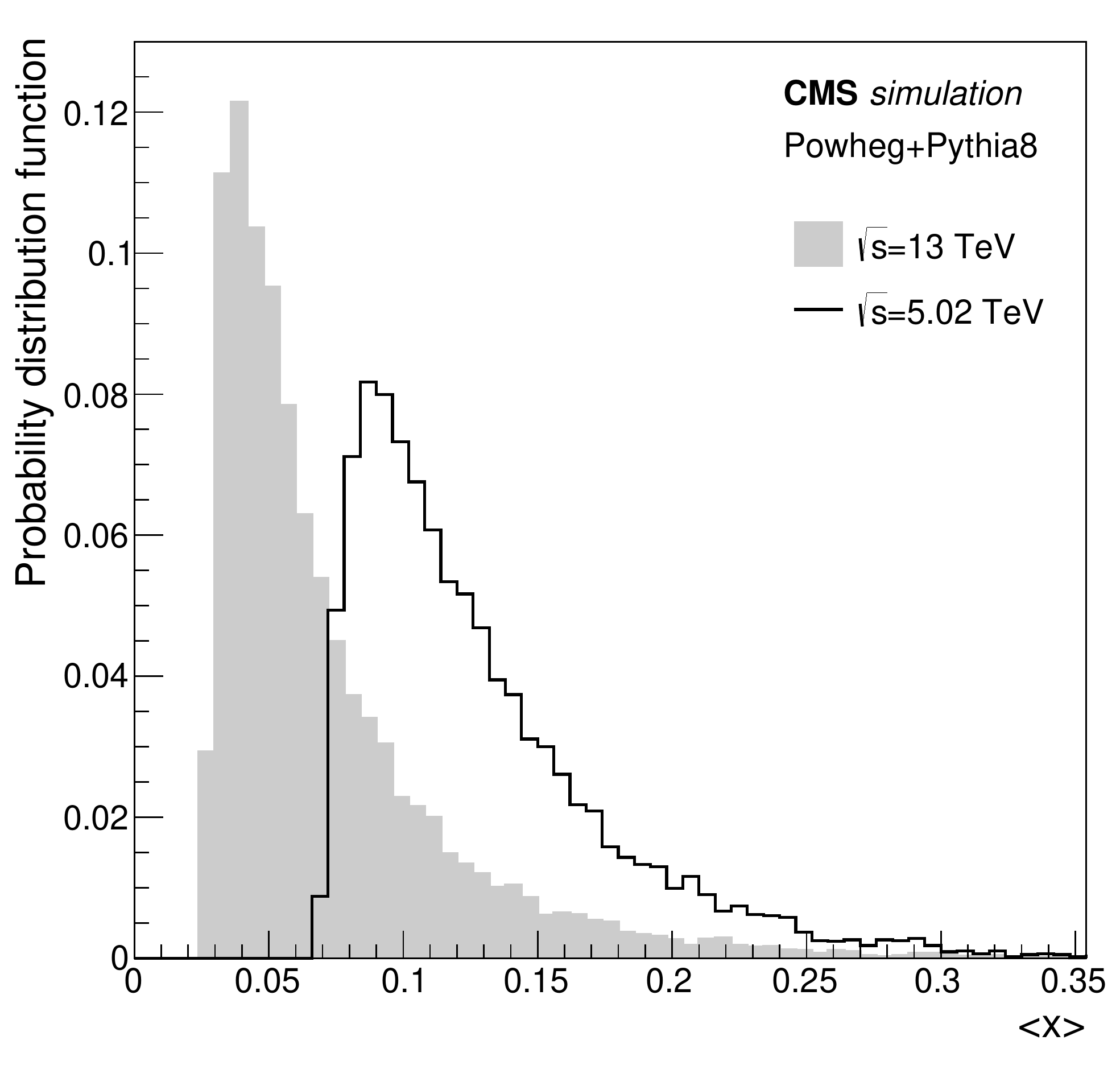}}
\end{minipage}\par\medskip

\caption{
(a) Representative fractions (in \%) of initial partonic states yielding $\rm{t\bar{t}}$ pairs in pp collisions at
$\sqrt{s}=5.02$ TeV simulated at NLO with \textsc{POWHEG}~\cite{powheg,hvq} (v2) and the NNPDF3.0~\cite{nnpdf30} NLO PDF set. 
(b) Expected average momentum $\langle x \rangle$ of the incoming partons measured with respect to the proton momentum,
for pp collisions at $\sqrt{s}=5.02$ and 13 (filled histogram) TeV~\cite{top16015}.  }
\label{fig:tt_frac}
\end{figure}

%The  most  abundant  production  process  for $\rm{t\bar{t}}$ pairs  at  hadron  colliders is QCD pair production.% 
At the Born level two processes with cross sections proportional to $\alpha_{\rm{s}}^2$ contribute leading 
to $\rm{t\bar{t}}$ final states; gluon-gluon  ($gg$)  fusion and  quark-antiquark ($\rm{q\bar{q}}$)  annihilation (Fig.~\ref{fig:tt_prod}).
The relative importance of $\rm{q\bar{q}}$- as opposed to $gg$-initiated processes is simply dictated by $\sqrt{s}$ 
and the PDFs of the initial-state hadrons.
The fraction of $\rm{t\bar{t}}$ events
initiated by $gg$ collisions grows monotonically with $\sqrt{s}$. It is around 73\% at 5.02 TeV,
as calculated with POWHEG~\cite{powheg,hvq} (v2) at NLO using the NNPDF3.0~\cite{nnpdf30} NLO PDFs, and increases to around 86\% 
at 13 TeV (Fig.~\ref{tt_frac:a}).
%This thus makes $\rm{q\bar{q}}$ annihilation more likely in $\rm{p\bar{p}}$ collisions at the Tevatron compared to
%pp collisions at the LHC.
%Most of the times, each  quark  or  gluon  
%carries a modest fraction $x$ of the total energy of the incoming hadron (Fig.~\ref{tt_frac:b}), meaning  the  collision  must  be  
%energetic enough  to  generate  top  quarks, i.e., the partonic center-of-mass energy $\sqrt{x_1x_2s}$
%should at least equal to twice the top quark mass.

At NLO, $\rm{t\bar{t}}$ production processes with cross sections proportional to $\alpha_{\rm{s}}^3$ become relevant, 
while new production channels, like ($\rm{q}g$) or ($\rm{\bar{q}}g$), open up.
Ultraviolet divergences occurring in  NLO  calculations  are  systematically treated by absorbing them into redefinitions
of the couplings in the Lagrangian, introducing the $\mu_{\rm{R}}$ scale into the calculation.  
Infrared and collinear divergences in the initial state are systematically canceled by absorbing them into the PDFs, 
introducing the $\mu_{\rm{F}}$ scale into the calculation (Section~\ref{sec:theo}).
The precision of $\rm{t\bar{t}}$ cross section calculations is further improved at NNLO, i.e., including processes up to $\alpha_{\rm{s}}^4$,
and at NNLL order, i.e., resumming contributions that are proportional to
$\alpha_{\rm{s}}^n \rm{ln(...)^{2n}}$ at the $n$-th order. 
The most precise prediction of the inclusive $\rm{t\bar{t}}$ production cross section to date (NNLO+NNLL) reaches an uncertainty of 
around 5\% (at a fixed top quark mass of 172.5 GeV), dominated by the uncertainties associated with the choice of the $\mu_{\rm{R}}$ and 
$\mu_{\rm{F}}$ scales as well as with the PDF sets~\cite{tt_NNLO}.
Comparisons of experimental measurements--already surpassed such a threshold--to the state-of-the-art theoretical predictions  
allow  to  test  the  perturbative  QCD calculations. 

\subsubsection{ Measuring it inclusively with its partner  }
\label{sec:tt_incl}

\begin{figure}[!ht]
  \centering
  \includegraphics[scale=0.55]{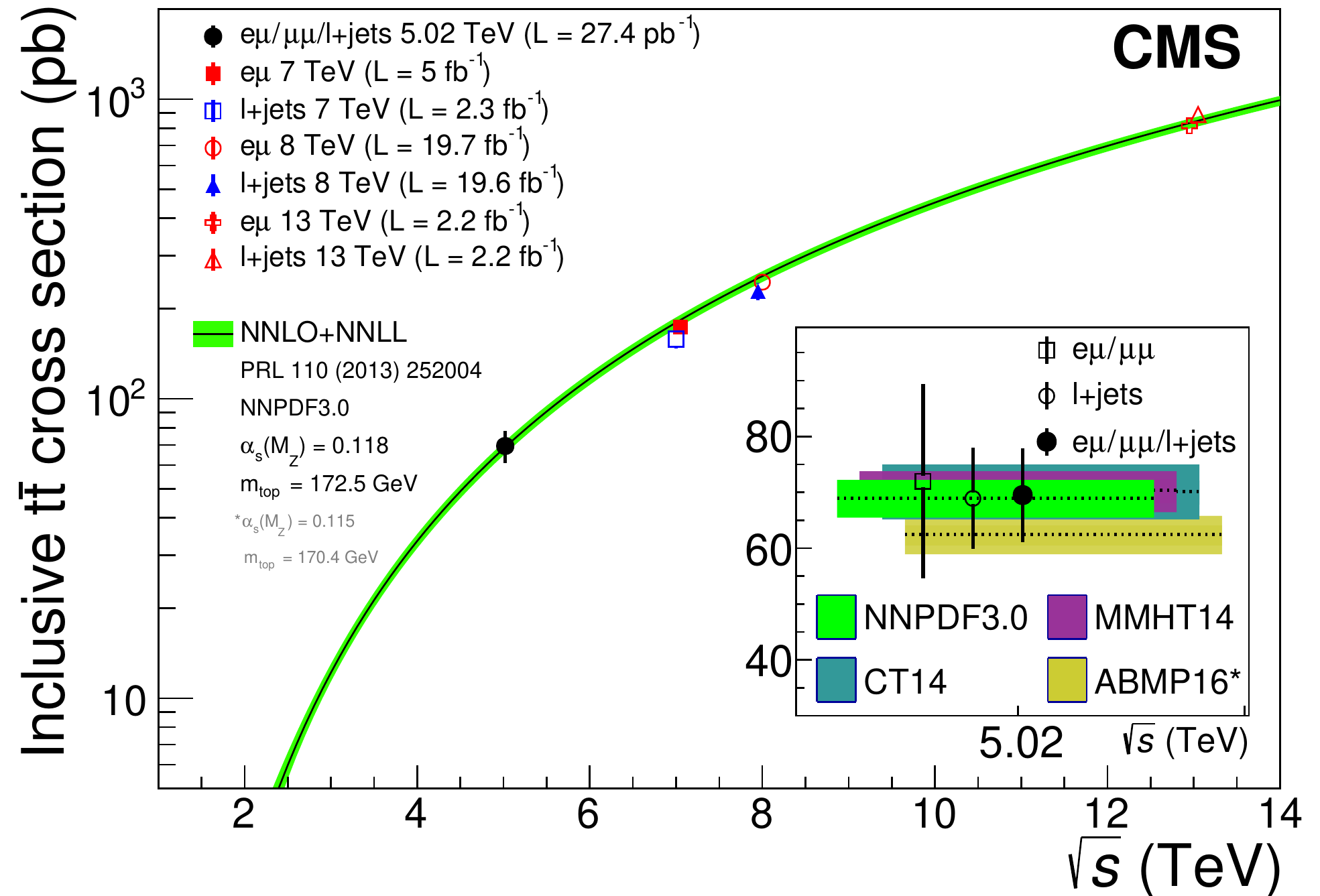}
    \caption{
       Inclusive $\sigma_{\rm{t\bar{t}}}$ in  pp  collisions  as  a  function  of  the  center-of-mass  energy;  
       the most precise CMS measurements at $\sqrt{s}=7$, 8~\cite{dilep_78,ljets_78}, and 13~\cite{dilep_13,ljets_13} TeV in the separate dilepton and $\ell$+jets 
       channels are displayed, along with the combined measurement at 5.02 TeV~\cite{top16023}.
       The NNLO+NNLL theoretical prediction~\cite{tt_NNLO} using the NNPDF3.0~\cite{nnpdf30} NNLO PDF set with $\alpha_{\rm{s}}(M_{\rm{Z}})=0.118$ and
       $m_{\rm{top}}=172.5$ GeV is shown in the main plot.  In the inset, additional predictions at $\sqrt{s}=5.02$ TeV
       using the MMHT14~\cite{mmht14}, CT14~\cite{ct14}, and ABMP16~\cite{abmp16} PDF sets, the latter with $\alpha_{\rm{s}}(M_{\rm{Z}})=0.115$ and $m_{\rm{top}}=170.4$,
       are compared, along with the NNPDF3.0 prediction, to the individual and combined results from the same analysis. 
       The vertical bars and bands represent the total uncertainties in the data and in the predictions, respectively~\cite{top16023}.
}
    \label{fig:sqrts}
\end{figure}

Precise measurements of inclusive $\sigma_{\rm{t\bar{t}}}$ have been performed at all $\sqrt{s}$ and in multiple collision systems 
delivered by the LHC. The most precise results in pp collisions are summarized in Fig.~\ref{fig:sqrts}.  
All reported measurements are in excellent agreement with the theoretical predictions.  
Thanks to the high integrated luminosity recorded, in particular for the higher $\sqrt{s}$, 
those measurements are not limited by the statistical uncertainty in the available data samples, 
but they are rather dominated by the systematic variations.
Less precise measurements in the all-hadronic channel~\cite{hadr_7,hadr_8,hadr_13} and in final states involving hadronically-decaying
$\tau$ leptons~\cite{tau_7,tau_8} have also been reported.
Despite the large branching fraction, the $\sigma_{\rm{t\bar{t}}}$ measurement in the all-hadronic channel
poses the challenge of an overwhelming QCD multijet background. 
This type of background bears a major contribution to final states containing hadronically decaying $\tau$ leptons,
whose reconstruction is also susceptible to misidentification.  

The  decay channels  with  the  best  precision so far are  the dilepton ($\rm{e}^\pm \mu^\mp$) and $\ell$+jets ($\ell=\rm{e},\mu$)  final  states  with  either  
one electron ($\rm{e}$) or muon ($\mu$). Although the branching ratio is small, the dilepton channel 
suffer less from  backgrounds  uncertainties;  Drell-Yan contribution can be heavily  suppressed  by
the  requirement  of  opposite  lepton  flavor and in the presence of high jet multiplicity.   
While  the initial $\sigma_{\rm{t\bar{t}}}$ measurements(e.g.~\cite{dilep_7_init1,dilep_7_init2,dilep_7_init3}) have been based  on simple counting  experiments,  
the  subsequent application of profile likelihood fits, 
simultaneously relying on the number of events counted in the different event categories, has considerably helped to reduce the impact of systematic variations.
Indeed,  by  accounting for  the  main  experimental uncertainties,  such  as  the  jet  energy  scale,  in  terms  of  auxiliary (``nuisance'')
fit parameters, the choice of appropriate categories (e.g. based on the number of b-tagged jets  and  the  number  of  additional non-b-tagged  
jets) and/or distributions (jet $p_{\rm{T}}$ spectrum of the non-b-tagged jet with the lowest $p_{\rm{T}}$) led to a reduction of 
the sensitivity to those parameters~\cite{dilep_78}.  

\begin{figure}
\begin{minipage}{.5\linewidth}
\centering
\subfloat[]{\label{ljets_13:a}\includegraphics[scale=0.4]{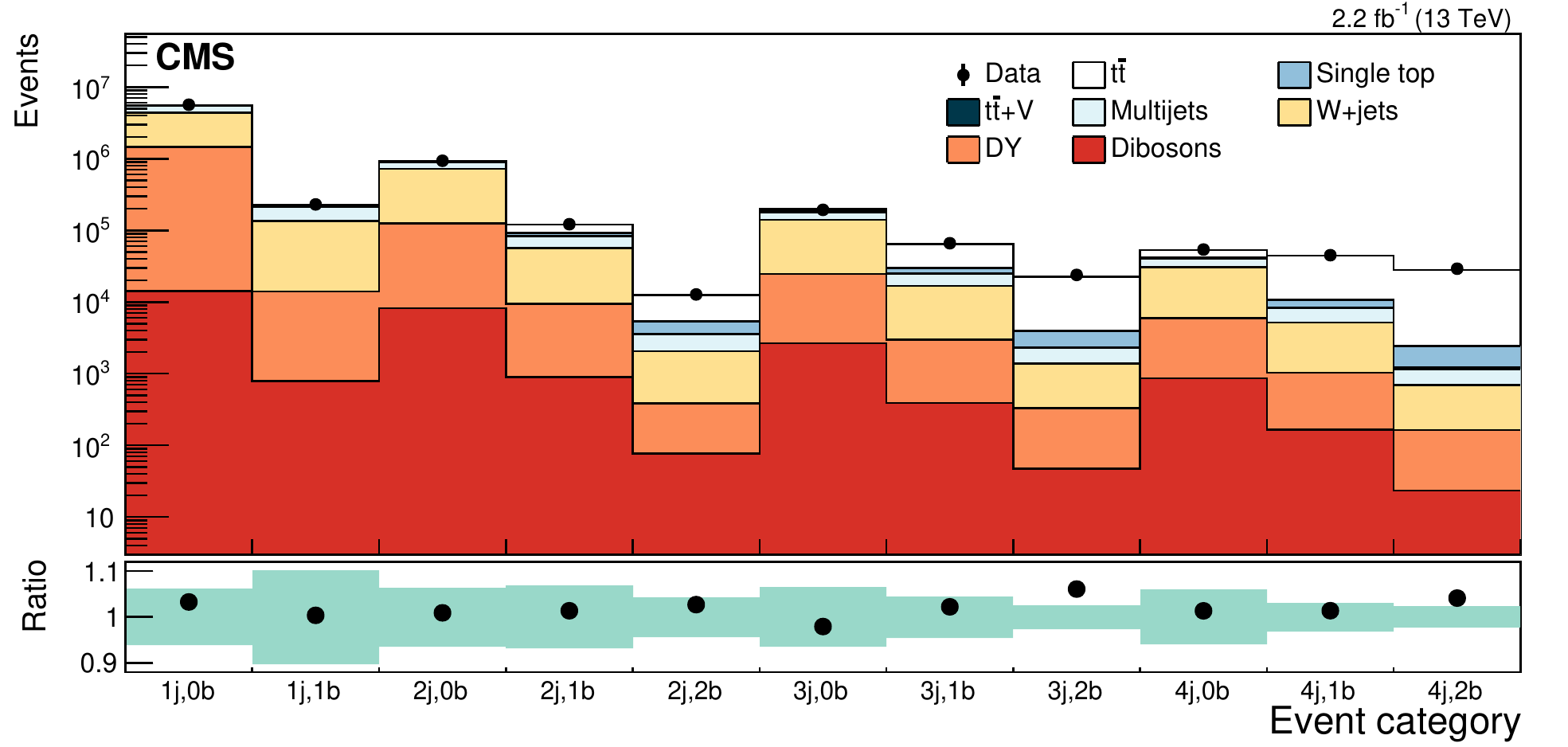}}
\end{minipage}%
\begin{minipage}{.5\linewidth}
\centering
\subfloat[]{\label{ljets_13:b}\includegraphics[scale=0.35]{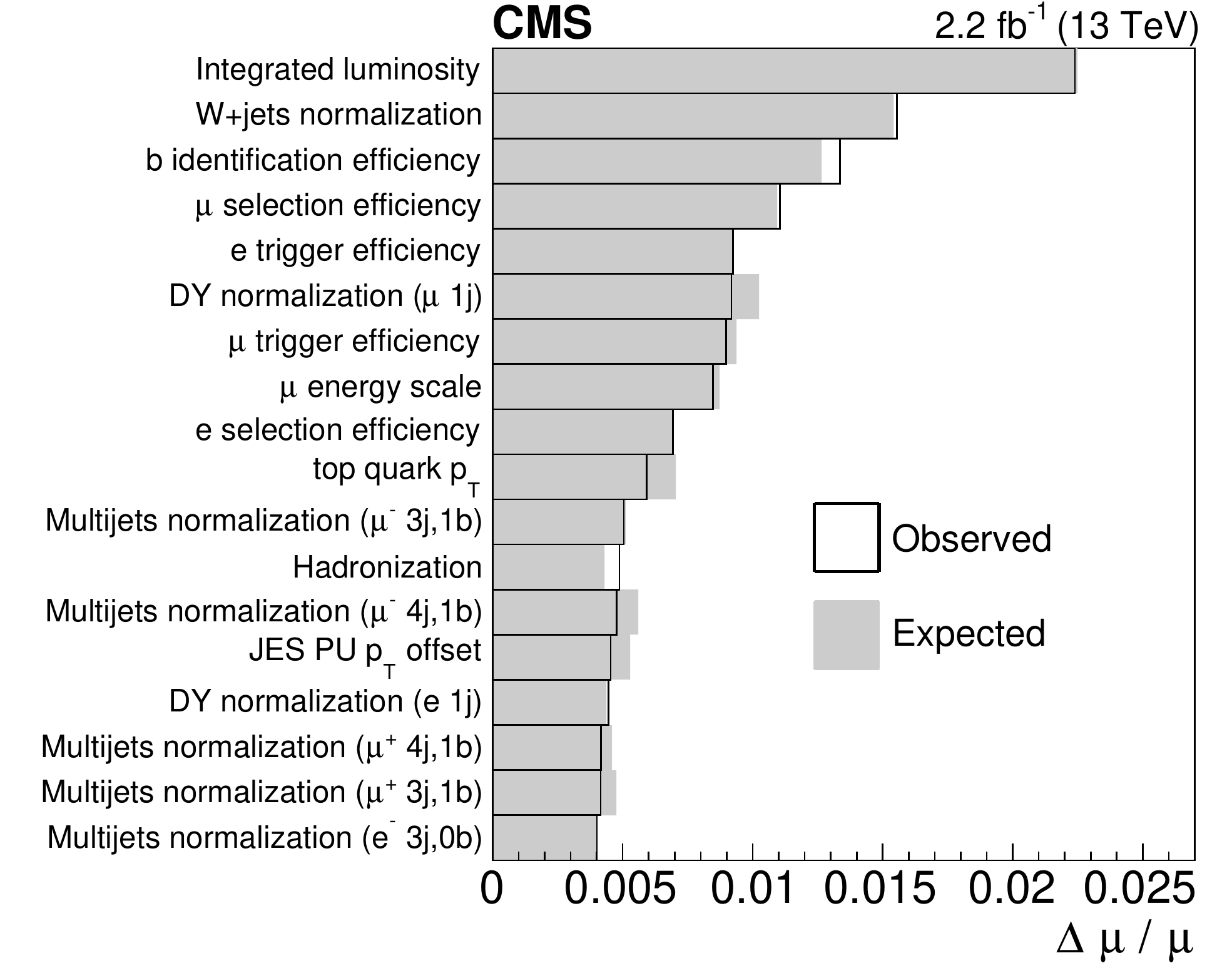}}
\end{minipage}\par\medskip

\caption{
(a) 
Event yields from data and the expected \ttbar signal and backgrounds for each of the independent event categories 
of jet and b tagged jet multiplicity considered in the latest $\ell$+jets analysis at $\sqrt{s}=13$ TeV.
Distributions are combined for the two lepton charges and flavors. 
The bottom panel shows the ratio between the data and the expectations with the relative uncertainty (shaded band)
owing to the statistical uncertainty in the simulations and the systematic uncertainty in the total integrated luminosity.
(b) 
Estimated change in the parametrized ratio of the measured cross section to the reference theoretical value, $\Delta\mu/\mu$,
stemming from the listed experimental and theoretical sources of uncertainties in the same analysis. 
The various contributions are shown from the largest to the smallest observed impact on $\mu$~\cite{ljets_13}.  }
\label{fig:ljets_13}
\end{figure}

The $\ell$+jets channel is less hampered by the statistical limitation of the data sample,  yet it brings additional
backgrounds from the EW production of top quarks, W+jets and QCD multijet production, that are typically estimated using 
data-driven techniques. The latest analysis at $\sqrt{s}=13$ TeV~\cite{ljets_13} relied on 
a profile likelihood fit to all events with one isolated lepton, significant missing transverse 
momentum and at least one reconstructed jet, while the jet and b-tagged jet multiplicity was used to categorize the events.  
As shown in Fig.~\ref{fig:ljets_13} (a), the low multiplicity categories are dominated by backgrounds, allowing
both their normalization and shape (e.g. the  invariant  mass  of  the  lepton  and b-tagged  jet) to be determined from the fit 
procedure. The large number of event categories with varying
signal and background contributions allowed the nuisance parameters to be strongly constrained by the data, 
rendering the uncertainty in the integrated luminosity as the largest single source of systematic uncertainty (Fig.~\ref{fig:ljets_13}, right).
The typical assumption that the model should correctly capture all the sources of systematic
uncertainties, along with their correlations across different event categories and distributions, is implicitly  made.

Theoretical predictions for ratios of $\sigma_{\rm{t\bar{t}}}$ at different $\sqrt{s}$ values of a and
b, $R(\rm{a}/\rm{b})$, can benefit from cancellations in the total uncertainty, e.g., $R(8/7)$ is calculated to be 
$1.430\pm0.013$ at NNLO+NNLL~\cite{tt_NNLO} accuracy. The total relative uncertainty is below 1\%, as the  uncertainties in $\sigma_{\rm{t\bar{t}}}$ at
the  two $\sqrt{s}$ values are  highly correlated. 
In addition to the quoted PDF~\cite{pdf4lhc}, $\alpha_{\rm{s}}$, and $\mu_{\rm{R}}$, $\mu_{\rm{F}}$ scale uncertainties, $R(8/7)$ varies
by less than 0.1\% for a $\pm1$ GeV variation in $m_{\rm{top}}$.
A series of systematic uncertainties also cancel in the experimentally-determined ratios, especially for measurements 
performed with the very same technique. The most precise CMS measurement of this kind is currently $R(8/7)=1.41\pm0.06$~\cite{dilep_78},
obtained after a partial cancellation of systematic uncertainties, consistent with the predicted ratio.

\clearpage
\subsubsection{ Closing the gap between Tevatron and LHC energies }
\label{sec:tt_incl_5TeV}

\begin{figure}[htp]
\centering
\includegraphics[width=0.3\textwidth]{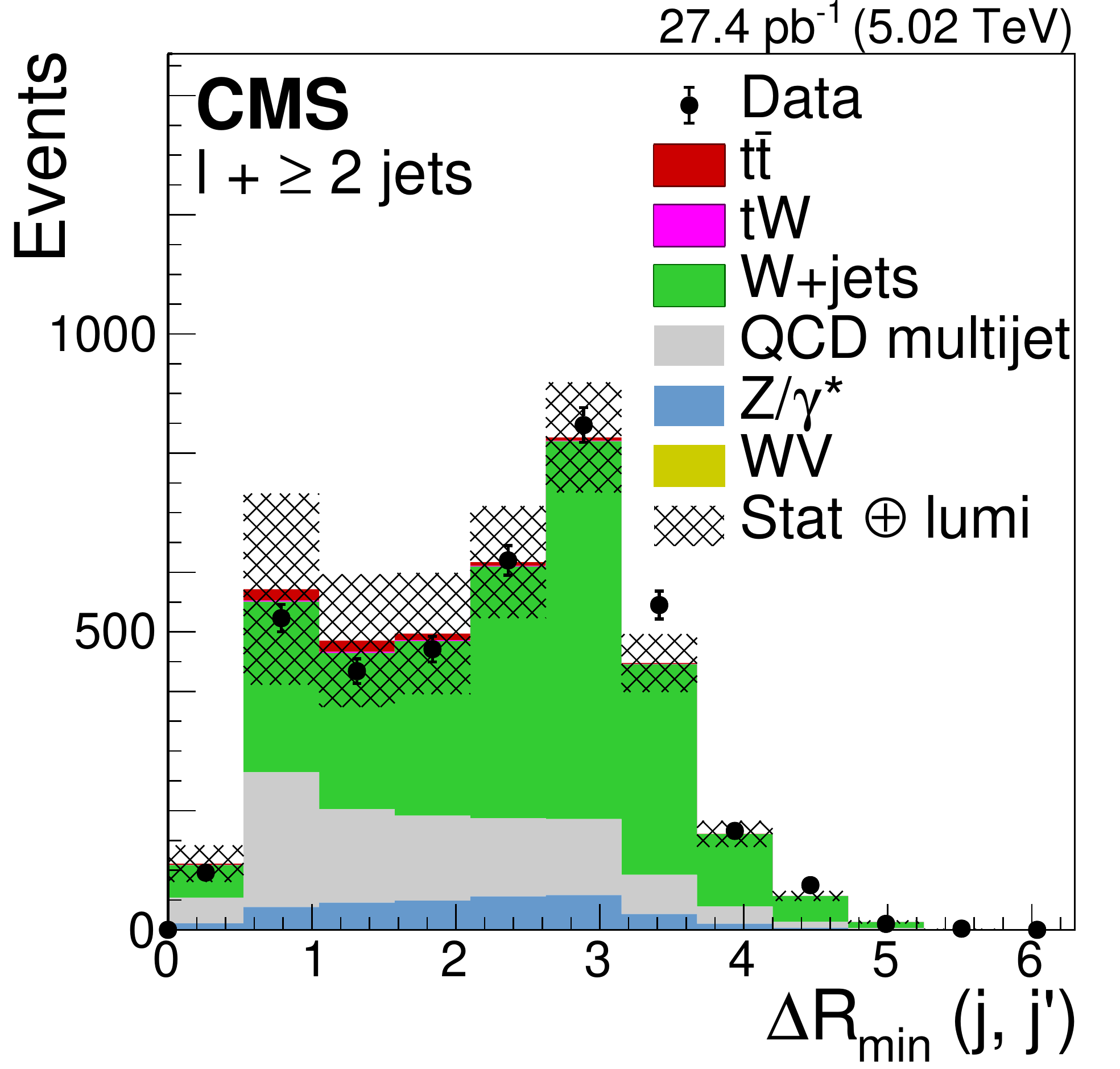}
\includegraphics[width=0.3\textwidth]{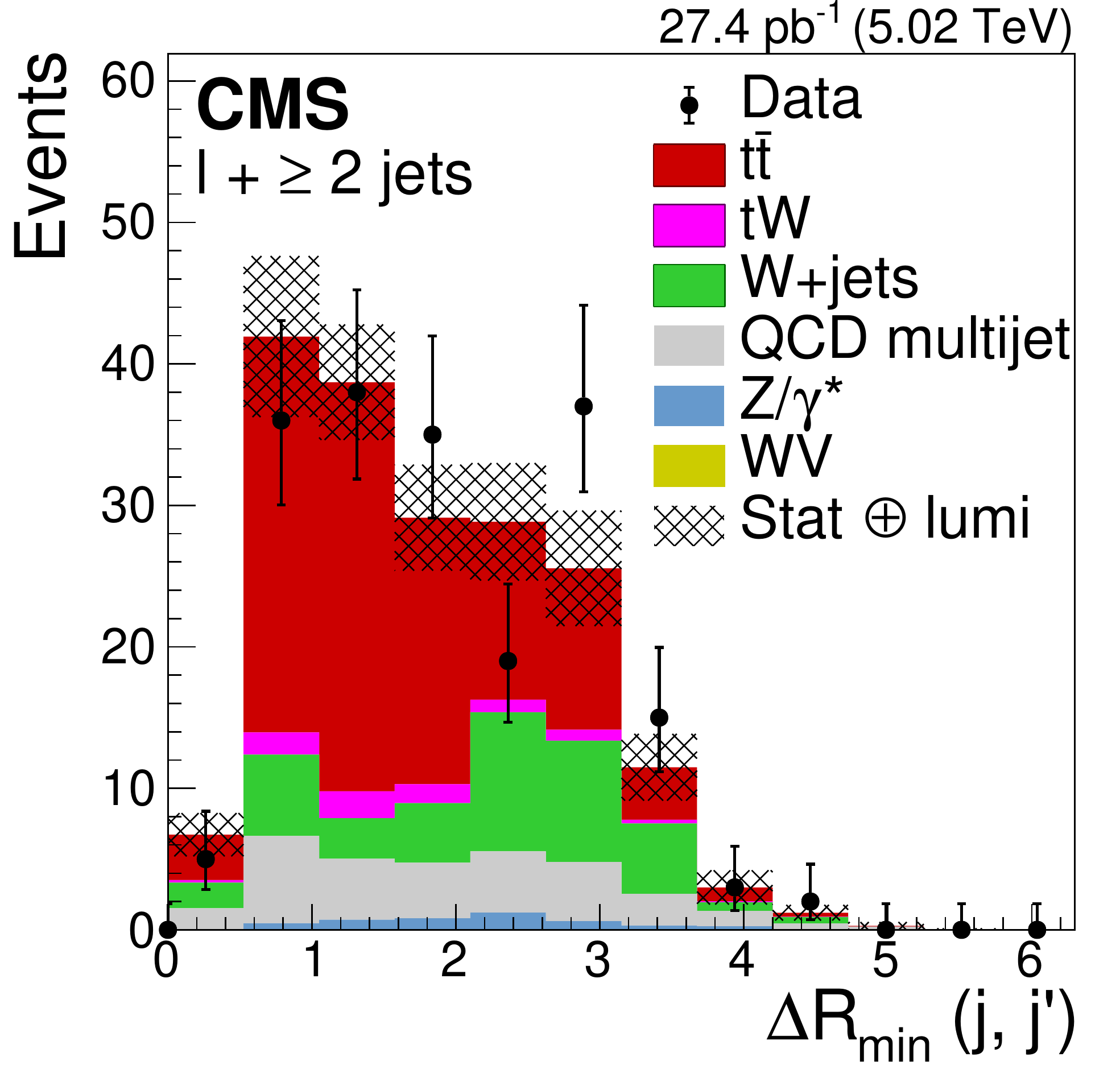}
\includegraphics[width=0.3\textwidth]{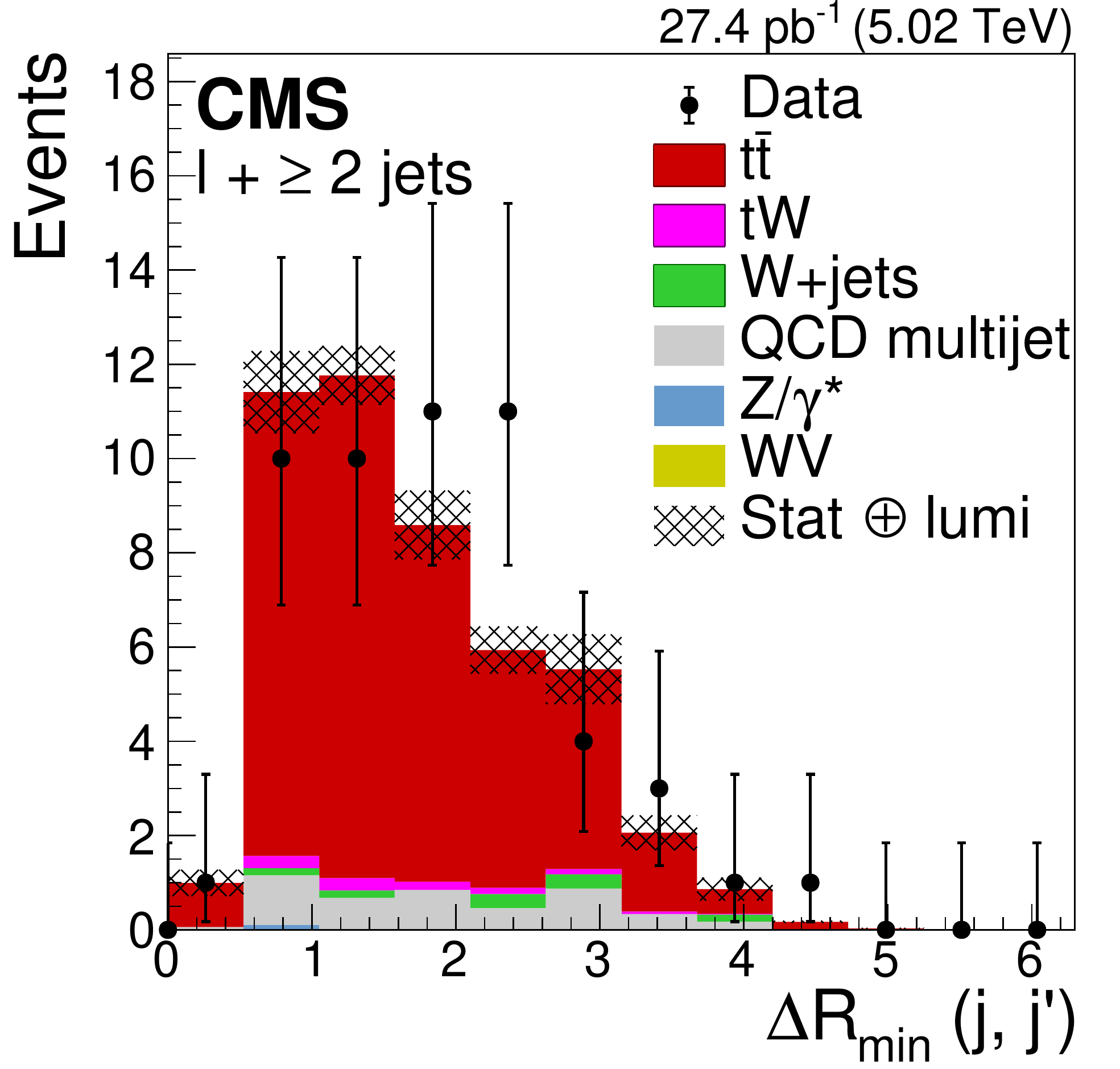}
\caption[]{Distribution of the angular distance $R=\sqrt{\eta^2+\phi^2}$ of the least separated jets, $\min\Delta R(j,j')$, for $\ell$+jets events
in 0b{}- (left), 1b{}- (center) and $\geq$2b{}- (b)  tagged jet categories.
The distributions observed in the data are compared to the sum of the expectations for the signal and backgrounds prior to the fit.
The shaded band represents the statistical and integrated luminosity uncertainties on the expected signal and background yields~\cite{top16023}.
}
\label{fig:top16023}
\end{figure}

In November 2015, the LHC delivered pp collisions at $\sqrt{s}=5.02$ TeV, producing an integrated luminosity of $27.4\pm0.6$ pb$^{-1}$~\cite{lum16001}.
Measurements of $\rm{t\bar{t}}$ production at various $\sqrt{s}$ probe different values of the fraction $x$ of the proton longitudinal 
momentum, and thus provide complementary information on the gluon PDF (Fig.~\ref{tt_frac:b}).
The combined measurement using dilepton and $\ell$+jets channels (at the level of measured values instead of likelihood functions) achieved 
a total relative uncertainty of 12\%~\cite{top16023}, that represents a remarkable achievement and a significant improvement relative to 
the first observation using the $\rm{e}^\pm \mu^\mp$ channel alone~\cite{top16015}. Indeed, the correlation in phase space of the light jets carries a distinctive hallmark with respect to the main backgrounds that are 
controlled by counting the number of b-tagged jets in the selected $\ell$+jets events. The signal extraction has been then
performed by maximizing a profile likelihood fit to the distribution of a kinematic variable (Fig.~\ref{fig:top16023}), sensitive to the resonant
behavior of the light jets, for different categories of lepton flavor and jet multiplicity.
Analogously to the most recent LHC studies of the inclusive $\sigma_{\rm{t\bar{t}}}$,
the measurement is first performed in a fiducial phase space--a restricted region of the phase space
that closely resembles the detector acceptance in $p_{\rm{T}}$ and $\eta$ of leptons and jets--and it is then extrapolated to the full phase space based on MC simulation.
The individual and combined results at $\sqrt{s}=5.02$ TeV are separately compared to the predictions for the ABMP16~\cite{abmp16}, CT14~\cite{ct14}, MMHT14~\cite{mmht14}, 
and NNPDF3.0~\cite{nnpdf30} PDF sets. The nice agreement among them can be traced back to consistent values of $\alpha_{\rm{s}}$ and $m_{\rm{top}}$, 
that are associated with the respective PDF set.
The limited-precision measurement can be complemented with the significantly larger data set recorded in 2017, 
equivalent to almost $0.3$ fb$^{-1}$~\cite{evian_arg}.

\subsubsection{ Hot off the press; now also in nuclear collisions }
\label{sec:tt_incl_pPb}

Until recently, top quark measurements remained out of reach in nuclear collisions~\cite{david} due to the reduced amount
of integrated luminosity produced during the first period at the LHC, and the relatively low nucleon-nucleon center-of-mass energies ($\sqrt{s_{\rm{NN}}}$)
available at the BNL RHIC. 
Novel studies of top quark cross sections have finally become feasible with the 2016 LHC proton-lead (pPb) run at $\sqrt{s_{\rm{NN}}}=8.16\ \rm{TeV}$,
surpassing in performance almost eight times the designed instantaneous luminosity and delivering to CMS 174~nb$^{-1}$ of pPb collision data.   

\begin{figure}
\begin{minipage}{.5\linewidth}
\centering
\subfloat[]{\label{pPb:a}\includegraphics[scale=0.25]{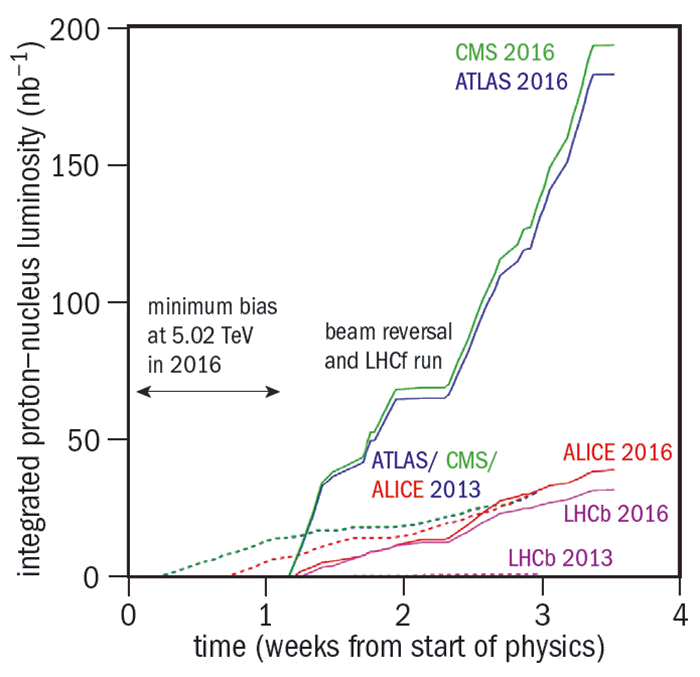}}
\end{minipage}%
\begin{minipage}{.5\linewidth}
\centering
\subfloat[]{\label{pPb:b}\includegraphics[scale=0.215]{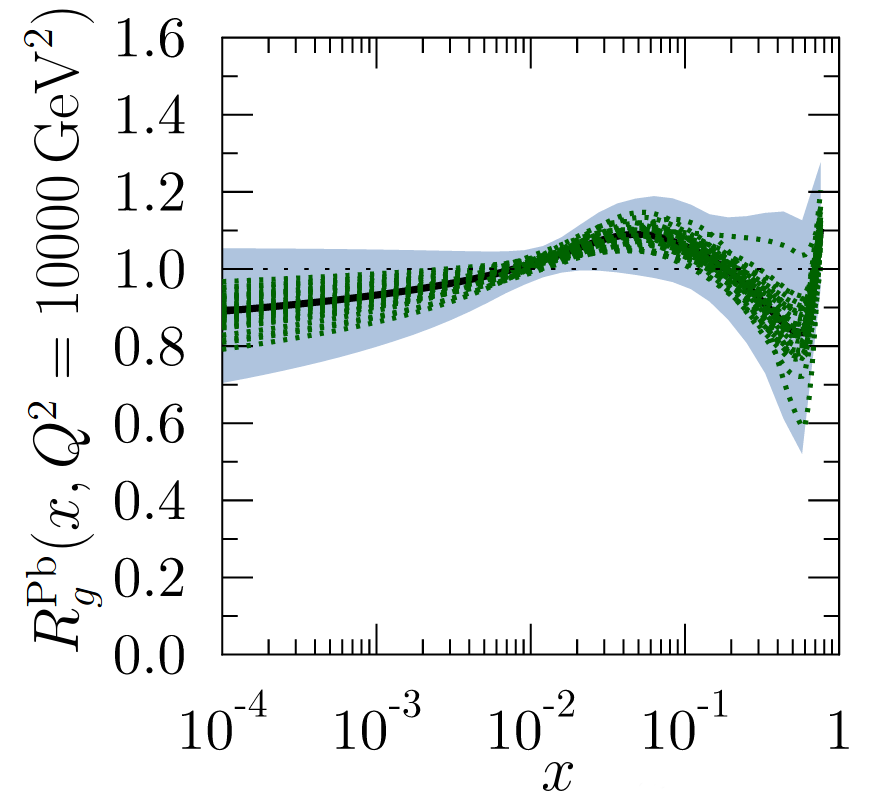}}
\end{minipage}\par\medskip

\caption{
(a) Accumulation of integrated luminosity in each LHC experiment in the pPb runs for years 2013 and 2016~\cite{Jowett:IPAC2017}.
(b) The EPPS16 nuclear modifications for the bound gluon PDF in Pb nucleus at the parametrization scale $Q^2=10^4$~$\rm{GeV^2}$.
The thick black curves correspond to the central fit values, the dotted curves to the individual error sets, and 
the total uncertainties are shown as blue bands~\cite{Eskola}.}
\label{fig:pPb}
\end{figure}

Asymmetric collisions of $^{208}\rm{Pb}^{82+}$ nuclei with protons had not been included in the initial LHC design.
However, unexpected discoveries in small collision systems, reminiscent of flow-like collective phenomena, engaged further investigations~\cite{Salgado}.
After the short, yet remarkable, pilot physics run in 2012, and the first full one-month run in early 2013, 
the second full pPb run took place in late 2016, in order to deliver data at $\sqrt{s_{\rm{NN}}}=8.16\ \rm{TeV}$ for each direction of the beams. 
Apart from complex bunch filling schemes due to the generation of the beams from two separate injection paths, 
the distinct feature of operation with asymmetric collisions at LHC is the difference in revolution and cavity frequencies.
Given that the colliding bunches have significantly different size and charge, both beams are displaced transversely, 
onto opposite-sign off-momentum orbits, and longitudinally, to restore collisions at the proper interaction points, 
during a process colloquially known as ``cogging''~\cite{Jowett:IPAC2017}. 
The long-term integrated luminosity goal of 100~nb$^{-1}$ has been clearly surpassed (Fig.~\ref{pPb:a}), 
rendering the 2016 pPb run the baseline for several years.

The binding energies of nucleons in the nucleus are several orders of magnitude smaller than the momentum transfers of deep-inelastic 
scattering (DIS). Therefore, such a ratio naively should be close to unity, except for small corrections for the Fermi motion of nucleons in the nucleus. 
Contrary to expectations, the EMC experiment discovered a declining slope to the ratio~\cite{EMC}, a result later confirmed with high-precision 
electron- and muon-scattering DIS data. 
The conclusions from the combined experimental evidence demonstrates the consistency in modifying the free nucleon PDFs, $f_{i}^{\rm{p}}(x,Q^2)$, at low $Q^2$ and 
letting the Dokshitzer-Gribov-Lipatov-Altarelli-Parisi evolution to take care of the momentum transfer dependence, i.e., 
$f_{i}^{\rm{p/A}}(x,Q^2)=R_{i}^{A}(x,Q^2)f_{i}^{\rm{p}}(x,Q^2)$, where $R_{i}^{A}(x,Q^2)f_{i}$ is  the  scale-dependent  nuclear 
modifications encoded in nPDFs (Fig.~\ref{pPb:b}).

\begin{figure}
\begin{minipage}{.5\linewidth}
\centering
\subfloat[]{\label{masses:a}\includegraphics[scale=0.38]{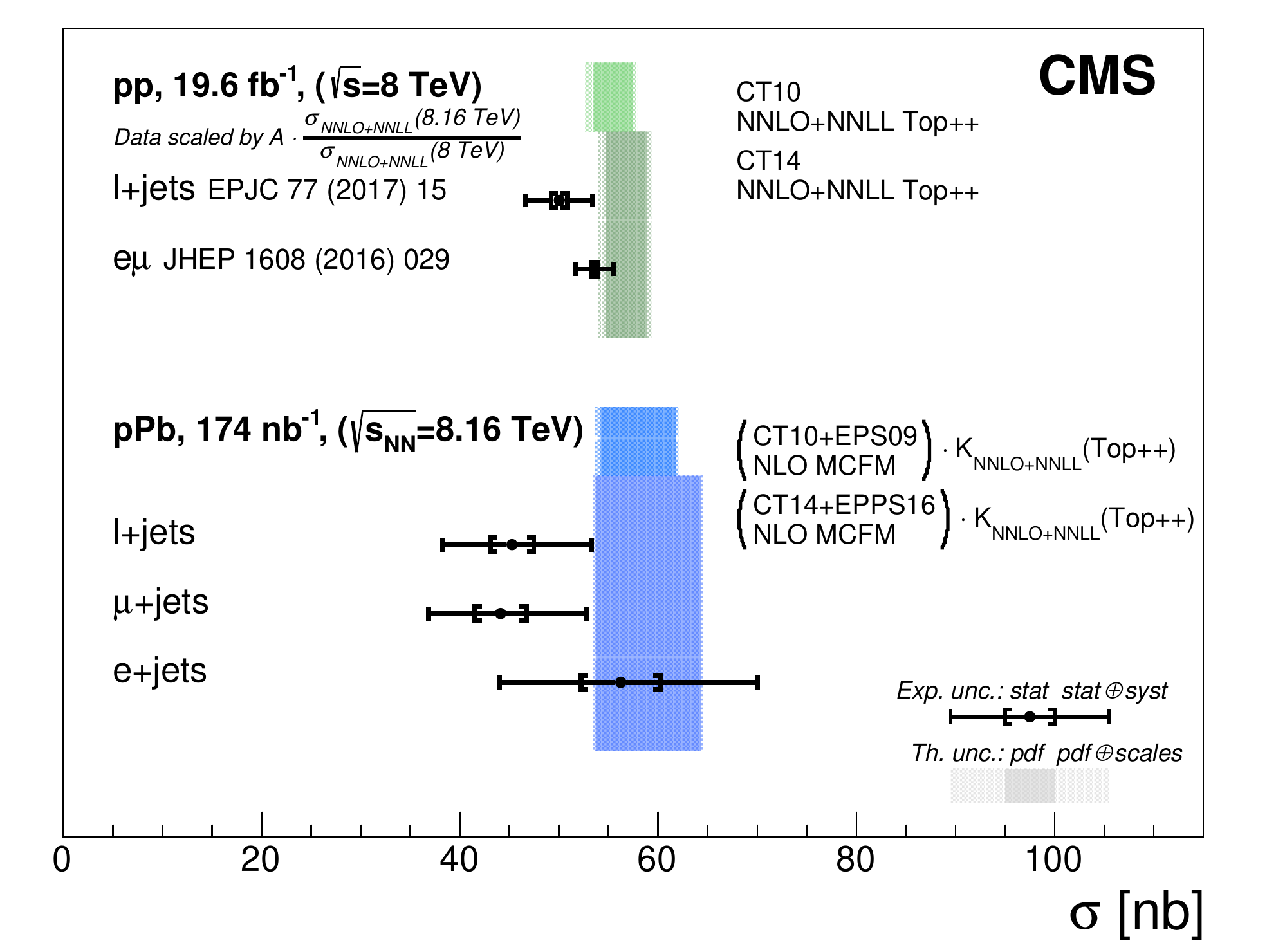}}
\end{minipage}%
\begin{minipage}{.5\linewidth}
\centering
\subfloat[]{\label{masses:b}\includegraphics[scale=0.3]{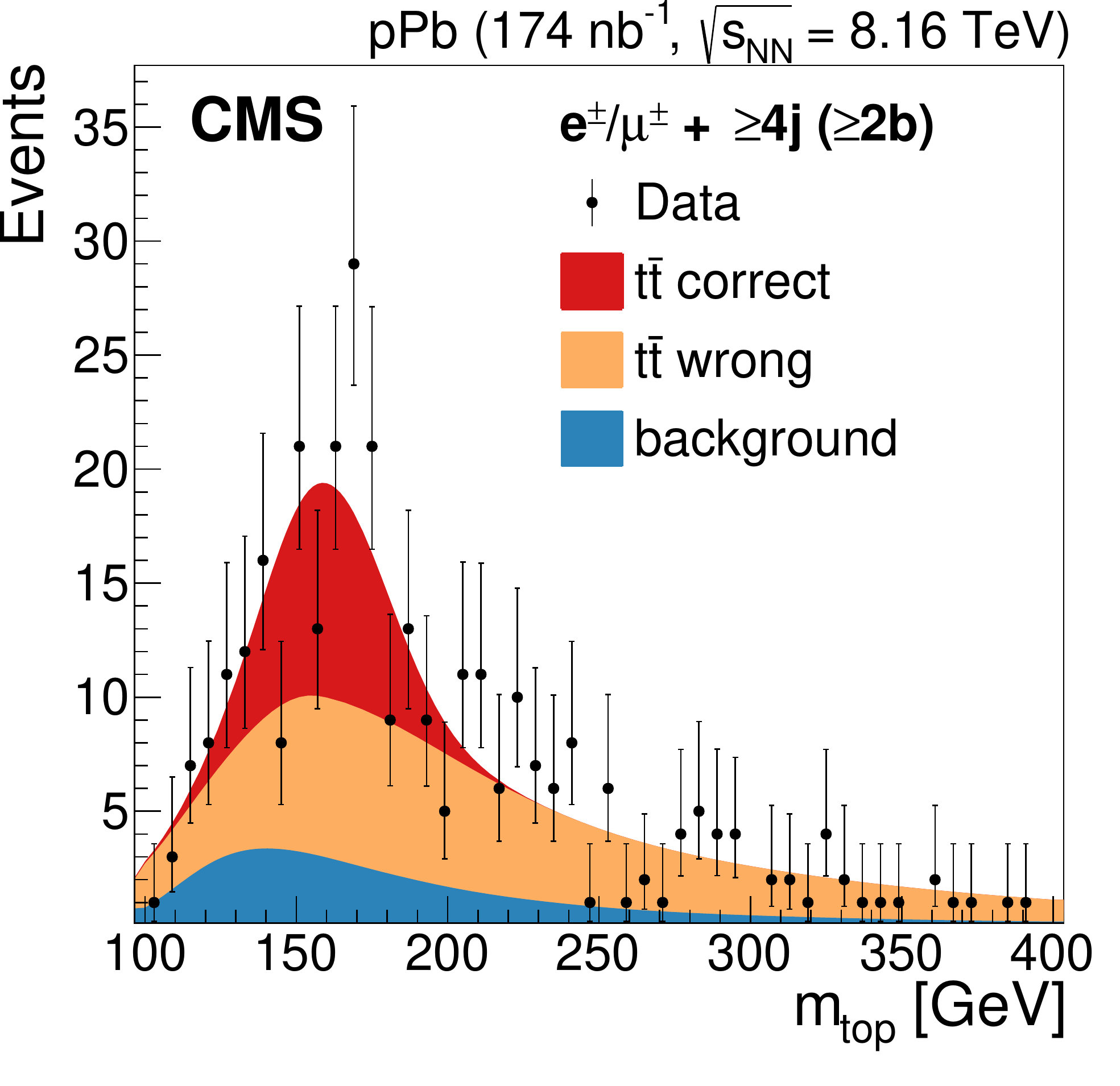}}
\end{minipage}\par\medskip

\caption{
(a) Top quark pair production cross section in scaled pp and pPb collisions collisions at $\sqrt{s_{\rm{NN}}}=8.16\ \rm{TeV}$; 
the CMS measurements~\cite{dilep_78,ljets_78,ttpPb} are compared to the NNLO+NNLL theory predictions~\cite{tt_NNLO} using state-of-the-art nPDFs~\cite{Eskola} with baseline proton PDFs~\cite{ct14}.}
(b) Invariant mass distributions of the $\rm{t} \to \mathrm{j} \mathrm{j}^{\prime} b$ candidates
in the $\geq$2 b-tagged jet category after all selections.
All signal and background parameters are kept fixed to the outcome of the $m_{\mathrm{j} \mathrm{j}^{\prime}}$ fit~\cite{ttpPb}.
\label{fig:masses}
\end{figure}

The top pair production cross section has been measured for the very first time in 
proton-nucleus collisions~\cite{ttpPb}, using pPb data at $\sqrt{s_{\rm{NN}}}=8.16\ \rm{TeV}$ with a total integrated luminosity of 174~nb$^{-1}$.
The measurement is performed by analyzing events with exactly one isolated lepton and at least four jets, 
and minimally relies on assumptions derived by simulating signal and background processes.
The resonant nature of the invariant mass of $\mathrm{j} \mathrm{j}^{\prime}$, $m_{\mathrm{j} \mathrm{j}^{\prime}}$, 
provides a distinctive feature of the signal with respect to the main backgrounds, i.e., from QCD multijet and W+jets processes.
The significance of the $\rm{t}\bar{\rm{t}}$ signal against the background-only hypothesis is above five standard deviations.
The measured cross section is $45\pm 8$~nb, 
consistent with the expectations from scaled pp data as well as perturbative quantum chromodynamics calculations (Fig.~\ref{masses:a}).
To further support the hypothesis that the selected data are consistent with the production of top quarks, 
a ``proxy'' of the top quark mass is constructed as the invariant mass of candidates formed by pairing the W candidate 
with a b-tagged jet (Fig.~\ref{masses:b}).
This first study clearly paves the way for further detailed investigations of top quark production in nuclear interactions (e.g.~\cite{liliana}),
providing in particular a new tool for studies of the hot and dense matter created in nucleus-nucleus collisions.

\clearpage
\subsubsection{ Just the scale has never been enough }
\label{sec:tt_diff}

\begin{figure}
\begin{minipage}{.3\linewidth}
\centering
\subfloat[]{\label{undfold:a}\includegraphics[scale=0.11]{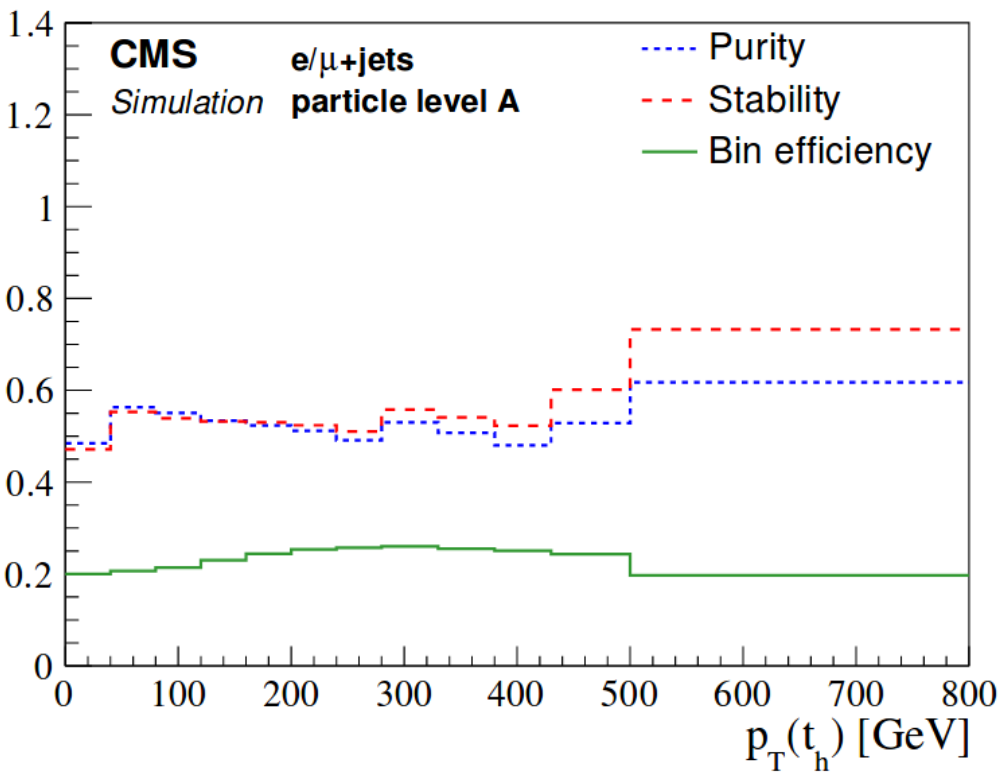}}
\end{minipage}%
\begin{minipage}{.3\linewidth}
\centering
\subfloat[]{\label{unfold:b}\includegraphics[scale=0.11]{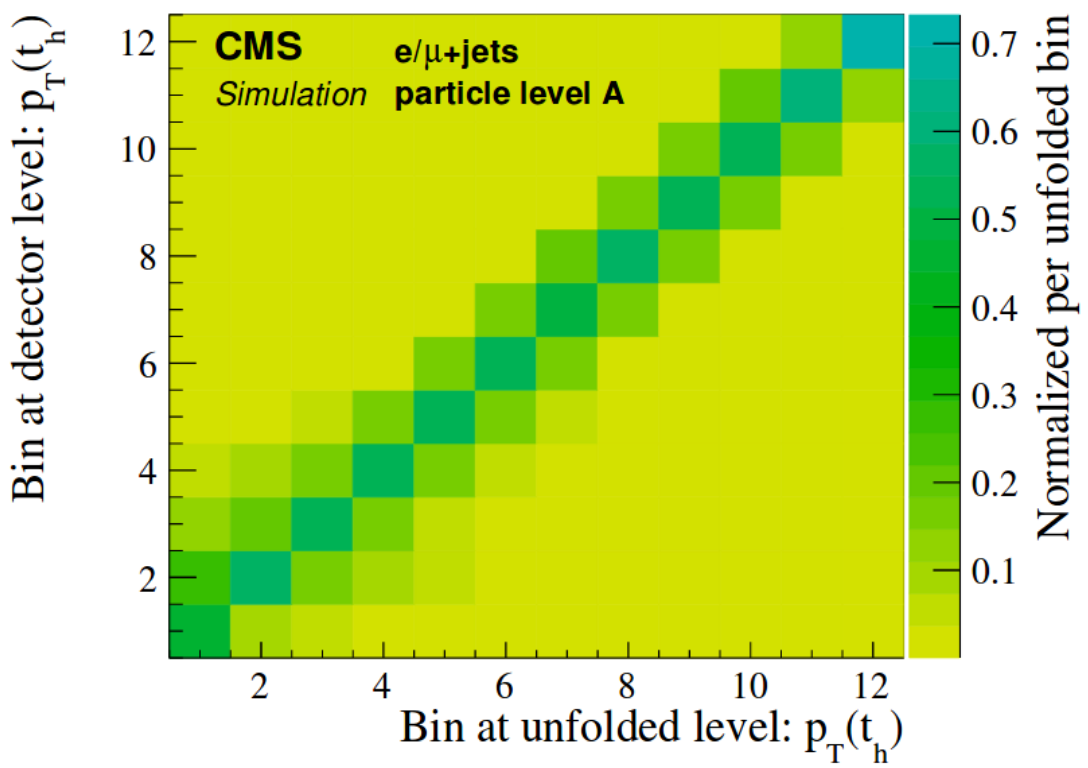}}
\end{minipage}
\begin{minipage}{.3\linewidth}
\centering
\subfloat[]{\label{unfold:c}\includegraphics[scale=0.11]{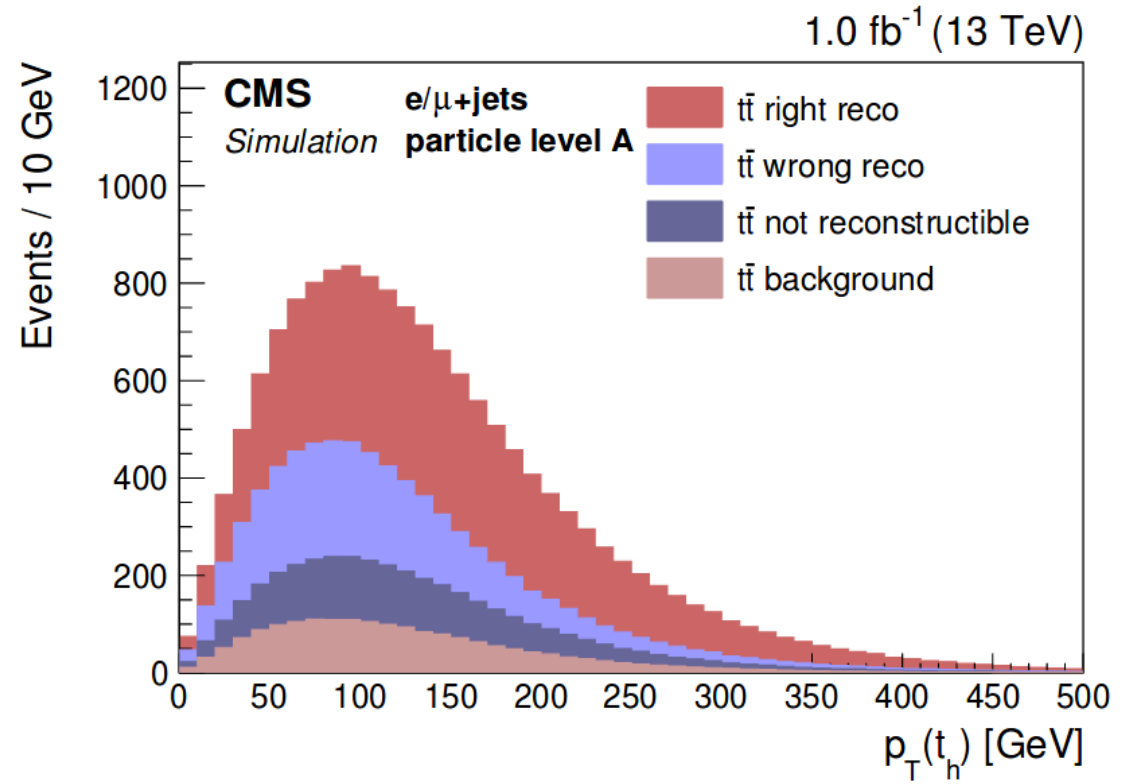}}
\end{minipage}%\par\medskip

\caption{
Representative metrics for estimating the sensitivity to the underlying top quark physics~\cite{particle}.
Ideally the most pure, stable, and efficient approach (a)
should reproduce as closely as possible the kinematic properties generated at parton level (b),
while it should be resilient against large corrections for the experimentally reconstructed quantities (c).
Exemplary distribution for the transverse momentum spectrum of the hadronically decaying top quark, $p_{\rm{T}}(\rm{t}_{\rm{h}})$, 
in the detector (smeared) phase space level in the $\ell$+jets channel,
extracted from a \textsc{POWHEG}~\cite{powheg,hvq} (v2) interfaced to \textsc{PYTHIA}~\cite{pythia8} (v8.205) full simulation~\cite{geant4} of the CMS detector.}
\label{fig:unfold}
\end{figure}

Although inclusive $\sigma_{\rm{t\bar{t}}}$ measurements already provide stringent tests of the theoretical dependence on the initial 
state and $\sqrt{s}$, the study of $\sigma_{\rm{t\bar{t}}}$ as  a  function  of  various  kinematic  properties
of analysis objects (differentially) allow for more detailed insights into the $\rm{t\bar{t}}$ production 
mechanism. Measurements of the distribution of kinematic observables can be translated into differential cross sections using
unfolding techniques (Fig.~\ref{fig:unfold}). Such procedures allegedly	correct the observed distributions for efficiency, acceptance and resolution effects. 
After all detector-related effects are alleviated by unfolding, differential cross sections from different experiments,
if performed in the same fiducial phase space, can be directly compared among each other and with predictions from 
MC event generators (at particle and  parton level)  or  from  theoretical  calculations (at parton level). 
The  directly  measured  observables, e.g., the  kinematic  distributions  of  leptons,  are  corrected 
back to the level of ``stable'' particles, which are accessible in MC generators in a fiducial region of the phase space.
On the contrary, the kinematics of the top quarks and antiquarks or of the $\rm{t\bar{t}}$ system are
well defined on parton, i.e, before decay but after gluon and photon radiation, rather than on particle level.

Differential  cross section results can be also normalized to the measured inclusive or fiducial $\sigma_{\rm{t\bar{t}}}$.
In this way normalization uncertainties, e.g., the luminosity uncertainty,  cancel and the sensitivity of
the measurement to the shape of kinematic distributions is improved:

\begin{align*}
\frac{1}{\sigma_{\rm{t\bar{t}}}}(\frac{d\sigma_{\rm{t\bar{t}}}}{dX})^i= \frac{1}{\sum_j(\Delta X^j\frac{x^j}{\Delta X^j\cdot \not{BR} \cdot \not{L}})}\frac{x^i}{\Delta X^i \cdot \not{BR} \cdot \not{L}} \ ,
\end{align*}
i.e., the calculation proceeded by having divided the number of signal (background subtracted) events in the data 
corrected for efficiency and possible migration events, $x^i$, by an optimal bin width, $\Delta X^i$.   

%\begin{figure}[!ht]
%  \centering
%  \includegraphics[scale=0.25]{parton_vs_particle.png}
%    \caption{
%	parton and particle level objects and their correspondence for t-
%channel single-top-quark production~\cite{cms_summary_SM}.
%}
%    \label{fig:cms_summary_SM}
%\end{figure}

\begin{figure}
\begin{minipage}{.5\linewidth}
\centering
\subfloat[]{\label{top17002_1:a}\includegraphics[scale=0.35]{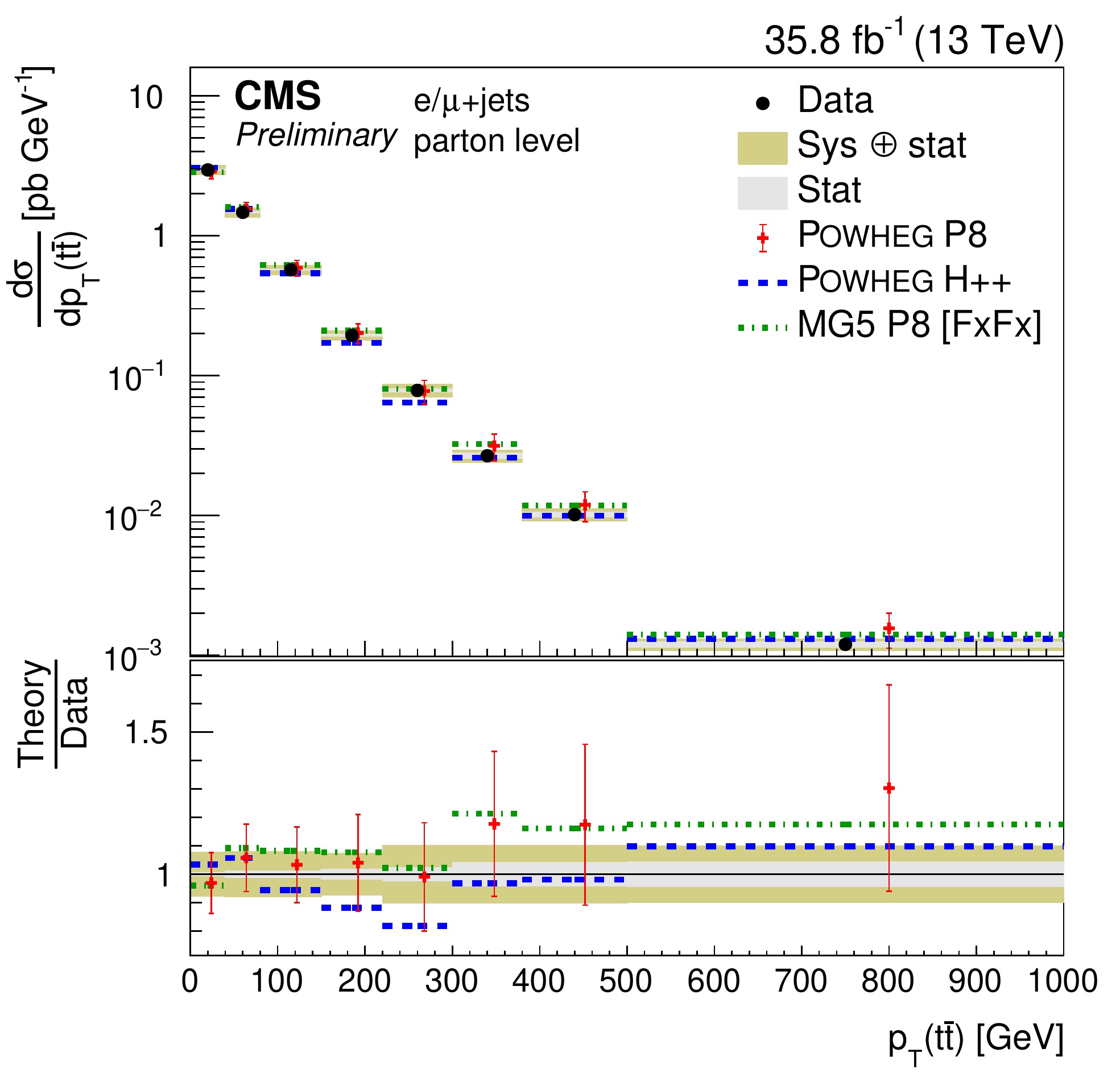}}
\end{minipage}%
\begin{minipage}{.5\linewidth}
\centering
\subfloat[]{\label{top17002_1:b}\includegraphics[scale=0.35]{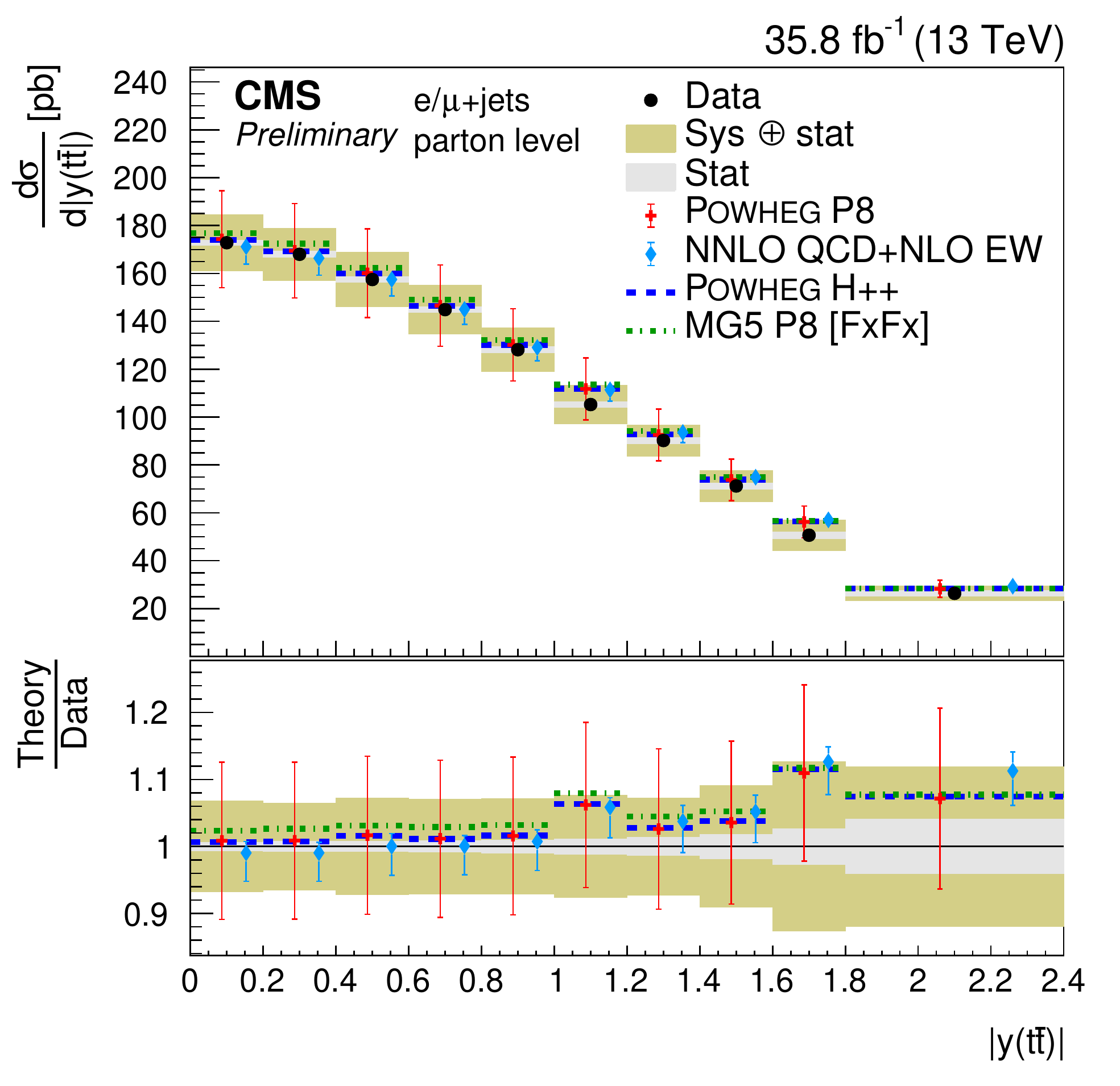}}
\end{minipage}\par\medskip
\begin{minipage}{.5\linewidth}
\centering
\subfloat[]{\label{top17002_1:c}\includegraphics[scale=0.35]{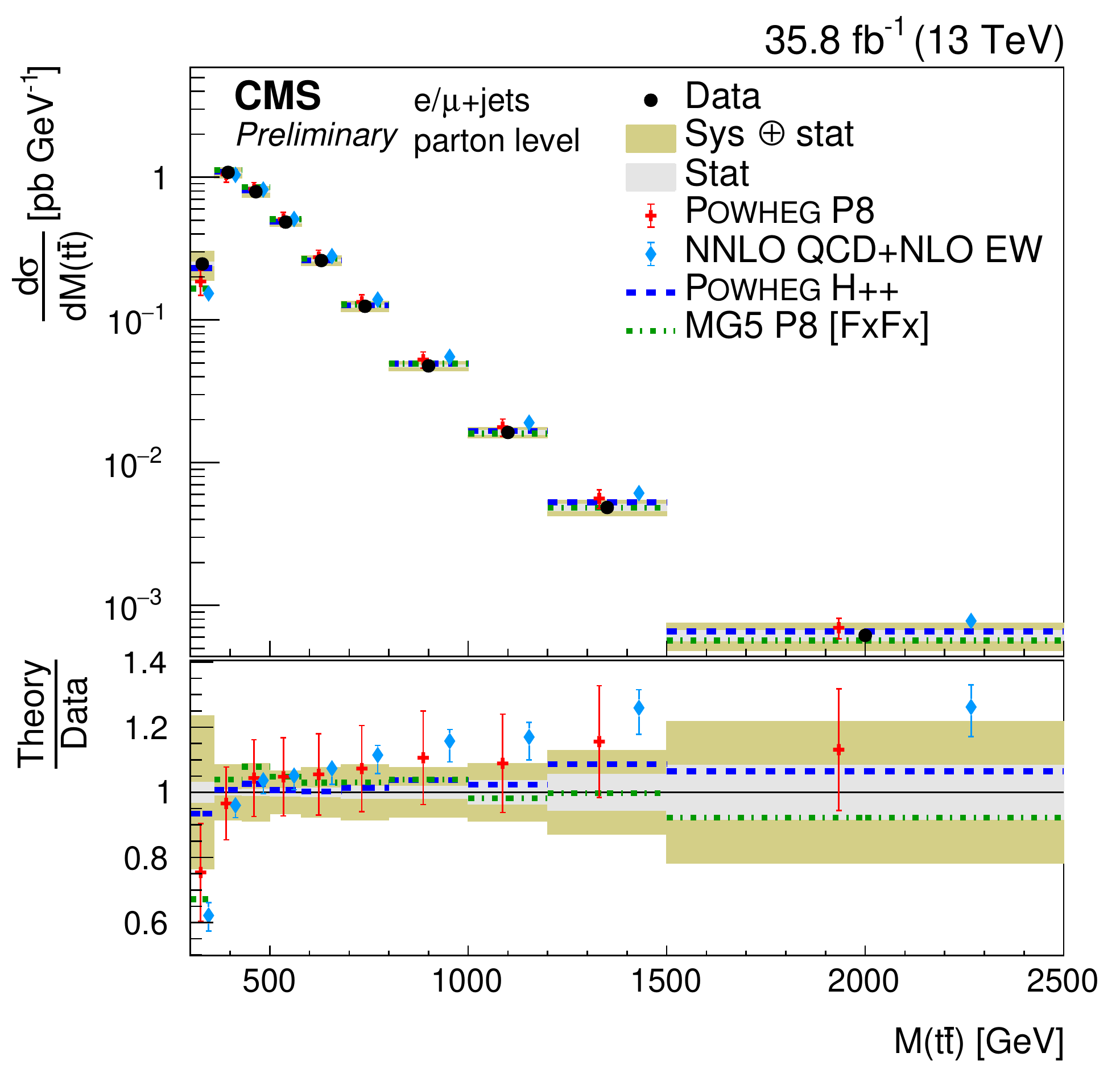}}
\end{minipage}%
\begin{minipage}{.5\linewidth}
\centering
\subfloat[]{\label{top17002_1:d}\includegraphics[scale=0.35]{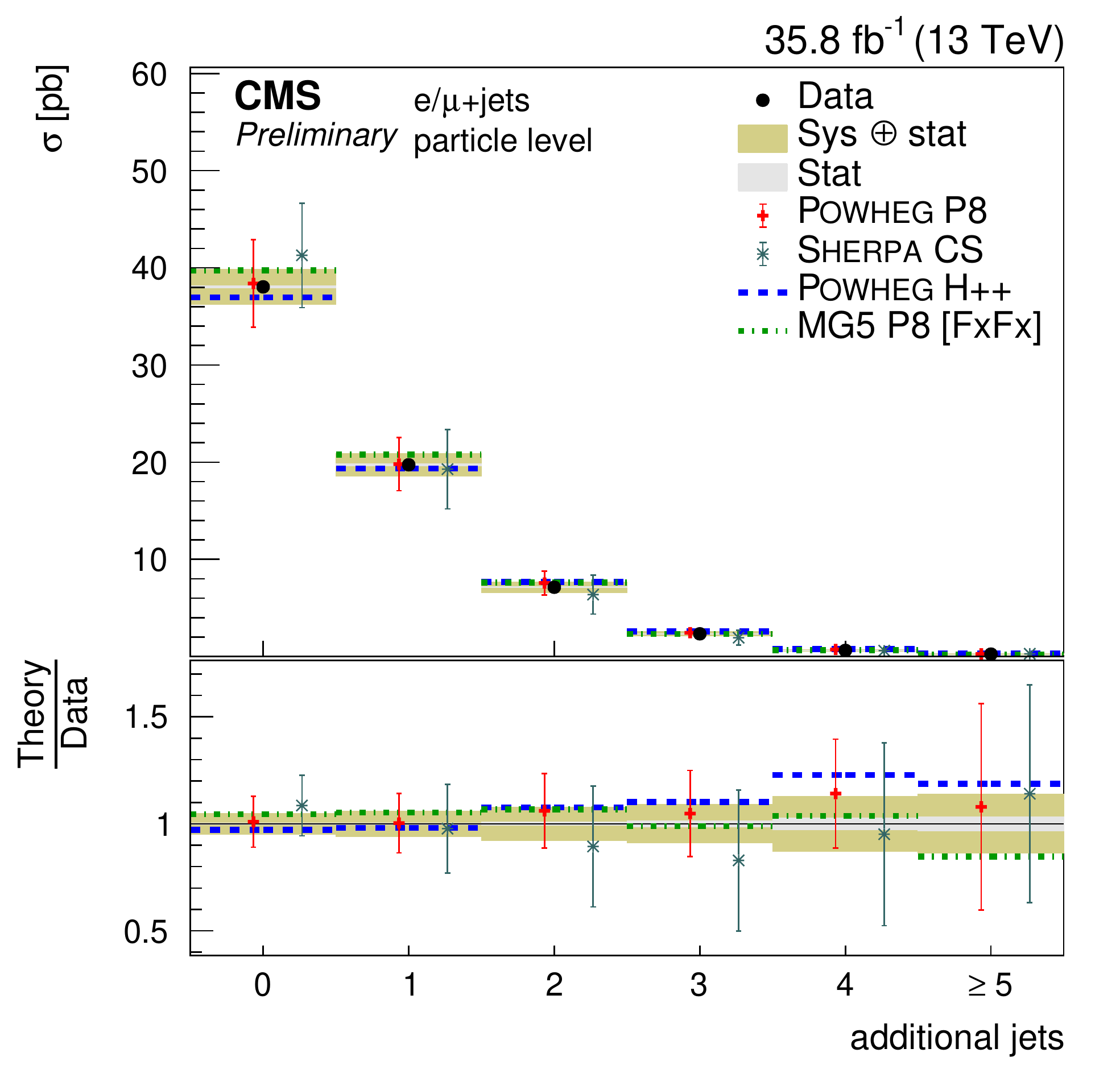}}
\end{minipage}\par\medskip

\caption{Differential cross section in $\ell$+jets channel at parton level as a function of 
$p_{\rm{T}}(\rm{t\bar{t}})$ (a), $|y(\rm{t\bar{t}})|$ (b),
and $M(\rm{t\bar{t}})$ (c).
The data are shown as points with light (dark) error bands indicating the statistical (statistical and systematic)
uncertainties. The cross section is compared to the predictions of \textsc{POWHEG}~\cite{powheg,hvq} (v2) combined with \textsc{PYTHIA}~\cite{pythia8} (v8.205) 
or \textsc{HERWIG++}~\cite{herwig}, and the simulations from \textsc{MG5\_aMC@NLO}~\cite{amc@nlo} interfaced to \textsc{PYTHIA}~\cite{pythia8} (v8.205) with \textsc{FxFx}
merging~\cite{fxfx}, and separately from \textsc{sherpa}~\cite{sherpa}. 
The ratios of the predictions to the measured cross sections are shown at the bottom of each panel.
In addition, the cross section is compared as a function of the additional jet multiplicity (d)~\cite{top17002}.}
\label{fig:top17002_1}
\end{figure}

The study of differential cross-sections therefore allow to test and constrain the MC generator and theoretical predictions. 
Figure~\ref{fig:top17002_1} shows differential cross-sections of the $\rm{t\bar{t}}$ system at parton level as function of  $p_{\rm{T}}(\rm{t\bar{t}})$, $|y(\rm{t\bar{t}})|$,
and $M(\rm{t\bar{t}})$, and at particle level as a function of the number of additional jets (not arising from the 
decay products of the $\rm{t\bar{t}}$ pair) compared to several MC generators. 
These studies~\cite{top17002} are crucial for BSM searches, where contribution from top quark production constitutes an important background, 
for it increases the confidence into the generation chain used or to define a domain of validity
in case of discrepancies with the data.
At the level of stable top quarks, differential cross section measurements
can be compared with SM predictions directly.
These predictions are available at various levels of QCD precision, that go beyond current NLO MC event generators, 
and even include electroweak corrections proportional to $\alpha_{\rm{s}}^2\alpha$, $\alpha_{\rm{s}}\alpha^2$, and $\alpha_{\rm{s}}\alpha^3$~\cite{tt_ew}. However, a visible trend at 
$M(\rm{t\bar{t}})\gtrsim 1$ TeV is revealed, a feature that has been initially covered by the limited statistical precision~\cite{cms_tuning_1} and thus needs further investigation.

\begin{figure}
\begin{minipage}{.5\linewidth}
\centering
\subfloat[]{\label{top17002_2:a}\includegraphics[scale=0.35]{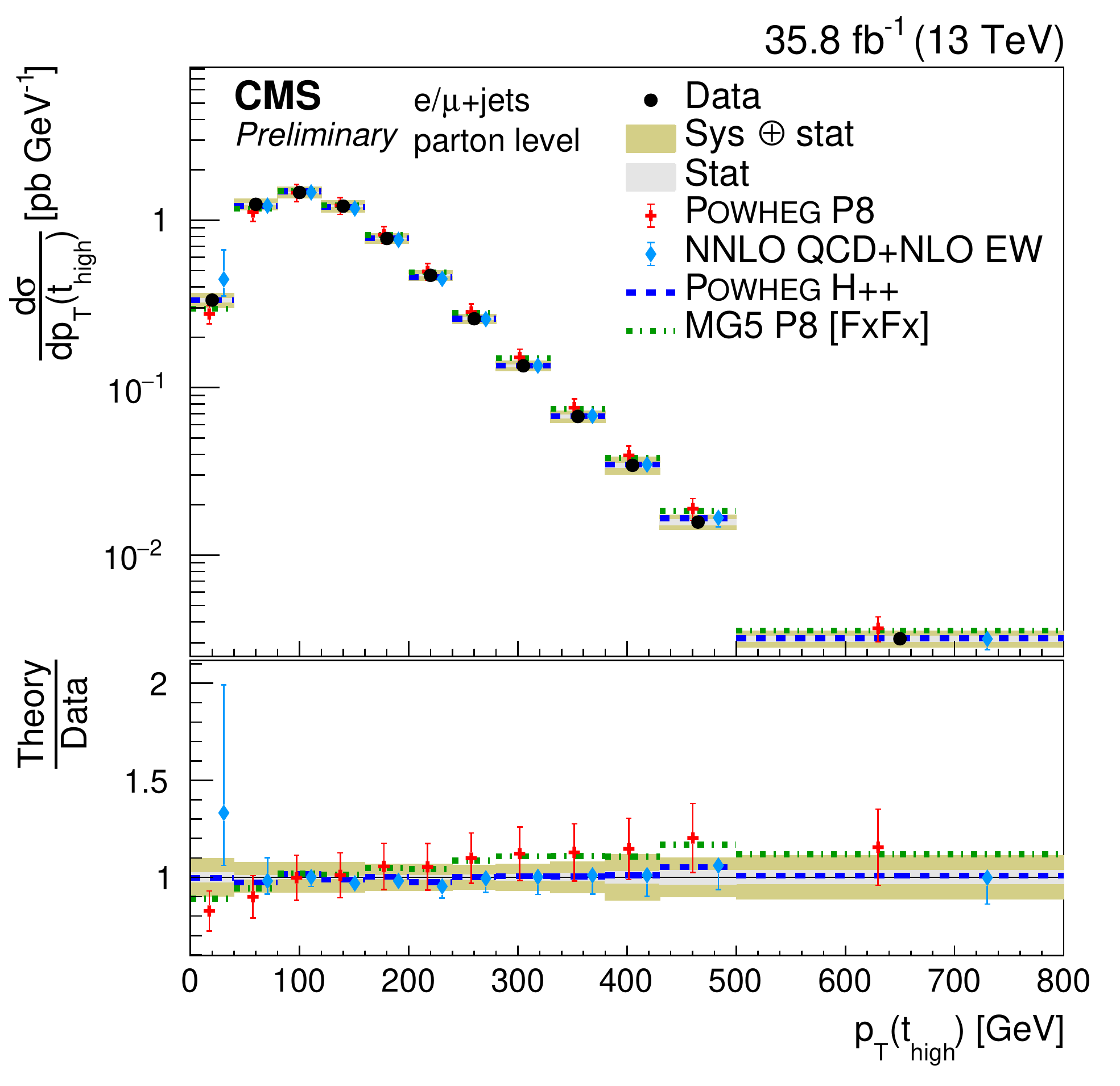}}
\end{minipage}%
\begin{minipage}{.5\linewidth}
\centering
\subfloat[]{\label{top17002_2:b}\includegraphics[scale=0.35]{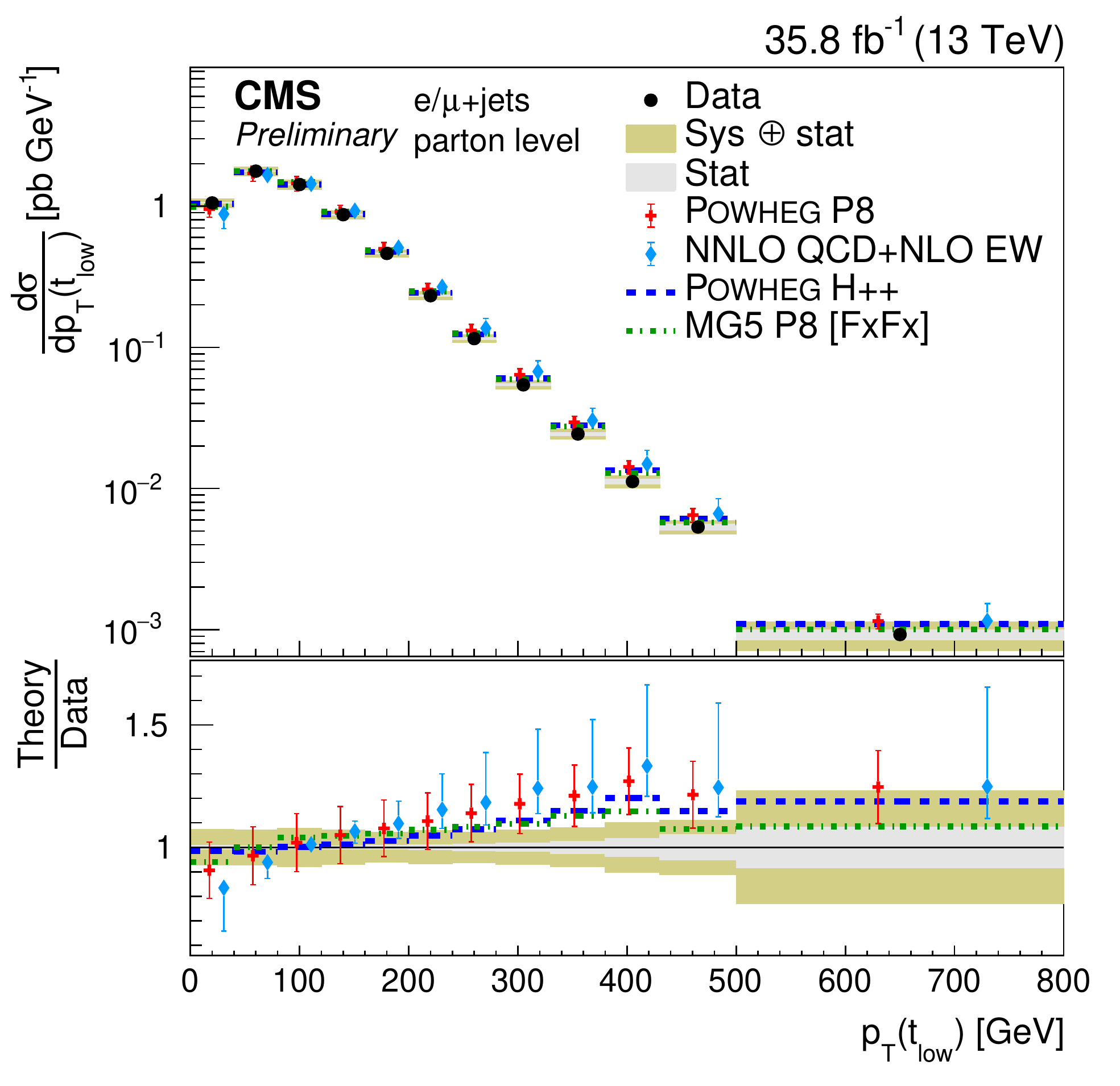}}
\end{minipage}\par\medskip

\caption{
Differential cross section in the $\ell$+jets channel at parton level as a function of the top quark with the 
higher (a) and lower (b) transverse momentum, $p_{\rm{T}}({\rm{t_{high}}})$ and $p_{\rm{T}}({\rm{t_{low}}})$~\cite{ljets_13}. 
The cross sections is compared to the predictions of \textsc{POWHEG}~\cite{powheg,hvq} (v2) combined with 
\textsc{PYTHIA}~\cite{pythia8} (v8.205) or \textsc{HERWIG++}~\cite{herwig}, and the simulations from \textsc{MG5\_aMC@NLO}~\cite{amc@nlo} 
interfaced to \textsc{PYTHIA}~\cite{pythia8} (v8.205) with \textsc{FxFx}
merging~\cite{fxfx}, and separately from the NNLO QCD+NLO EW~\cite{tt_ew} calculations. 
The ratios of the various predictions to the measured cross section are shown at the bottom of each panel~\cite{top17002}.}
\label{fig:top17002_2}
\end{figure}

Another observed deviation concerns about the reconstructed top quark $p_{\rm{T}}$, that
displays  a  softer  spectrum  in the data  than  it  is  predicted  by  MC simulation.  
An improved description of the data is achieved by an ad-hoc reweighting of $\rm{t\bar{t}}$ events, with the weight 
depending on the transverse top quark and antiquark momenta at parton level.
Since the reweighting mitigates the observed deviation of the top quark $p_{\rm{T}}$ spectrum,
yet it could well introduce distortions of any correlated variables, it is treated as an additional systematic uncertainty.
The recent full NNLO QCD+NLO EW calculation~\cite{tt_ew} of the top-quark $p_{\rm{T}}$ shows improved agreement with the 
measured spectrum, as compared to calculations with decreased accuracy. As it has been documented in the
literature,  although  EW  effects  are  rather  small  at  the  level  of  inclusive  $\sigma_{\rm{t\bar{t}}}$,  they
can have a sizable impact on differential distributions. However, after separately inspecting the differential cross section at parton level
as a function of the top quark with the higher (Fig.~\ref{top17002_2:a}) and lower (Fig.~\ref{top17002_2:b}) transverse momentum, 
$p_{\rm{T}}({\rm{t_{high}}})$ and $p_{\rm{T}}({\rm{t_{low}}})$,
it has been observed that the former is very well modeled, while the later is predicted to be harder.

The study of the jet multiplicity allows to probe the whole simulation chain. 
The lower multiplicities are mainly sensitive to the matrix-element description,
while the higher multiplicities are mainly sensitive to the parton shower description (Section~\ref{sec:theo}). 
As seen in Fig.~\ref{top17002_1:d}, the predictions of $\rm{t\bar{t}}$ + $\geq$ 4 jets could vary to up-to approximately 20\% 
depending on the generation chain, while they agree at lower jet multiplicity. 
The differential measurements thus help to disambiguate which generators are able to correctly describe the 
dynamics of the production, though currently there seems to be no MC event generator to simultaneously grasp 
the production mechanism over the wide kinematic range.

\clearpage
\subsubsection{ A factory of gluons over broad momenta}
\label{sec:tt_PDF}
\begin{figure}
\begin{minipage}{.5\linewidth}
\centering
\subfloat[]{\label{pdfs_cms:a}\includegraphics[scale=0.25]{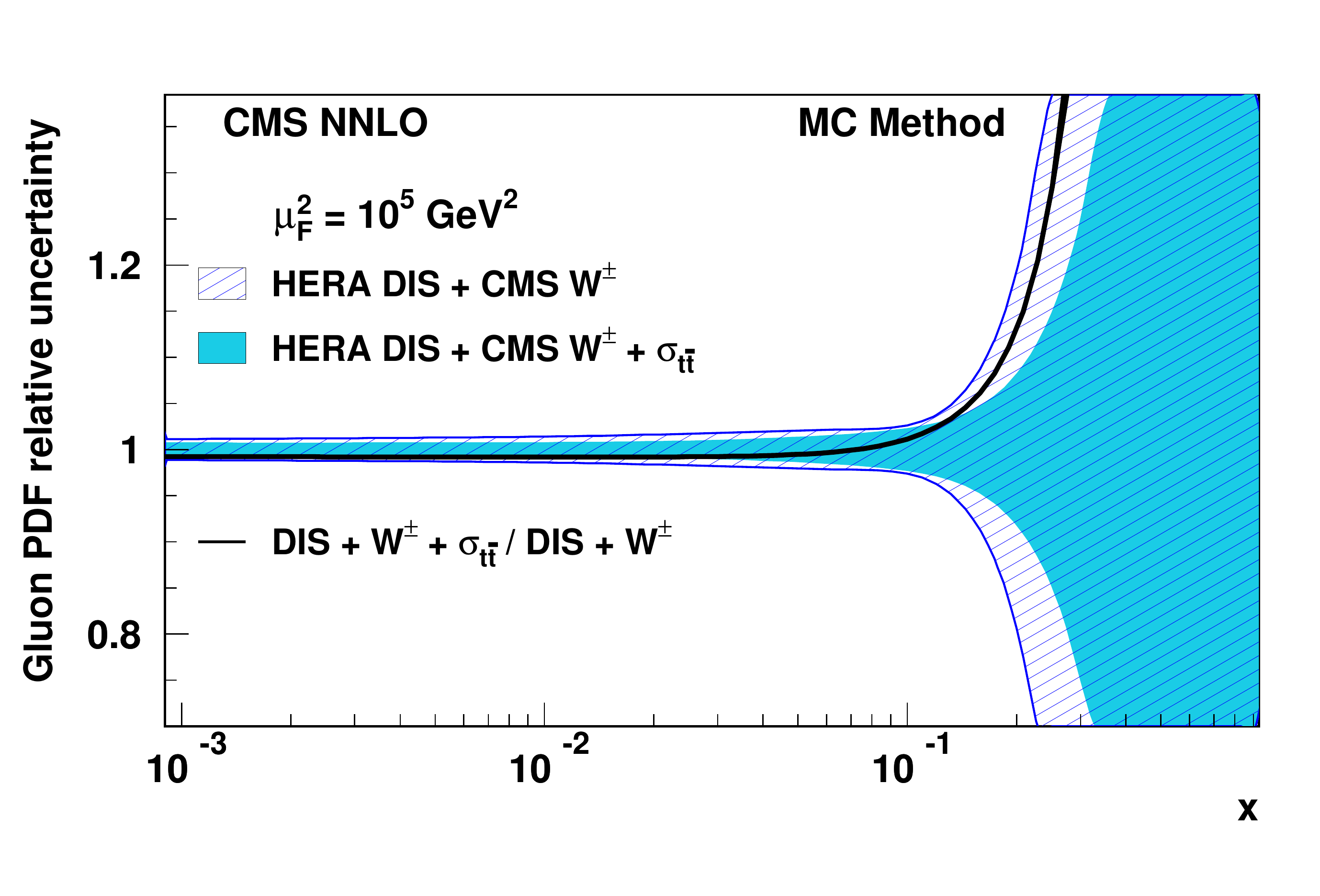}}
\end{minipage}%
\begin{minipage}{.5\linewidth}
\centering
\subfloat[]{\label{pdfs_cms:b}\includegraphics[scale=0.6]{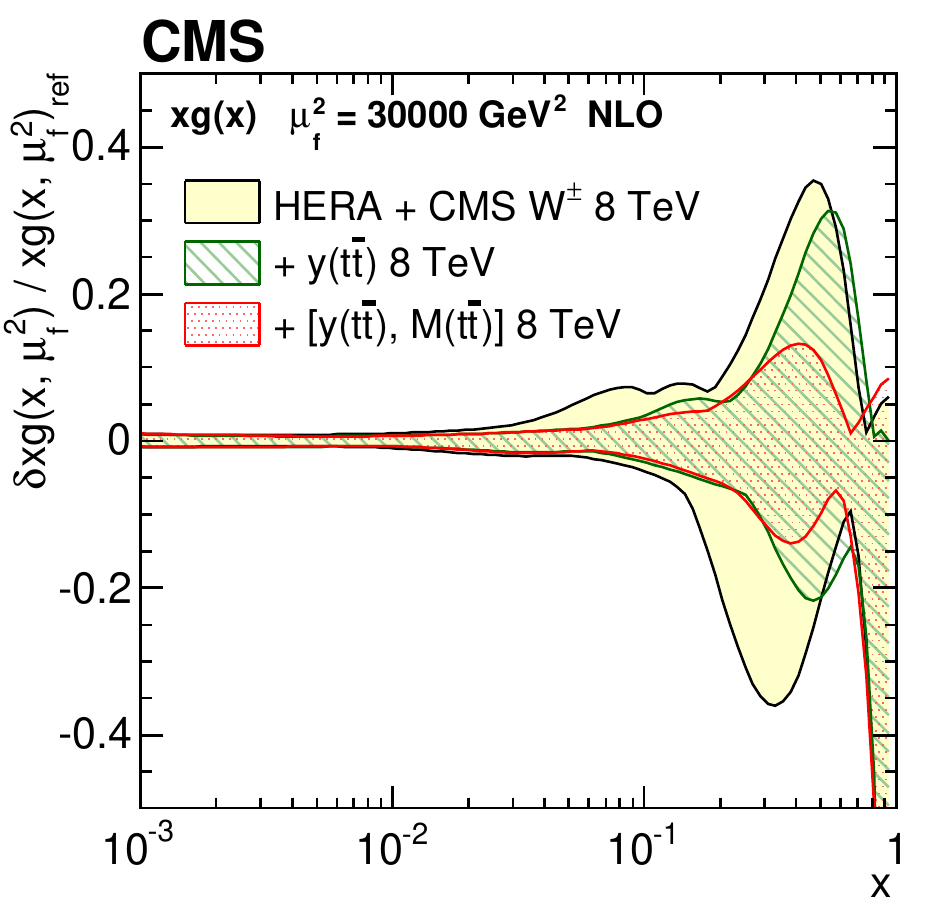}}
\end{minipage}\par\medskip

\caption{
The relative fractional uncertainties in the gluon PDF as functions of $x$ at the scale of (a)~\cite{top16023} 100000
and (b)~\cite{CMSdoublediff} 30000 GeV$^2$. 
The results of the fit including top quark measurements, and without those are compared.}
\label{fig:pdfs_cms}
\end{figure}

The impact of the $\sigma_{\rm{t}\bar{\rm{t}}}$ measurements on the knowledge of the
proton PDFs can be well illustrated based on the example at $\sqrt{s}=5.02$ TeV~\cite{top16023}. 
The evaluation represents the first QCD analysis at NNLO, 
together with the combined measurements of neutral- and charged-current cross sections for deep inelastic electron- and
positron-proton scattering (DIS) at HERA~\cite{hera_pdf_data}, and the CMS measurement of the muon charge asymmetry in W boson production~\cite{cms_pdf_data}.
The precise HERA DIS data, obtained from the combination of the individual H1 and ZEUS results, are directly sensitive
to the valence and sea quark distributions and probe the gluon distribution through scaling violations, rendering these data the core of all PDF fits.
Parametrized PDFs are fitted to the data in a procedure that is referred to as the ``PDF fit''.  
The experimental uncertainties in the measurement are propagated to the 14-parameter fit using the MC method.
Indeed, a moderate reduction of the uncertainty in the gluon PDF at high $x\gtrsim 0.1$ is observed, once
the measured values for $\sigma_{\rm{t}\bar{\rm{t}}}$ are included in the fit (Fig.~\ref{pdfs_cms:a}).  
All changes in the central values of the PDFs are well within the fit uncertainties, 
while the uncertainties in the valence quark distributions remain unaffected.

Due to the large $\rm{t\bar{t}}$ data samples at the LHC, also the first double differential cross sections have been
published, for example as a function of $M(\rm{t\bar{t}})$ and $|y(\rm{t\bar{t}})|$.
Their study focuses on correlations between kinematic properties of the top quarks and provides insights into extreme regions
of the phase space. Thereby the measurement of the double differential cross-section of $M(\rm{t\bar{t}})$ and $|y(\rm{t\bar{t}})|$
has been used to constrain the proton PDFs and especially to improve the gluon PDF~\cite{CMSdoublediff}. 
The uncertainties in the gluon distribution at $x > 0.01$ are significantly reduced,
once the $\rm{t\bar{t}}$ data are included in the fit, with the largest improvement in gluon PDF uncertainty (by more than a factor of two) 
to be observed at $x \approx$ 0.3 (Fig.~\ref{pdfs_cms:b}), as expected from the LO kinematic relation $x=(M(\rm{t\bar{t}})/\sqrt{s})exp[\pm y(\rm{t\bar{t}})]$.

\clearpage
\subsubsection{ Breaking the 1 TeV barrier }
\label{sec:tt_boost}

\begin{figure}
\begin{minipage}{.3\linewidth}
\centering
\subfloat[]{\label{boosted:a}\includegraphics[scale=0.4]{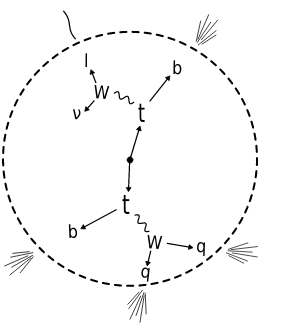}}
\end{minipage}%
\begin{minipage}{.3\linewidth}
\centering
\subfloat[]{\label{boosted:b}\includegraphics[scale=0.42]{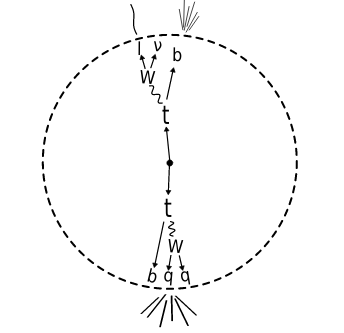}}
\end{minipage}%\par\medskip
\begin{minipage}{.3\linewidth}
\centering
\subfloat[]{\label{boosted:c}\includegraphics[scale=0.32]{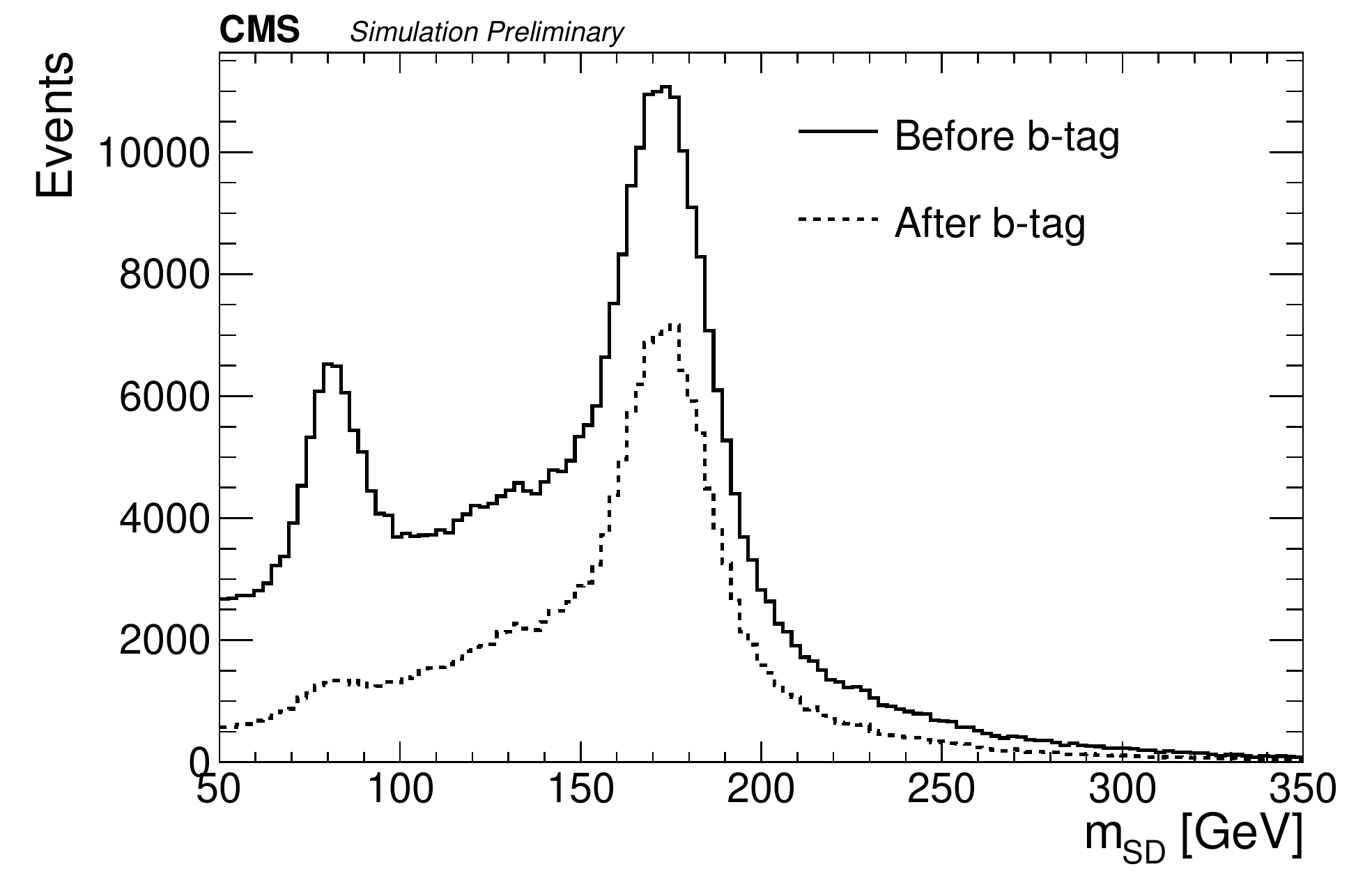}}
\end{minipage}%

\caption{
Illustration of resolved (a) and boosted (b) \ttbar event topologies in the $\ell$+jets decay channel~\cite{husemann}.
The corrected mass (c) for radiation, underlying event, and pileup effects, $m_{\rm{SD}}$, of the leading jet, 
before (solid line) and after (dashed line) the requirement of at least one b tagged substructure, in \ttbar simulated events~\cite{tt_boost_13}.
}
\label{fig:boosted}
\end{figure}

Top quarks produced in high-energy collisions may  receive  large  momenta,  either in  regular SM  processes  or  by
hypothetical high-mass particles  decaying  to  top  quarks.
To study the production of top quarks with large transverse momenta (``boosted'' top quarks), differential cross section measurements using boosted-top 
reconstruction techniques have been devised.
While at low $p_{\rm{T}}$ the events appear in resolved
topology (Fig.~\ref{boosted:a}), where all jets are visible, at high $p_{\rm{T}}\gtrsim$ 400 GeV the  decay  products (leptons and jets) 
start  becoming collimated, such that they begin to overlap in the $\eta$-$\phi$ space (Fig.~\ref{boosted:b}).
In the last decade, a large number of algorithms was conceived to analyze
boosted-jet topologies.  In these algorithms, jets are first reconstructed with
a typical large radius parameter (``fat'' jets), and their structure is subsequently
either declustered (jet-declustering algorithms) to identify potential substructures in the jet related to the decay of the
mother particle or assigned a probability of having stemmed from a given number of overlapping jets (jet-shape algorithms).
Combinations of several techniques can also be used that also allow for b-tagging in the dense environment of fat
jets (Fig.~\ref{boosted:c}), while the exact choice of algorithm depends on the expected event topology and momentum range of the boosted object.

\begin{figure}
\begin{minipage}{.5\linewidth}
\centering
\subfloat[]{\label{diff_boosted:a}\includegraphics[scale=0.4]{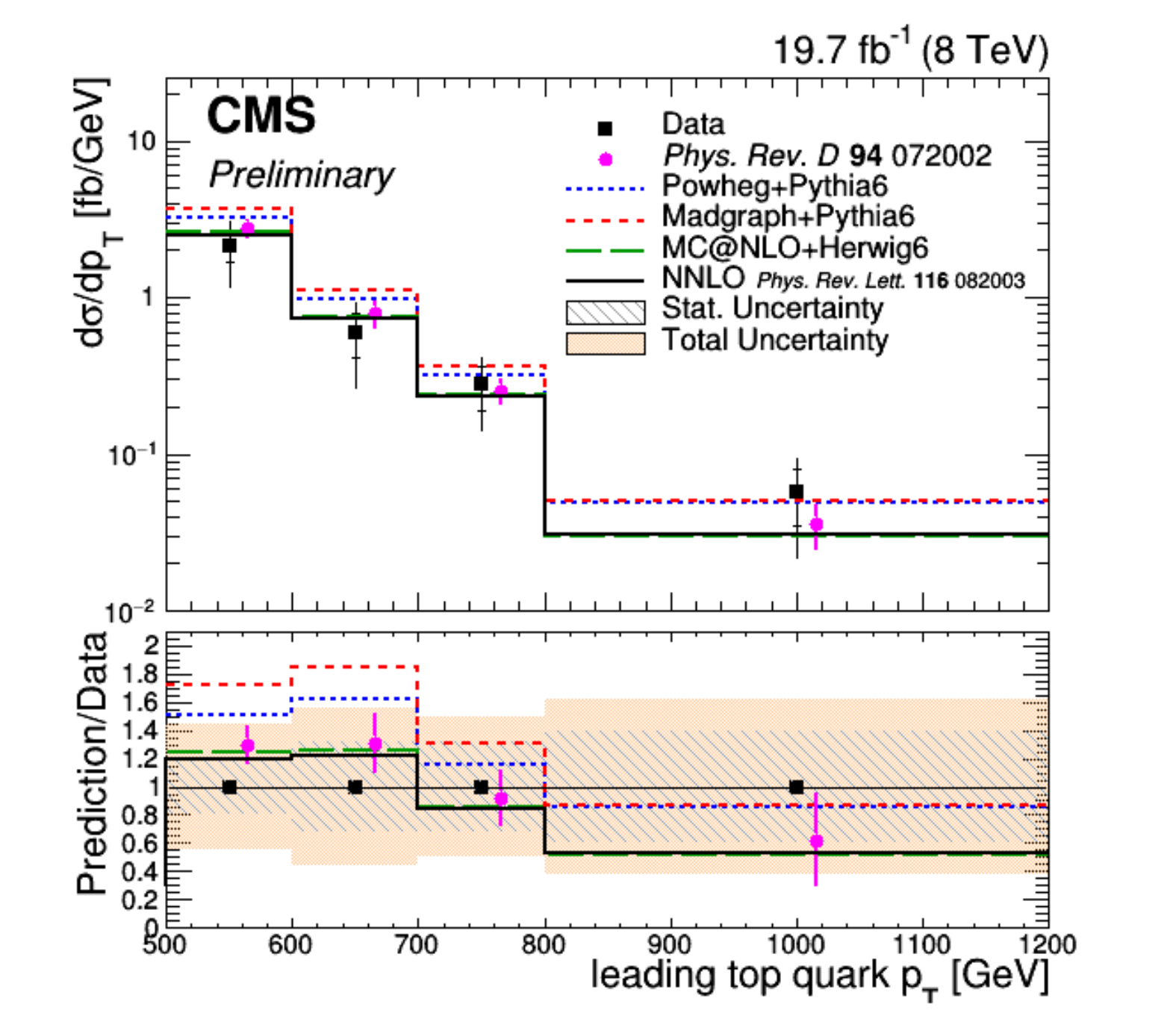}}
\end{minipage}
\begin{minipage}{.5\linewidth}
\centering
\subfloat[]{\label{diff_boosted:b}\includegraphics[scale=0.4]{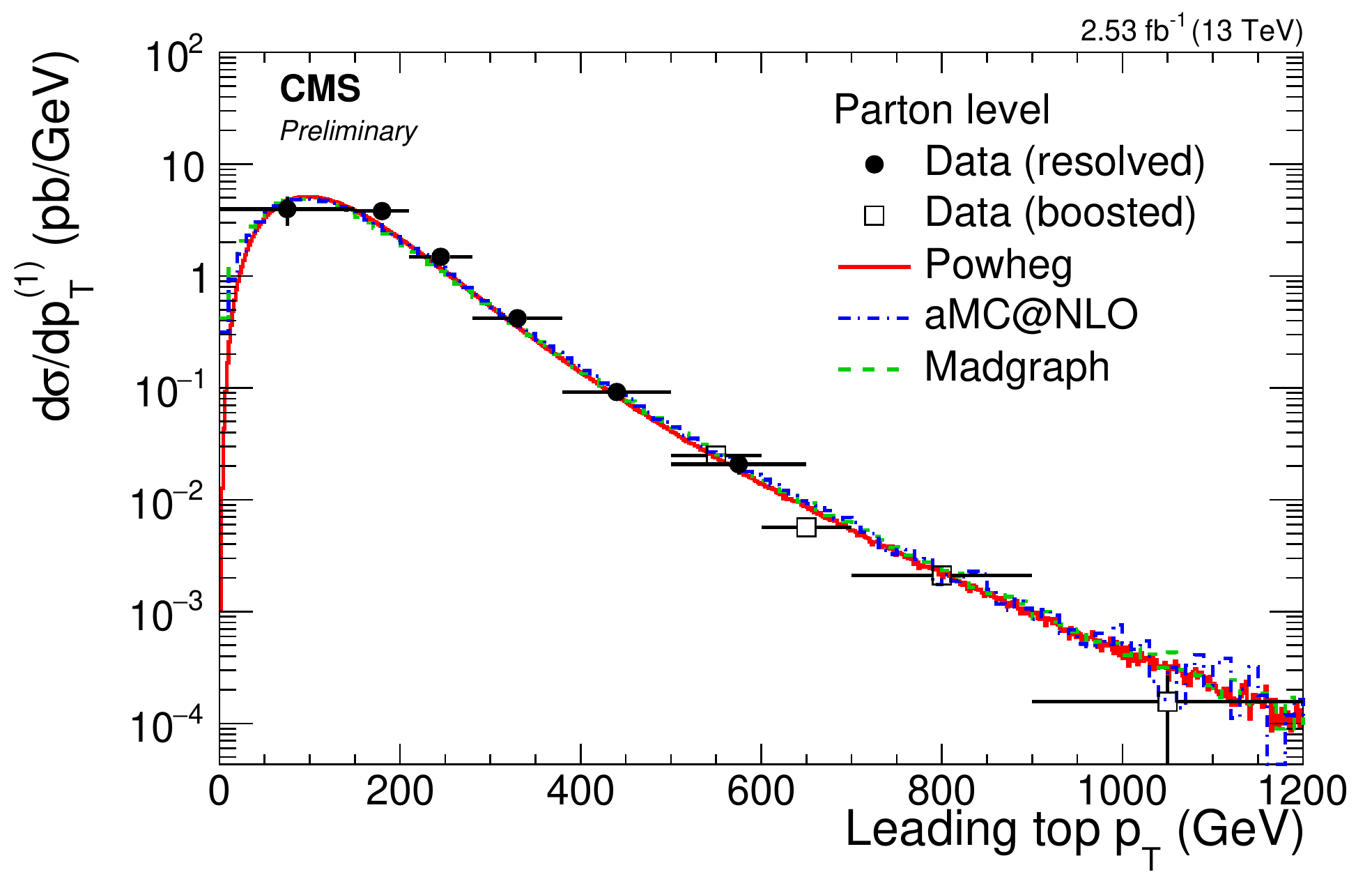}}
\end{minipage}%\par\medskip

\caption{
Differential cross section in the fully hadronic decay channel at $\sqrt{s}=8$~\cite{tt_boost_hadr_8} 
(a, boosted regime) and 13~\cite{tt_boost_13} (b, resolved and boosted regimes) TeV, respectively.
In addition, the unfolded result at $\sqrt{s}=8$ TeV is compared to the measurement in the $\ell$+jets~\cite{tt_boost_ljets_8} channel, also in the boosted regime.}
\label{fig:diff_boosted}
\end{figure}

Boosted top quark events can effectively double the reach in transverse momentum relative to the resolved topology, 
allowing to cross the TeV boundary in the measurement of top quark $p_{\rm{T}}$. 
Jet-shape algorithms have been proven to be robust for discriminating between boosted top quarks (a three-prong jet) 
and ordinary QCD multijet background that tends to have fewer substructures.
Given that the three-prong jet mass should be similar to that of $m_{\rm{top}}$, 
with a pair combination close to the W boson mass,
it is crucial the effects of radiation, underlying event (hadronic activity not attributed to the hard scattering), and pileup contamination to be mitigated. 
Soft and collinear particles can be removed from the fat jet clustering in a recursive procedure regulated by two parameters;
the first (typically denoted as ``$z_{\rm{cut}}$'') controls the strength of the fractional $p_{\rm{T}}$ selection, and the second 
(typically denoted as ``$\beta$'') suppresses the collinear radiation.
For the present studies at $\sqrt{s}=13$ TeV, the values of $z_{\rm{cut}}=0.1$ and $\beta=0$ are used, 
which have been found to yield the best performance for fat jets~\cite{jme15002}.
For the extraction of the cross section (inclusively and differentially as a function of the $p_{\rm{T}}$ of the leading top quark)
a template fit to the distribution of the corrected jet mass has been performed in the fully hadronic channel~\cite{tt_boost_13}. 
The signal and QCD multijet templates are taken from the MC simulation and a control sample in the data, respectively.
Similar measurements have been performed at $\sqrt{s}=8$ TeV in the fully hadronic~\cite{tt_boost_hadr_8} 
and $\ell$+jets~\cite{tt_boost_ljets_8} channels, respectively (Fig.~\ref{fig:diff_boosted}).

\begin{figure}
\begin{minipage}{.5\linewidth}
\centering
\subfloat[]{\label{diff_boosted_mass:a}\includegraphics[scale=0.36]{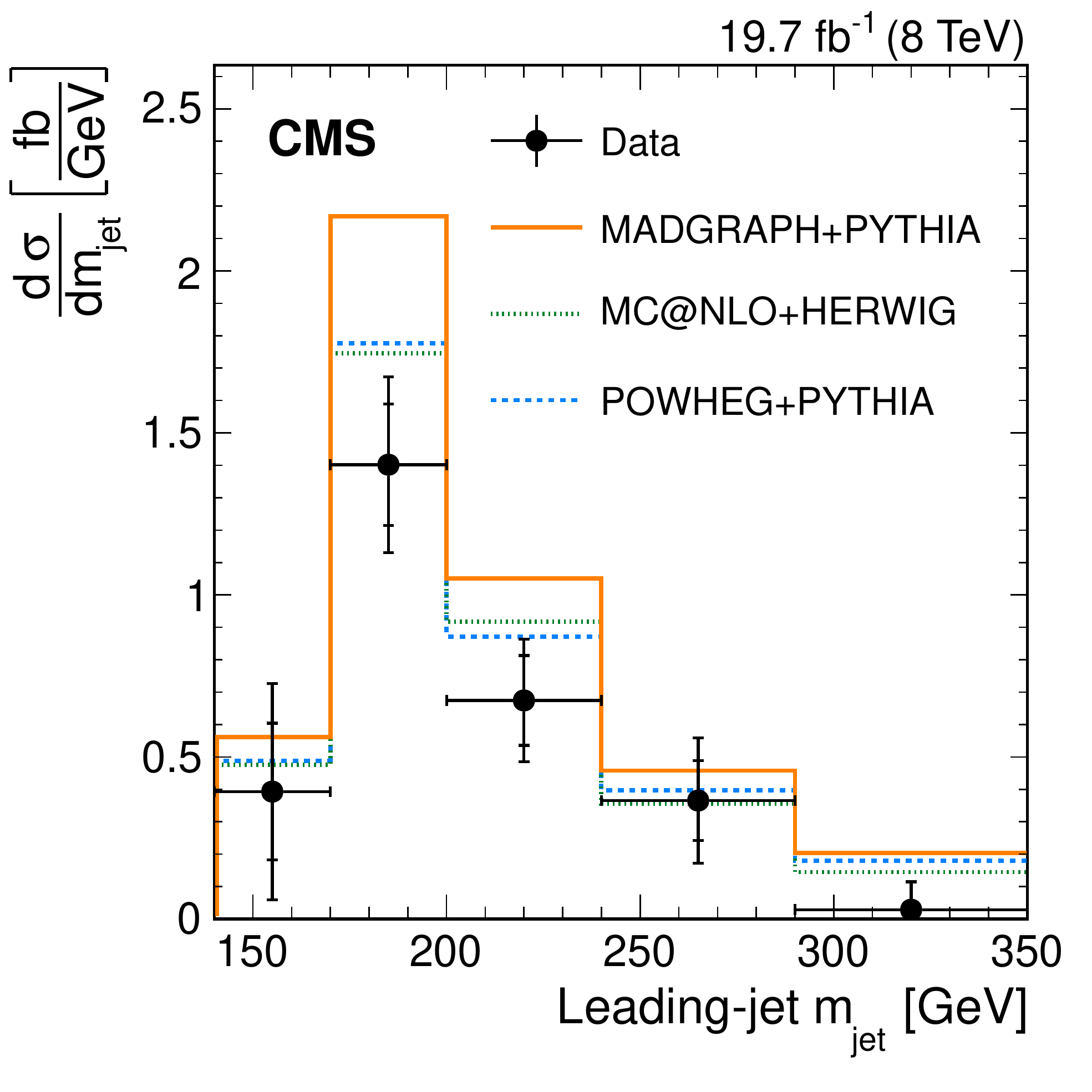}}
\end{minipage}
\begin{minipage}{.5\linewidth}
\centering
\subfloat[]{\label{diff_boosted_mass:b}\includegraphics[scale=0.4]{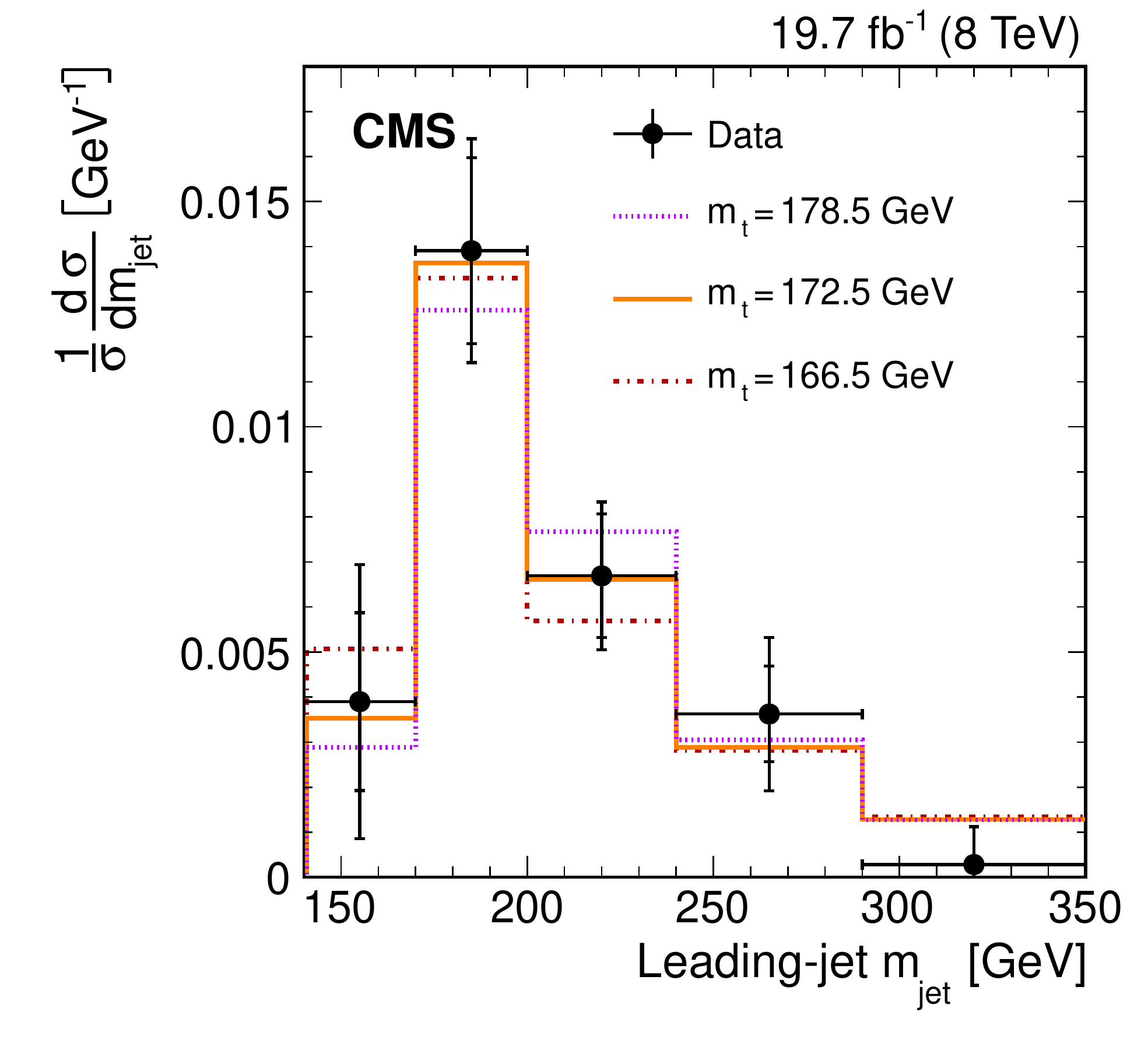}}
\end{minipage}%\par\medskip

\caption{
(a) Differential cross section in the $\ell$+jets channel at particle level as a function of the mass of the leading jet.
The cross sections from the combined electron and muon channels (points) are compared to LO predictions from 
\textsc{MadGraph}~\cite{madgraph} (v.5.1.5.11) interfaced to PYTHIA~\cite{pythia6} (v6.424) with \textsc{MLM}~\cite{mlm} merging, 
and NLO predictions separately using 
\textsc{POWHEG}~\cite{powheg,hvq} (v1) combined with \textsc{PYTHIA}~\cite{pythia6} (v6.424), and \textsc{MC@NLO}~\cite{mc@nlo} (v3.41) 
combined with \textsc{HERWIG++}~\cite{herwig}.
The vertical bars represent the statistical (inner)  and the total (outer) uncertainties. 
The horizontal bars show the bin widths.
(b) The differential cross section is further normalized to unity and compared to LO predictions 
from \textsc{MadGraph}~\cite{madgraph} (v.5.1.5.11)+\textsc{PYTHIA}\cite{pythia6} (v6.424) for 
three distinct values of $\rm{m_{top}}$~\cite{tt_boost_ljets_mass_8}.
}
\label{fig:diff_boosted_mass}
\end{figure}

The first differential $\rm{t\bar{t}}$ cross section measurement~\cite{tt_boost_ljets_mass_8} as a function of 
the mass of boosted top quarks  has  been  
also performed  as  a  proof  of  principle  to  measure $m_{\rm{top}}$ in the boosted final state (Fig.~\ref{fig:diff_boosted_mass}).
More specifically, the peak position in the three-prong jet mass distribution is sensitive to $m_{\rm{top}}$. 
The uncertainty in the top quark mass measurement is so far limited ($\pm$ 9 GeV), dominated by the statistical
and jet energy scale uncertainty in the boosted regime. Although, potential improvements and more constraints
in systematic uncertainties are anyway possible in future application of this method on larger data samples, 
the prospect of reaching a more reliable correspondence between $m_{\rm{top}}$ in any well-defined renormalization
scheme and the top quark mass parameter in MC event generators should be faced with care. The first attempts 
to retain a universal description of the $m_{\rm{top}}$ scheme with decay effects, and to remove soft contamination indicate
significantly different values for the allowed region in both $z_{\rm{cut}}$ and $\beta$ parameters~\cite{softdrop_mass}.
%https://arxiv.org/pdf/1708.02586.pdf%
Related to that, pileup mitigation could be also effectively handled experimentally with reverting to a per-particle-level 
suppression techniques, that were found to maintain top tagging performance up to very high number of simultaneous pp interactions~\cite{jme16003}.
%https://cds.cern.ch/record/2256875/files/JME-16-003-pas.pdf%

\clearpage

%\subsection{``If we be doomed to remain single, we do''\protect\footnote{T. Hardy, Under the Greenwood Tree.}}
\subsection[``If we be doomed to remain single, we do'']{``If we be doomed to remain single, we do''\ \footnote{T. Hardy, \textit{Under the Greenwood Tree.}}}
\label{sec:tt_st}

%\begin{figure}[!ht]
%  \centering
%  \includegraphics[scale=0.35]{t-channel.png}
%    \caption{A representative diagram for the electroweak production of top quarks via \textit{t}-channel.}
%    \label{fig:t_channel}
%\end{figure}

\begin{figure}
\begin{minipage}{.33\linewidth}
\centering
\subfloat[]{\label{st_prod:a}\includegraphics[scale=0.25]{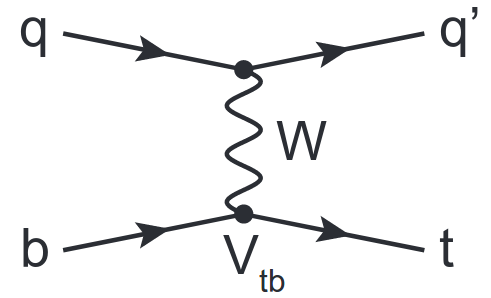}}
\end{minipage}%
\begin{minipage}{.33\linewidth}
\centering
\subfloat[]{\label{st_prod:b}\includegraphics[scale=0.35]{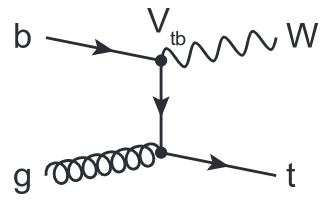}}
\end{minipage}
\begin{minipage}{.33\linewidth}
\centering
\subfloat[]{\label{st_prod:c}\includegraphics[scale=0.35]{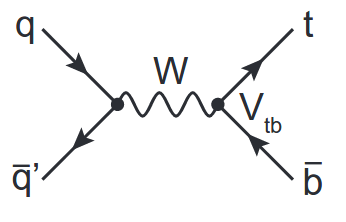}}
\end{minipage}%\par\medskip

\caption{Representative diagrams~\cite{husemann} for electroweak single top-quark production at LO:
\textit{t}-channel (a), associated tW (b), and \textit{s}-channel production (c).}
\label{fig:st_prod}
\end{figure}

In addition  to  \ttbar  production,  top quarks can also be produced singly via electroweak processes. 
Although the Born-level diagrams, classified by the virtuality of the exchanged W boson as shown in Fig.~\ref{fig:st_prod}, 
exhibits a clear separation of three distinctive mechanisms, NLO corrections to tW diagram render these contributions 
to interfere with the \ttbar production. 																																												
In fixed-order calculations, the interference can be controlled by excluding the resonant region,
where the invariant mass of the Wb system is close to the top quark mass, or by vetoing additional b quarks based on \pt spectrum. 																																			
Due to inital- and final-state effects, such approaches are not directly applicable in MC generators. 
%and thus processes in which the intermediate top quark is on its mass shell (``double resonant'' processes)
%are consistently removed.
The overlap between tW and \ttbar production is treated either by diagram removal (DR), where doubly-resonant diagrams are removed from the calculation, 
or diagram subtraction (DS), that cancels \ttbar contributions locally by implementing a subtraction term.
  Both techniques should lead to comparable results, 																									
  though in specific regions of the phase space discrepancies of $\mathcal{O}(30\%)$ exist for more complicated processes~\cite{olga}.

Typically, the single top-quark production measurements focus on the leptonic top quark decays, 
rendering a generic basis for signal scouting a relatively high-$p_{\rm{T}}$ lepton, a significant amount of missing momentum as well as a
b-tagged jet from the top-quark decay.
The additional signatures of single top-quark production are: a light flavor jet with large $|\eta|$ in \textit{t}-channel,
an additional b-jet in \textit{s}-channel, and two oppositely charged high-$p_{\rm{T}}$ leptons in associated tW production.
For all single top-quark production channels, \ttbar production is thus a major background.

%\subsubsection{ W boson is not always real }
%\label{sec:tt_st_main}

\begin{figure}
\begin{minipage}{.5\linewidth}
%\left
\subfloat[]{\label{st_tw:a}\includegraphics[width=.9\textwidth,left]{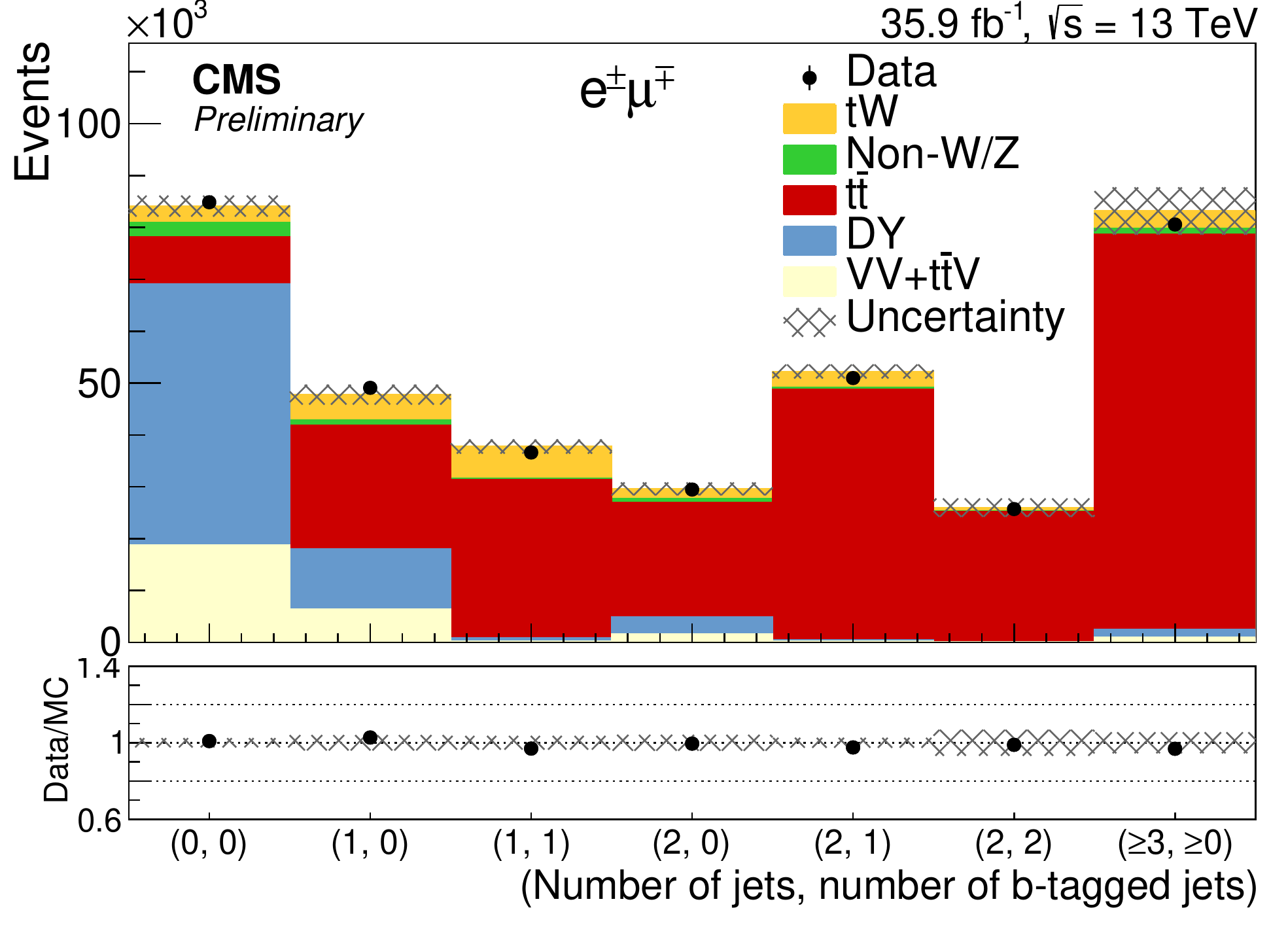}}
\end{minipage}%
\begin{minipage}{.5\linewidth}
%\centering
\subfloat[]{\label{st_tw:b}\includegraphics[width=.9\textwidth,center]{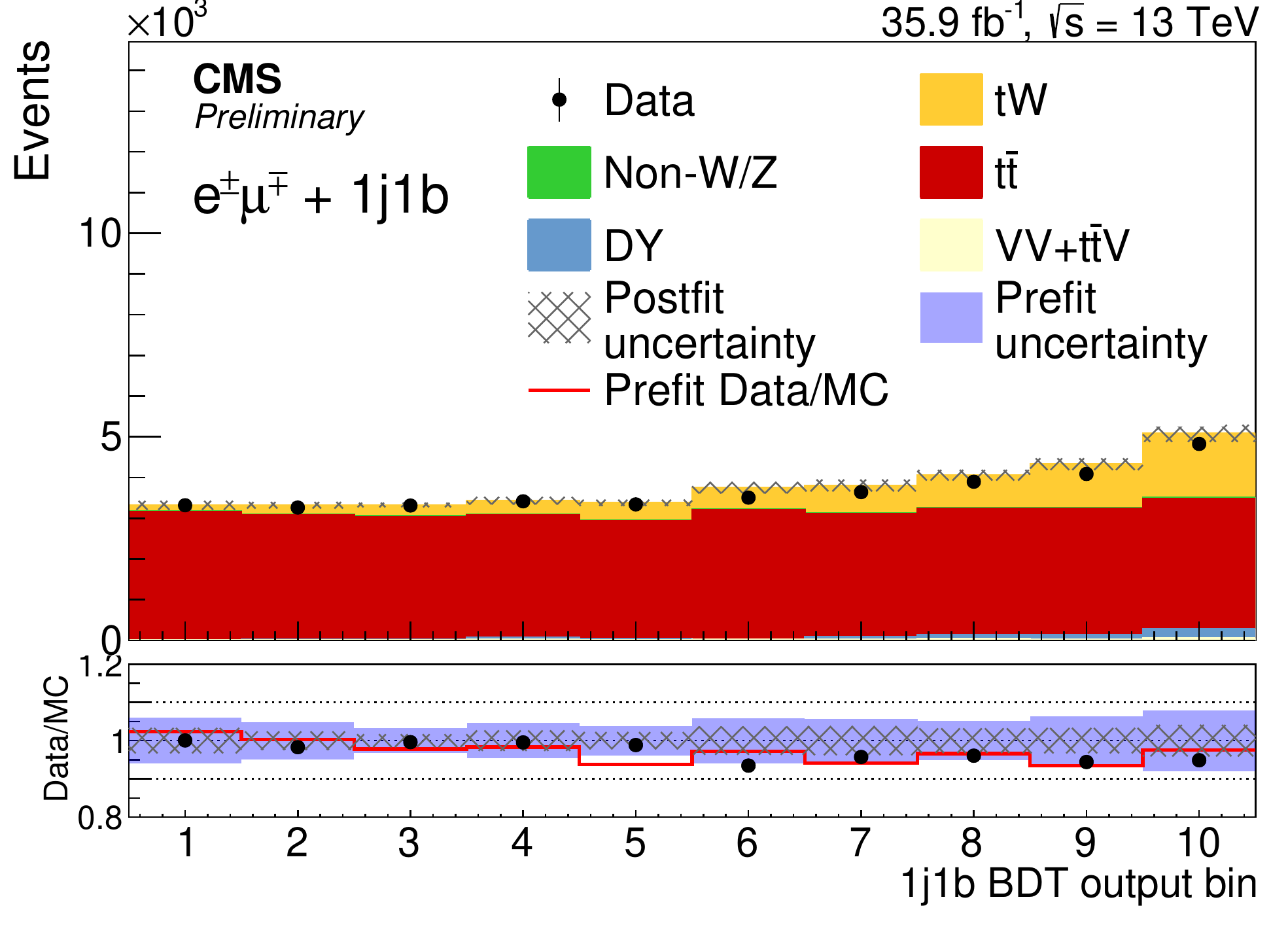}}
\end{minipage}\par\medskip
\begin{minipage}{\linewidth}
\centering
\subfloat[]{\label{st_tw:c}\includegraphics[width=.5\textwidth,center]{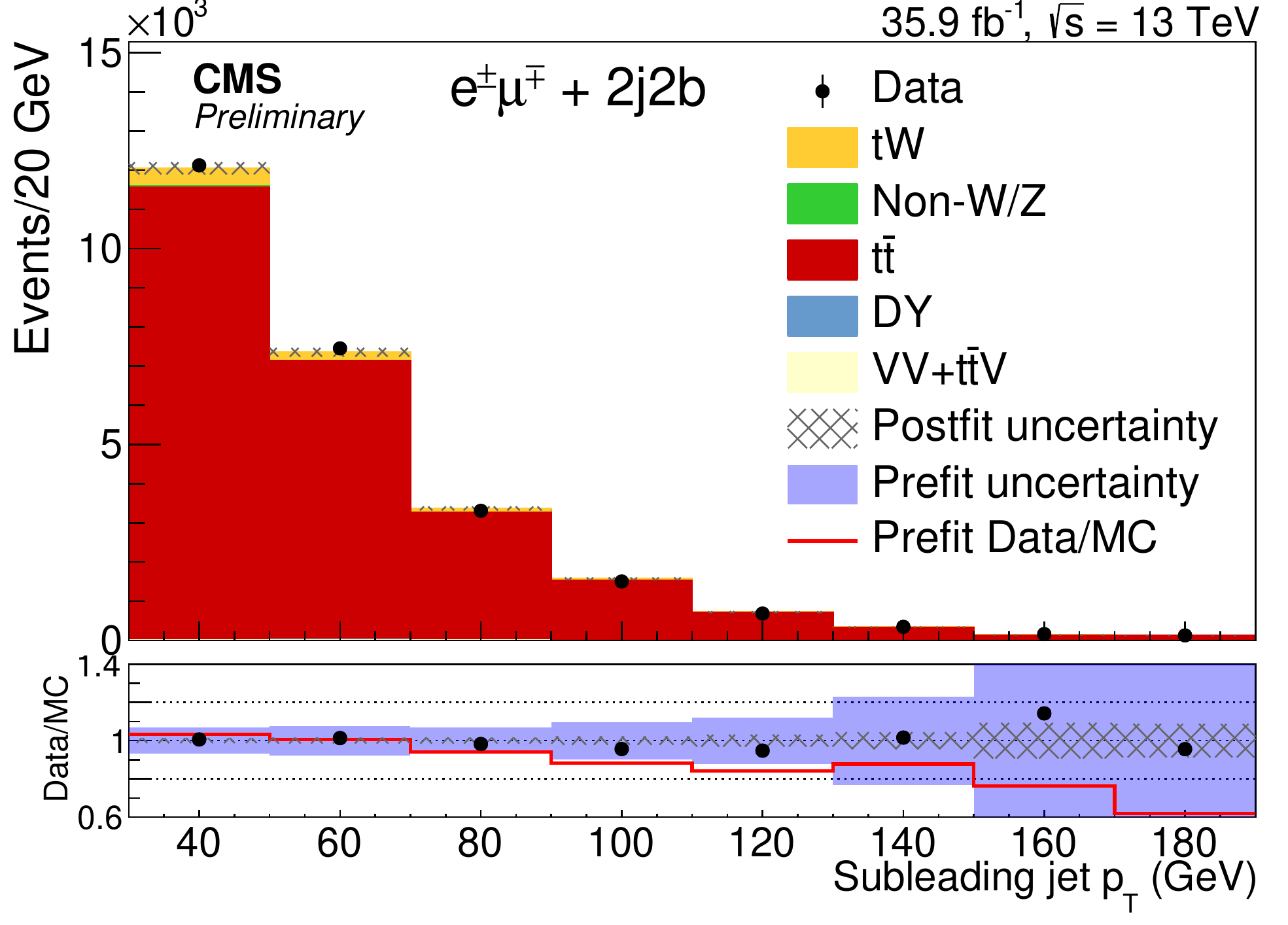}}
\end{minipage}

\caption{ 
(a) Yields observed in data compared with those expected from simulation as a function of the number of jets and b-tagged jets. 
The error band includes the statistical and all systematic uncertainties excluding the ones related to background normalization
and integrated luminosity.
Post-fit comparison between the observed data and simulated events in the classifier output (b) and $p_{\rm{T}}$ of the subleading jet (c)  
for events with exactly one jet which is tagged as a b jet and for events with exactly two jets which are tagged as b jets, respectively.
The error band includes the statistical and systematic uncertainties.  
In all cases, the bottom of each panel shows the ratios of the data to the sum of the expected yields~\cite{st_tW_run2}.
}
\label{fig:st_tw}
\end{figure}

The electroweak \textit{t}-channel production of top quarks (Fig. \ref{st_prod:a}) is the most abundant single top
production mode in the SM with a predicted cross section at NNLO 
that equals to $\sigma_{t\textrm{-ch.}}=214.5^{+1.0}_{-1.2}\ (\rm{scale})$$ $ pb at $\sqrt{s}=13$ TeV \cite{st_table} 
(approximately 70\% of the total single top quark cross section). 
It is also known fully differentially to NNLO accuracy, first calculated assuming stable top quarks~\cite{st_table}, and recently for 
production and decay~\cite{st_t_decay}. The NLO corrections to the LO \textit{t}-channel cross section  are  accidentally  small--of  the  order  of  a  few
percent--and the NNLO corrections are of the same order (Ref.~\cite{andrea} for an overview). 
%This should not appeared complacent about NNLO QCD corrections to t-channel single-top production cross section; they must be applied to enable studies 
%of electroweak production of top quarks at the LHC with a percent accuracy.
Since single top-quark production in the \textit{t}-channel has a moderately large cross section, 
it was established early in LHC Run 1~\cite{st_t_early_run1} (the first single top measurement that did not depend on multivariate techniques)
and Run 2~\cite{st_t_early_run2}. 

In associated tW production (Fig. \ref{st_prod:b}) the W boson  is  real.  This  process  is  known  to  approximate NNLO QCD accuracy~\cite{st_tW_approxNNLO}, and it was
observed for the first time at the LHC~\cite{st_tW_run1_8}, for its cross section was negligible at the Tevatron.
In \textit{s}-channel  single  top-quark  production (Fig. \ref{st_prod:c})  a  time-like  virtual W boson
is  exchanged.   The  process  is  also known  to  approximate  NNLO QCD accuracy~\cite{st_s_approxNNLO}, and it is characterized by the smallest single top-quark 
production cross section at the LHC. So far it remains elusive~\cite{st_s_run1}, and it was only
established as a separate single top-quark production channel at Tevatron~\cite{st_s_tevatron}. 

%E. L. Berger et al, NNLO QCD Corrections to t-channel Single Top-Quark Production and Decay, Phys. Rev. D 94, 071501 (2016)%
%The  process  can  be calculated  in  a  scheme  in  which  the  initial state b quark either originates from flavor excitation 
%in the proton (``5 flavor'' scheme), or from the initial-state gluon splitting into a pair of b quarks (``4 flavor'' scheme). 
%N. Kidonakis, “Theoretical results for electroweak-boson and single-top production”,PoS DIS2015 (2015) 170, arXiv:1506.04072.%

\begin{figure}[!ht]																	
  \centering
  \includegraphics[scale=0.55]{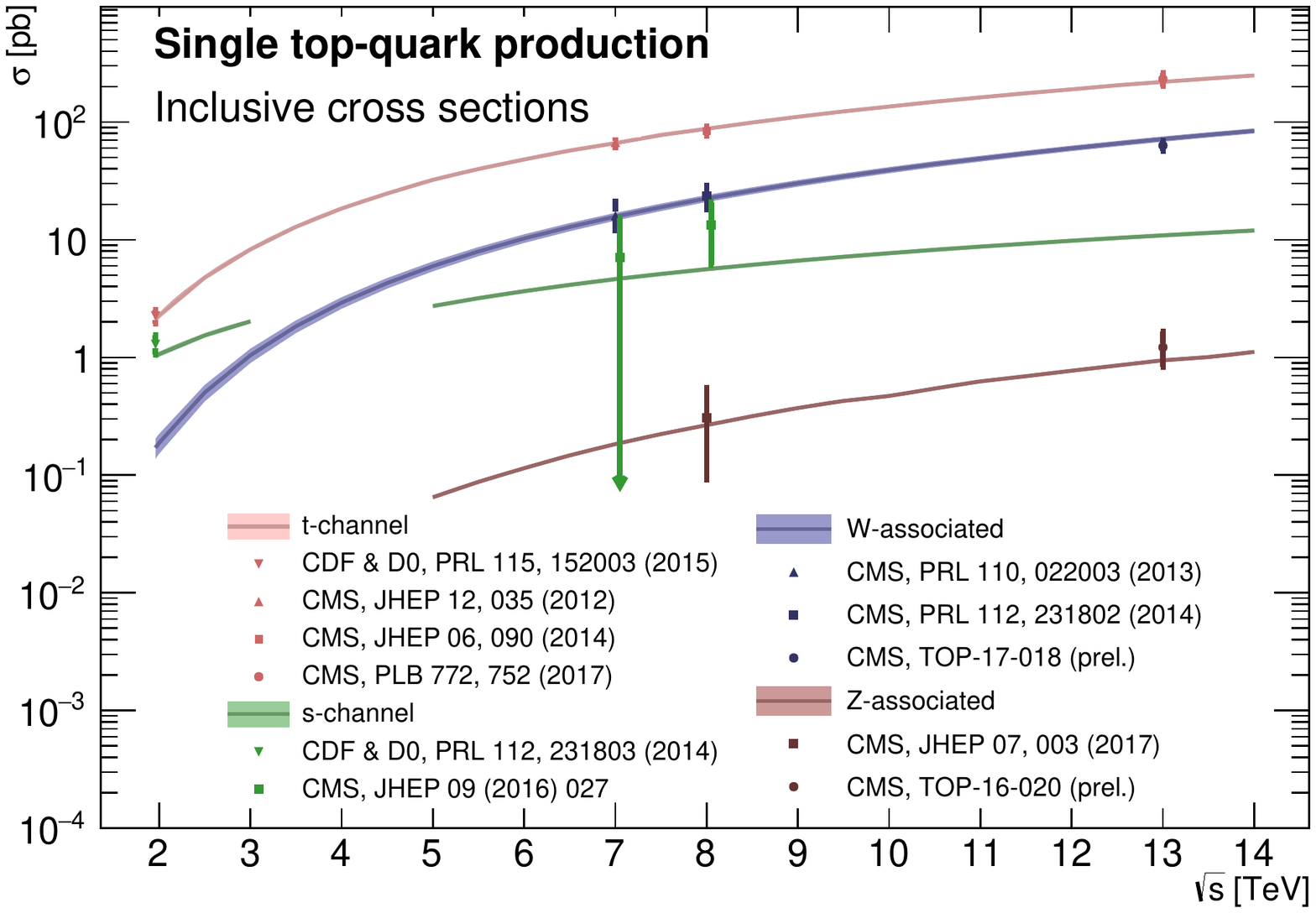}
    \caption{
       Summary of LHC (and Tevatron as on the legend) measurements of the inclusive single top quark production cross section in 
       \textit{t}-~\cite{st_t_run1_7,st_t_run1_8,st_t_run2} and \textit{s}-channels~\cite{st_s_run1}, and in association with 
       a W~\cite{st_tW_run2,st_tW_run1_7,st_tW_run1_8} 
       or Z~\cite{st_tZ_run2,st_tZ_run1} boson.
       The measurements are compared to theoretical calculations based on approximate 
       NNLO QCD accuracy~\cite{st_tW_approxNNLO,st_s_approxNNLO,st_t_approxNNLO}, 
       expect for the Z-associated production that is based on \textsc{MG5\_aMC@NLO}~\cite{amc@nlo} (v.254).
       For all predictions the NNPDF3.0~\cite{nnpdf30} PDF set has been used and the uncertainties include variations due to $\mu_{\rm{F}}$, $\mu_{\rm{R}}$ scales, and PDF parameterization.   
       The predictions for the \textit{s}-channel production is given up to $\sqrt{s}=3$ TeV for $\rm{p\bar{p}}$ collisions 
       and for pp collisions starting from $\sqrt{s}=5$ TeV, while for the rest of the production channels the 
       curves corresponding to both pp and $\rm{p\bar{p}}$ collisions coincide at the considered accuracy~\cite{andrea,cms_summary_st}.
}
    \label{fig:st_summary}
\end{figure}

Multivariate separation of signal and background, and profile likelihood fits to control the amount \ttbar background and to extract 
the cross section are adequate, if not mandatory. A characteristic example of measuring the associated tW production
has been recently released~\cite{st_tW_run2}, which also takes advantage of 36 fb$^{-1}$ of pp collision data.
Events have been categorized depending on the number of jets and b tagged jets 
as presented in Fig.~\ref{st_tw:a}. The cross section has been then estimated through a profile likelihood fit to the distributions of the classifier
output (Fig.~\ref{st_tw:b}) and $p_{\rm{T}}$ of the subleading jet (Fig.~\ref{st_tw:c}) 
simultaneously in signal- and background-enriched regions, respectively.
A summary of inclusive single top-quark cross section measurements at
the LHC in all production channels and for different center-of-mass energies
is presented in Fig.~\ref{fig:st_summary}.  
With the large data sets available at the LHC, also the first differential cross sections
for \textit{t}-channel single top-quark production as a function of the top quark transverse
momentum and rapidity became feasible~\cite{st_t_diff}. 
All measurements are in good agreement with the SM prediction.

\begin{figure}
\begin{minipage}{.5\linewidth}
\centering
\subfloat[]{\label{rt_results:a}\includegraphics[scale=0.3]{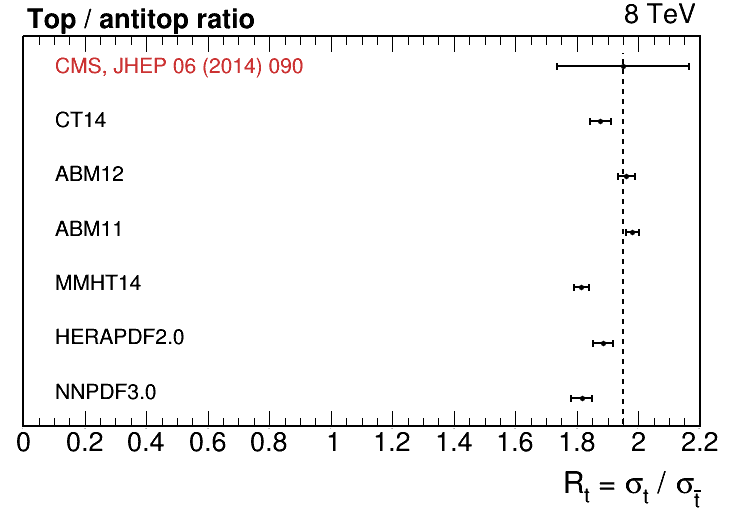}}
\end{minipage}
\begin{minipage}{.5\linewidth}
\centering
\subfloat[]{\label{rt_results:b}\includegraphics[scale=0.3]{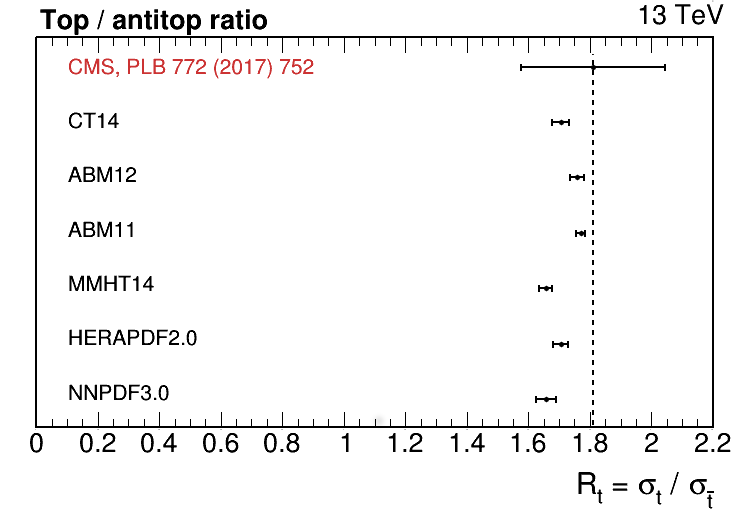}}
\end{minipage}%\par\medskip

\caption{
Measurements of $R_{t\text{-ch.}}$ $\sqrt{s}=8$ (a) and 13 (b) TeV, compared with theoretical expectations at NLO
employing the CT14~\cite{ct14}, MMHT14~\cite{mmht14}, NNPDF3.0~\cite{nnpdf30}, and
HERAPDF2.0~\cite{hera20} PDF sets. Former versions of ABMP16~\cite{abmp16} have been used. 
Error bars for the different PDF sets represent the quadratic sum of the 68\% confidence level interval
of the predictions of the eigenvectors in the respective set, and the uncertainty due to $\mu_{\rm{F}}$, $\mu_{\rm{R}}$ scales. 
An uncertainty of $\pm2$ GeV in the top quark mass is also considered, while the statistical uncertainty due to the finite number of 
iterations employed for the integration is less significant (adapted from Ref.~\cite{andrea}).
}
\label{fig:rt_results}
\end{figure}

An interesting observable in \textit{t}- and \textit{s}-channel single top-quark production is  the  ratio  of  inclusive production  rates  
for single top quark to top antiquark, $R_{t\text{-ch.}}$. More specifically, the top quark is produced with an up-type 
quark (or down-type antiquark) in the initial state, whereas the top antiquark is produced with a  down-type quark  
(or  up-type  antiquark).  Hence, in pp collisions $R_{t\text{-ch.}}$ is  sensitive  to the ratio of PDFs for up-type and
down-type quarks (and down-type and up-type antiquarks), with a naive expectation of $R_{t\text{-ch.}}=2$ 
for up and down valence quarks only. The exact predictions of $R_{t\text{-ch.}}$ depend on $\sqrt{s}$ though; 
the higher the $\sqrt{s}$ the greater the impact of sea quarks, and thus the more the deviation from the naive expectation.
Experimentally, $R_{t\text{-ch.}}$ is  a  robust  observable  in  which many uncertainties cancel.  
Measurements of $R_{t\text{-ch.}}$ from LHC Run 1 (Fig.~\ref{rt_results:a}) and Run 2 (Fig.~\ref{rt_results:a}) are compatible
with the SM prediction within the presently large uncertainties.
The updated ABMP16 (NLO version retrieved from Ref.~\cite{abmp16_nlo}) PDF sets predict ratios similar to their ABM12 predecessor,
while the latest NNPDF3.1~\cite{nnpdf31} version lies within $+1\sigma$ relative to NNPDF3.0. 
The values for the new ATLAS-epWZ16~\cite{atlas-epWZ16} PDF set is compatible with the ones from the MMHT14~\cite{mmht14} PDFs.
The relative importance of $R_{t\text{-ch.}}$ in PDF extractions is expected 
to grow with more integrated luminosity. 
Another complementary constraint on PDFs can be obtained from the ratio
of \textit{t}-channel cross sections at $\sqrt{s}=7$ and 8 TeV~\cite{st_t_run1_8}.
%Measurement of the t-channel single-top-quark production cross section and of the |Vtb| CKM matrix element in pp collisions at sqrt(s) = 8 TeV JHEP 1406 (2014) 090%

\clearpage

\clearpage
\subsection{ It pins down to two numbers; the mass and the width }
\label{sec:prop}

\begin{figure}
\begin{minipage}{.5\linewidth}
\centering
\subfloat[]{\label{tt_mass:a}\includegraphics[scale=0.4]{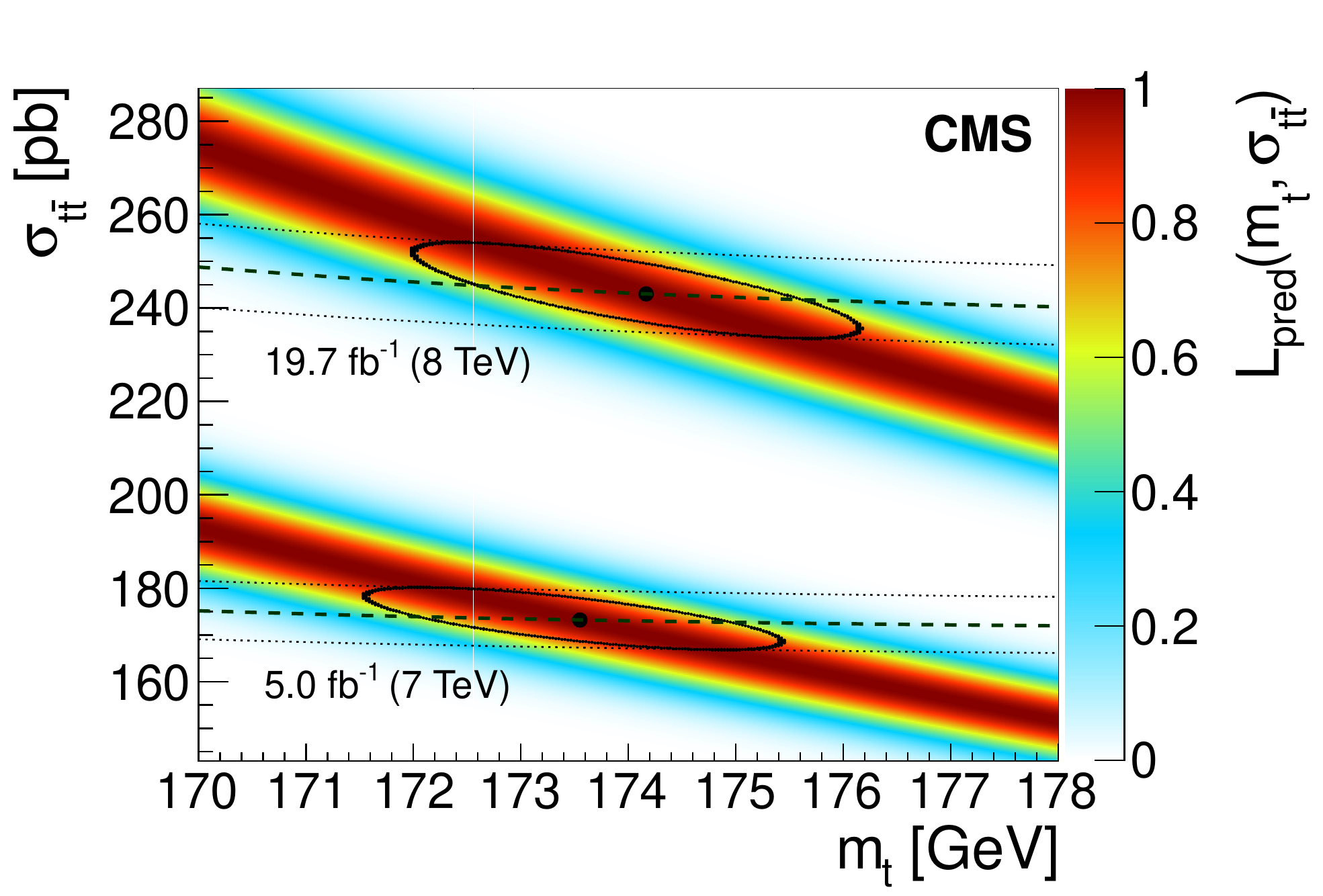}}
\end{minipage}
\begin{minipage}{.5\linewidth}
\centering
\subfloat[]{\label{tt_mass:b}\includegraphics[scale=0.25]{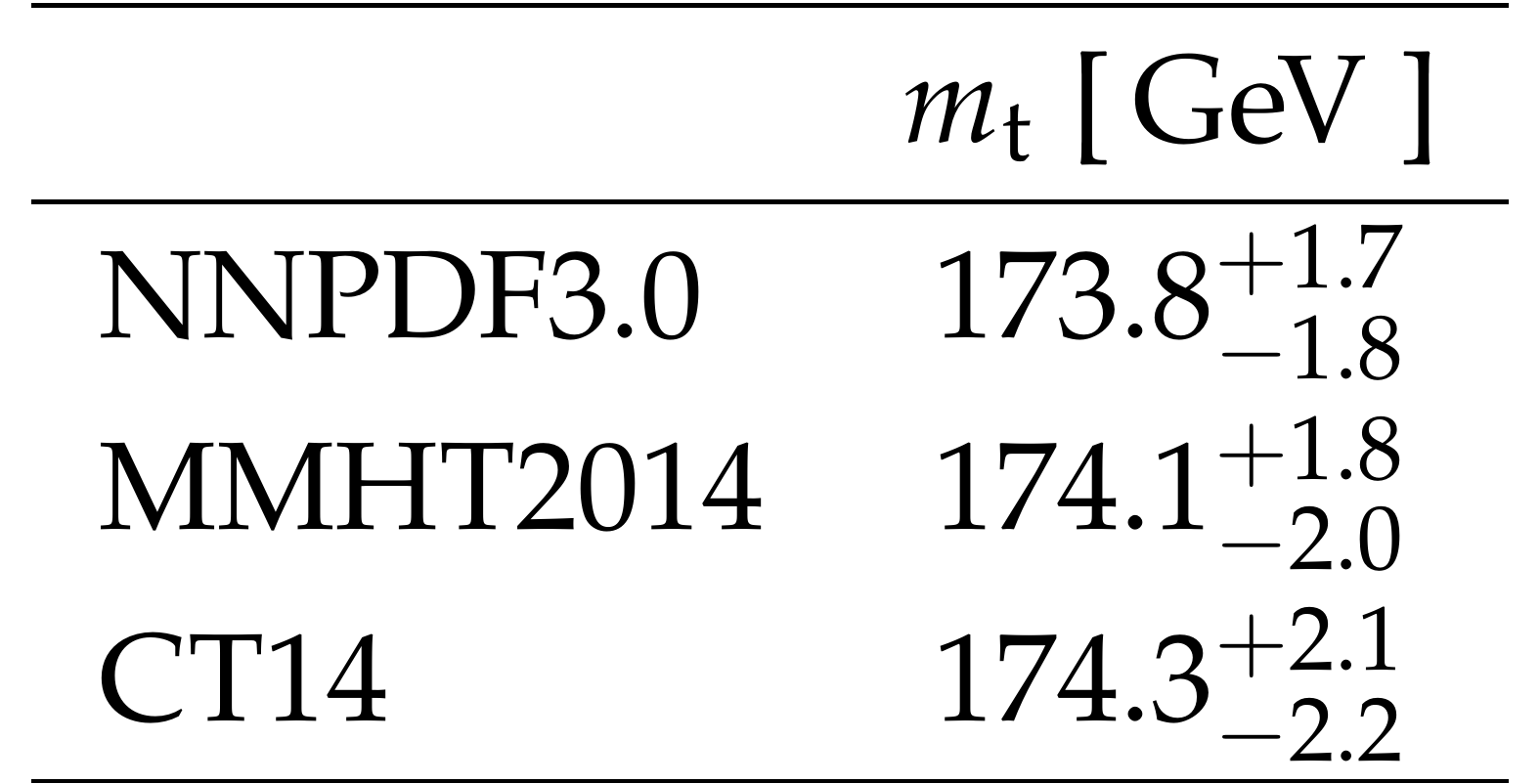}}
\end{minipage}%\par\medskip

\caption{
(a) 
Likelihood for the predicted dependence of the \ttbar production cross section on the top quark pole mass 
for $\sqrt{s}=7$ and 8 TeV, employing the NNPDF3.0~\cite{nnpdf30} PDF set. 
The measured dependences on the mass are given by the dashed lines, 
along with their $\pm 1\sigma$ uncertainties represented by the dotted lines. 
The extracted mass at each value of $\sqrt{s}$ is indicated by a black point, with its $\pm 1\sigma$ uncertainty as the continuous
contour~\cite{dilep_78}. 
(b) 
Top quark pole mass at NNLO+NNLL~\cite{tt_NNLO} accuracy extracted by comparing the combined measured \ttbar production 
cross section at $\sqrt{s}=7$ and 8 TeV~\cite{dilep_78} with predictions employing the CT14~\cite{ct14}, MMHT14~\cite{mmht14}, and NNPDF3.0~\cite{nnpdf30} PDF sets~\cite{dilep_78}.
}
\label{fig:tt_mass}
\end{figure}

Moreover $\sigma_{\rm{t\bar{t}}}$ measurements  can  be  reinterpreted  in  terms  of  constraints  
%either  
on the  top  quark mass.
% or on $\alpha_{\rm{s}}$.
The inclusive $\rm{t\bar{t}}$  production cross section, as predicted by perturbative QCD,
is a (steeply) falling function of $m_{\rm{top}}$. In a given renormalization scheme, 
e.g., in the  on-shell  or  the $\rm{\overline{MS}}$ scheme, the mass parameter is defined unambiguously~\cite{lepenik}. 
The measured inclusive cross section also exhibits a  (weak)  dependence  on  the mass parameter  used  in  the  MC  simulation~\cite{hoang};
the  higher  the momenta  transferred  to  the decay  products,
the higher the acceptance for $\rm{t\bar{t}}$ events, hence the
measured $\sigma_{\rm{t\bar{t}}}$ decreases slightly with $m_{\rm{top}}$. 
The top-quark mass can be thus determined from the intersection of the curves describing the $m_{\rm{top}}$ dependence
of the theoretical and the measured $\sigma_{\rm{t\bar{t}}}$, as illustrated in Fig.~\ref{tt_mass:a}.
This method has been inherited from Tevatron~\cite{d0_polemass_1,d0_polemass_2}, 
and it has been subsequently implemented at LHC~\cite{cms_polemass_7TeV} 
reaching a precision on the top quark pole mass of $\mathcal{O}(1\%)$ ~\cite{dilep_78}.

%The  strong coupling  constant  has  been  extracted  from  the $\rm{t\bar{t}}$ production  cross  section simultaneously with 
%the pole mass of the top quark~\cite{dilep_78}. Interestingly, while NNLO computations, e.g., of QCD jet production at hadron colliders 
%have only become available in 2017~\cite{dijets_nnlo},  $\rm{t\bar{t}}$ production  has been  already  known  to  NNLO  precision 
%since  2013~\cite{tt_NNLO}.
%Therefore the extraction of $\alpha_{\rm{s}}$ from $\sigma_{\rm{t\bar{t}}}$ measurement constitutes the first NNLO
%measurement of the strong coupling constant at a hadron collider.  The resulting value of $\alpha_{\rm{s}}$--evaluated
%at the energy scale of Z boson mass--tends to be lower than values from other sources, though it has been included 
%into the most recent world average value~\cite{pdg_2017}. %(Fig.~\ref{fig:alphas}).

%\begin{figure}[!ht]
%  \centering
%  \includegraphics[scale=0.25]{alphas.png}
%    \caption{
%       Summary of measurements of $\rm{t\bar{t}}$ as a function of the energy scale $Q$.
%       The respective degree of QCD perturbation theory used in the extraction of $\rm{t\bar{t}}$ is indicated in 
%       the brackets~\cite{pdg_2017}.
%}
%    \label{fig:alphas}
%\end{figure}

\begin{figure}
\begin{minipage}{.5\linewidth}
\centering
\subfloat[]{\label{tt_width:a}\includegraphics[scale=0.35]{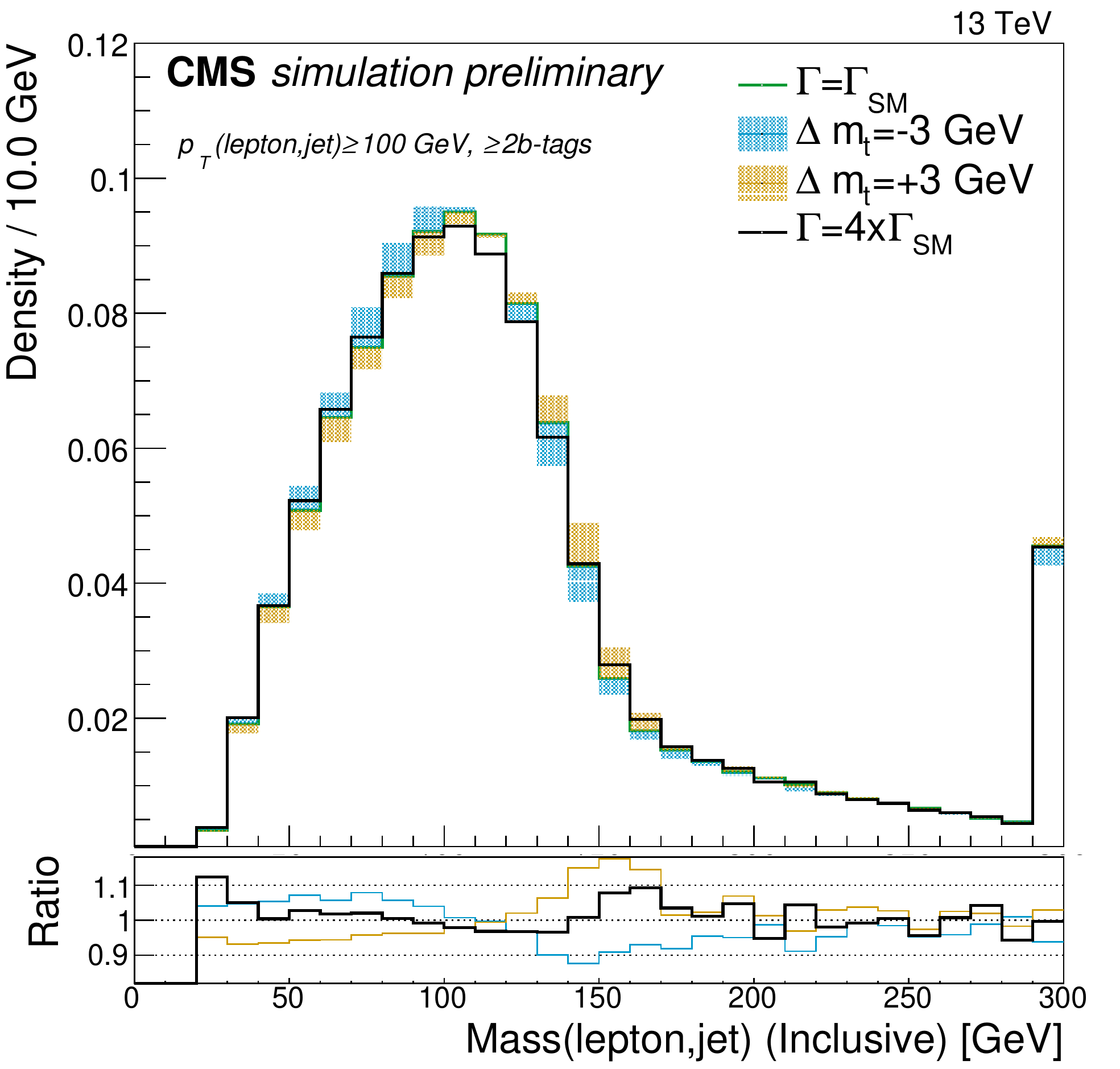}}
\end{minipage}
\begin{minipage}{.5\linewidth}
\centering
\subfloat[]{\label{tt_width:b}\includegraphics[scale=0.35]{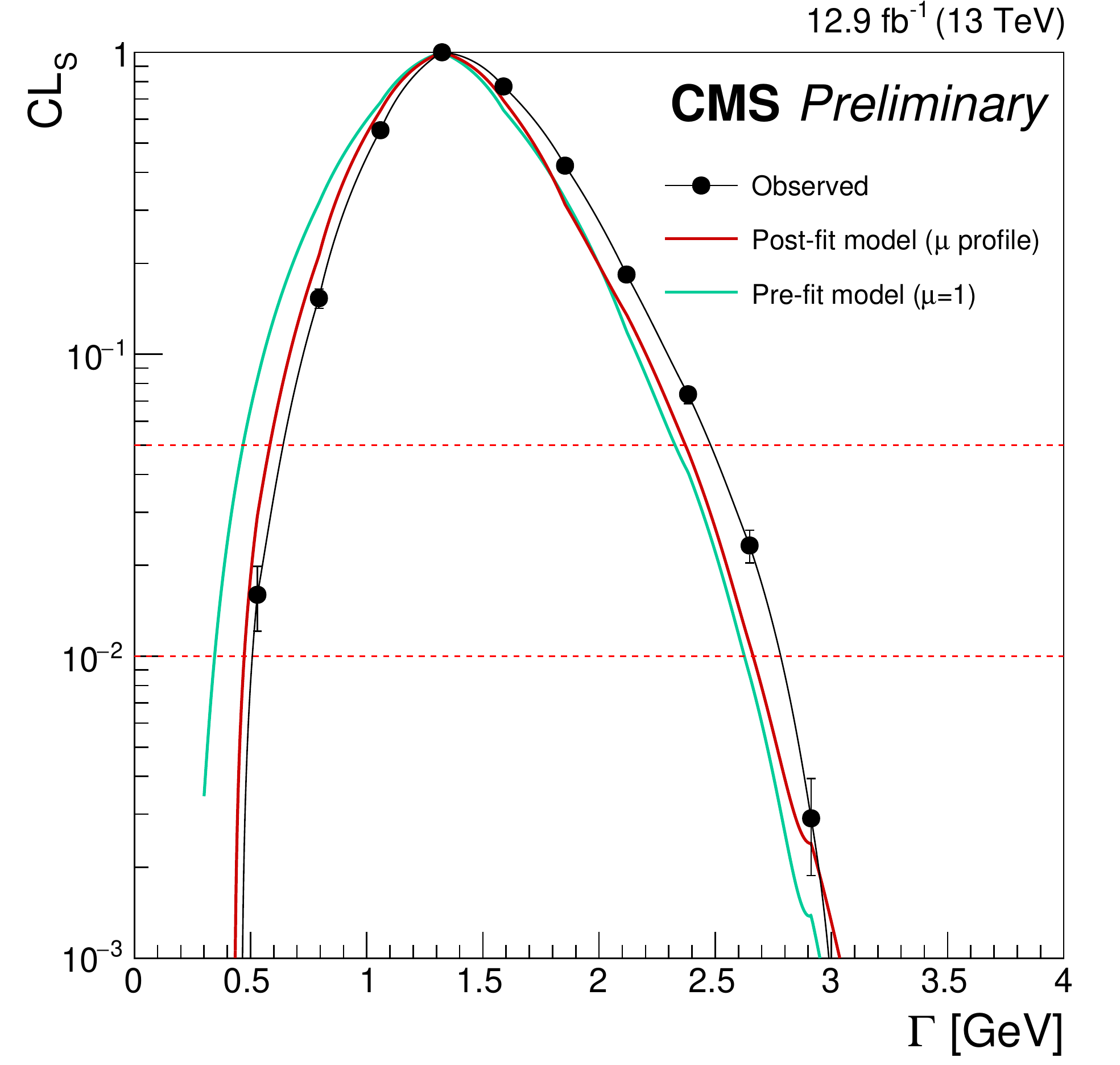}}
\end{minipage}%\par\medskip

\caption{
(a) 
Expected sensitivity for the distribution of the invariant mass of charged lepton and b jet pairs, $M_{\ell \rm{b}}$, in the dilepton $\rm{t\bar{t}}$ 
channel for simulated \textsc{POWHEG}~\cite{powheg,hvq} (v2) combined with \textsc{PYTHIA}~\cite{pythia8} (v8.205) events, where the top quark mass 
is varied by $\pm3$ GeV with respect to the default value of 172.5 GeV, and the width is varied by a factor of 4 with respect
to the SM value. The plot describes boosted ($p_{\rm{T}}\geq100$) events, with at least two b tagged jets ($\geq$2b category). 
The top panel shows the distribution with the last bin displaying the overflow of the histograms,
while the bottom plot shows the ratio with respect to the $m_{\rm{top}}=172.5$ GeV and $\Gamma_{\rm{t}}=\Gamma_{\rm{SM}}$  scenario.
(b) 
Evolution of the confidence limits as a function of the top quark width. 
The derived limits at the 95\% confidence level are derived from the intersection of the fits 
to the horizontal line at 0.05~\cite{tt_width}.
}
\label{fig:tt_width}
\end{figure}

%\clearpage
The  width  of  the  top-quark, $\Gamma_{\rm{t}}$,  can  be  determined  directly  from  the kinematic reconstruction of its decay products.
Similar to measurements of $m_{\rm{top}}$, an observable sensitive to the top quark width is built from reconstructed quantities.
Such an observable is the invariant mass of charged lepton and b jet pairs in dilepton $\rm{t\bar{t}}$ events (Fig.~\ref{tt_width:a}).   
In  a  series  of  binary  hypotheses,  the  SM value  of $\Gamma_{\rm{t}}$, $\Gamma_{\rm{t}}=\Gamma_{\rm{SM}}$, has been tested~\cite{tt_width} 
against different BSM width hypotheses to extract an observed 95\% confidence interval of $0.6 \leq \Gamma_{\rm{t}} \leq 2.5$ GeV (Fig.~\ref{tt_width:b})
to be compared with the NLO value of $\Gamma_{\rm{SM}}=1.330$~\cite{tt_width_NLO} assuming $m_{\rm{top}}=172.5$ GeV and  $\alpha_{\rm{s}}=0.118$.
The sensitivity of this direct method~\cite{tt_width} (including single resonant tW production) is lower than the sensitivity of indirect methods (e.g.~\cite{tt_width_indir}), 
but both are still far away from challenging its $< 1\%$ theoretical uncertainty.

\clearpage

\subsection{ Rare and rarer signatures }
\label{sec:tt_rare}

\begin{figure}
\begin{minipage}{.5\linewidth}
\centering
\subfloat[]{\label{ttg:a}\includegraphics[scale=0.35]{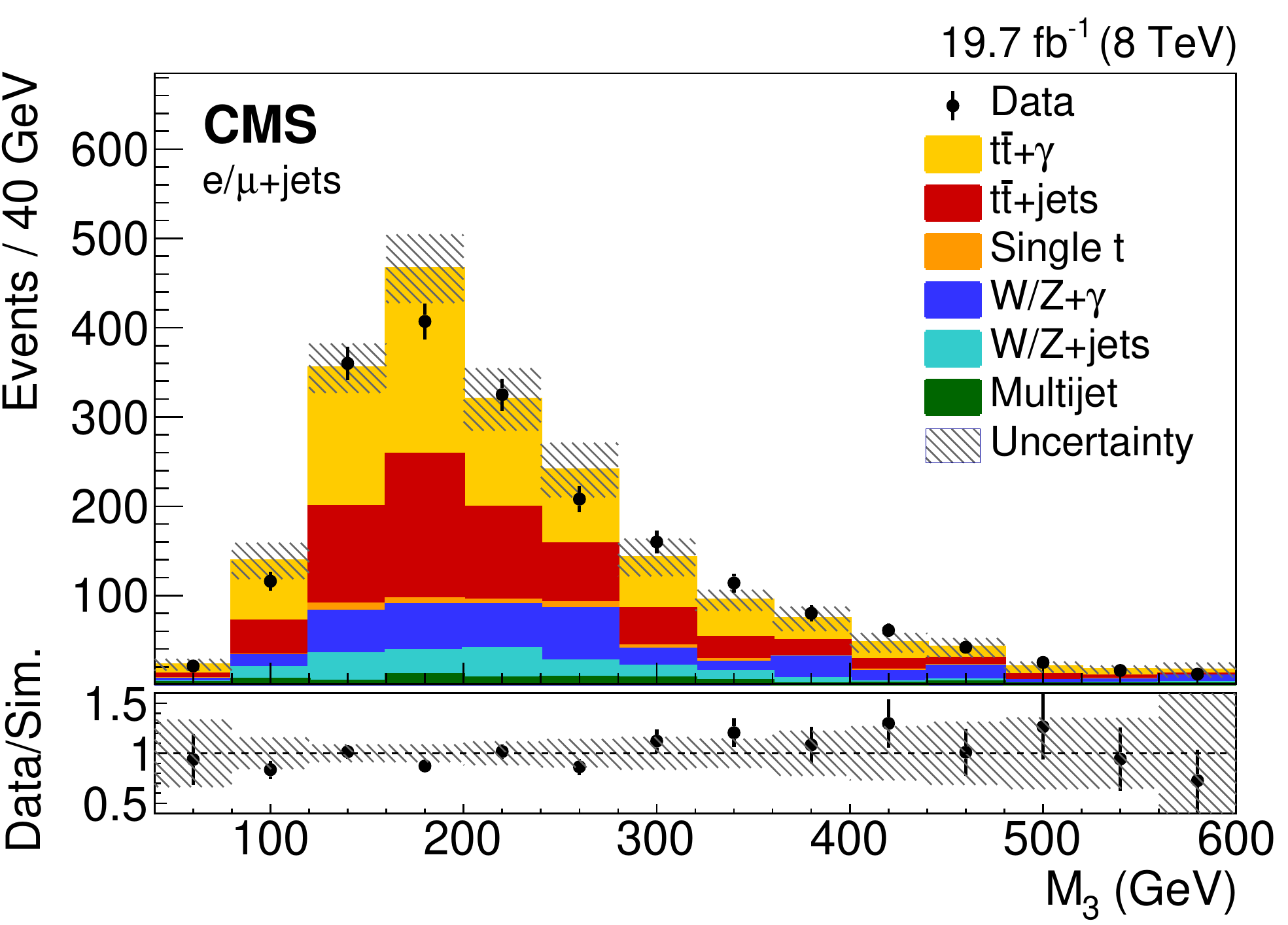}}
\end{minipage}
\begin{minipage}{.5\linewidth}
\centering
\subfloat[]{\label{ttg:b}\includegraphics[scale=0.35]{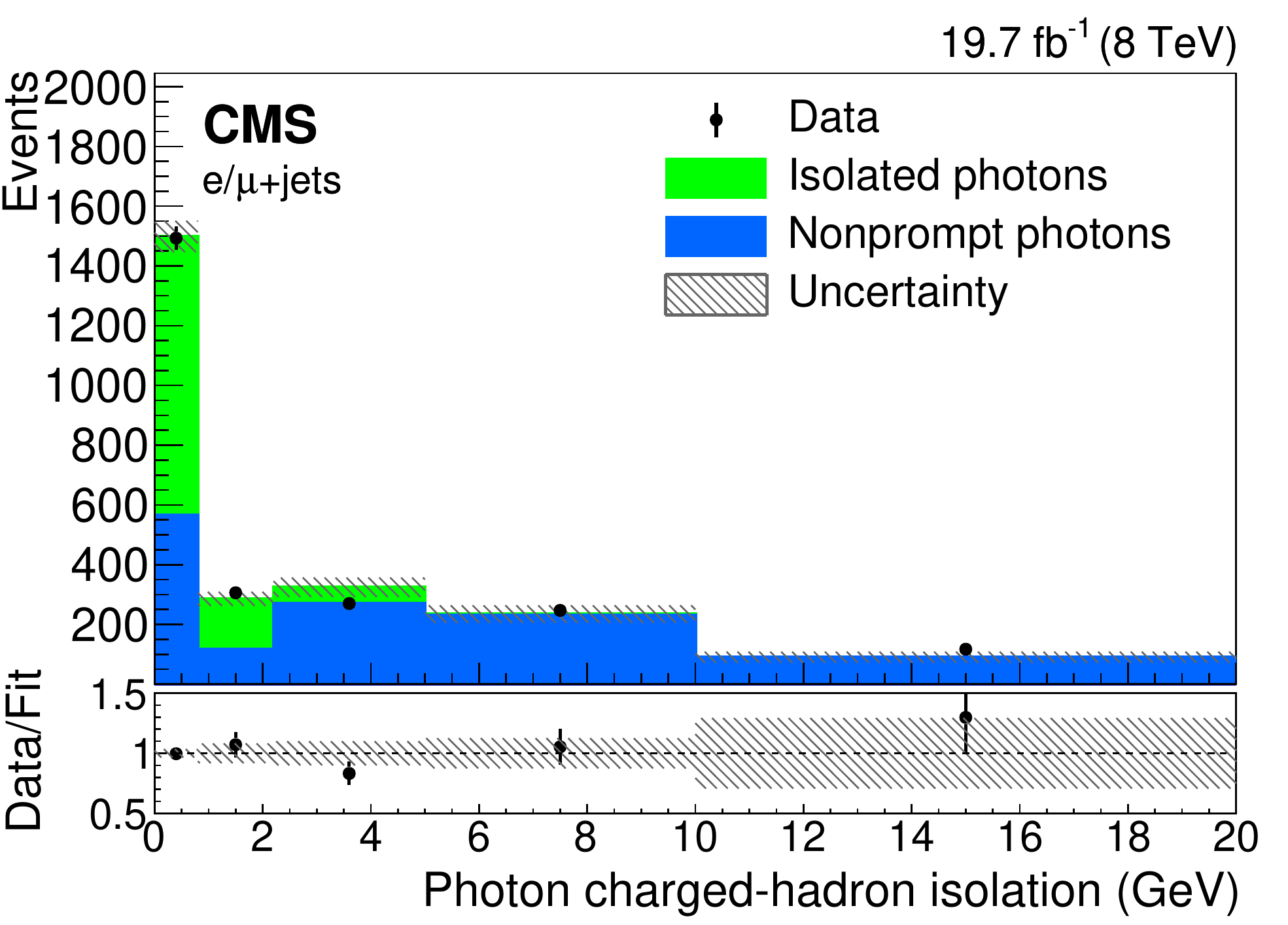}}
\end{minipage}%\par\medskip

\caption{
(a) Distribution of the invariant mass of the three jet combination that
gives the highest vectorial sum of individual jet transverse momentum, $M_3$, in data and simulation.
The fraction of events passing the photon selection containing top quark pairs, referred to as
the ``top quark purity'', is used to distinguish \ttbar pair events from other backgrounds. 
The lower panel shows the ratio of the data to the prediction from simulation. 
The uncertainty band is a combination of statistical and systematic uncertainties in the simulation.
(b) Result of the fit to the photon charged-hadron isolation in the $\ell$+jets decay channel. 
The uncertainty band shows the statistical uncertainties in the templates derived directly from the data. 
The measurement of the ``photon purity'', defined as the fraction of reconstructed photons 
in the selection region,  is used to discriminate between  the  genuine  photons  expected  from $\ttbar\rm{\gamma}$ 
and the nonprompt photons from the the \ttbar background.
The lower panel shows the ratio of the distribution observed in data to the sum of the templates scaled to the fit result~\cite{cms_ttg}.
}
\label{fig:ttg}
\end{figure}

The production of top quarks in association with the electroweak gauge bosons, or even with the Higgs
boson,  is  predicted  to  be  rare  in  the  SM,  with  inclusive  production  cross sections in pp
collisions at $\sqrt{s}=13$ TeV as predicted at NLO QCD below 10 pb.
However, measuring these processes gives direct access to the couplings of the top quark, meaning that any 
deviation of the measured values from the SM prediction would be an indication of BSM physics.
Although evidence for $\ttbar\rm{\gamma}$ production had been already reported at Tevatron~\cite{cdf_ttg},
the main challenge lies in separating  signal photons
from genuine (hadron decays into photon pairs) or misidentified (hadrons and electrons) photons. 
By simultaneously reconstructing a top quark in the event (Fig.~\ref{ttg:a}) and rejecting photons close to the tail of the photon isolation,
a high fraction of signal events could be retrieved (Fig.~\ref{ttg:b}). 
The production cross section, defined in a fiducial region of phase space (exactly one charged lepton and the corresponding
neutrino, at least three jets, and a photon), is compatible with the SM prediction at NLO~\cite{cms_ttg}.

\begin{figure}[!ht]
  \centering
  \includegraphics[scale=0.65]{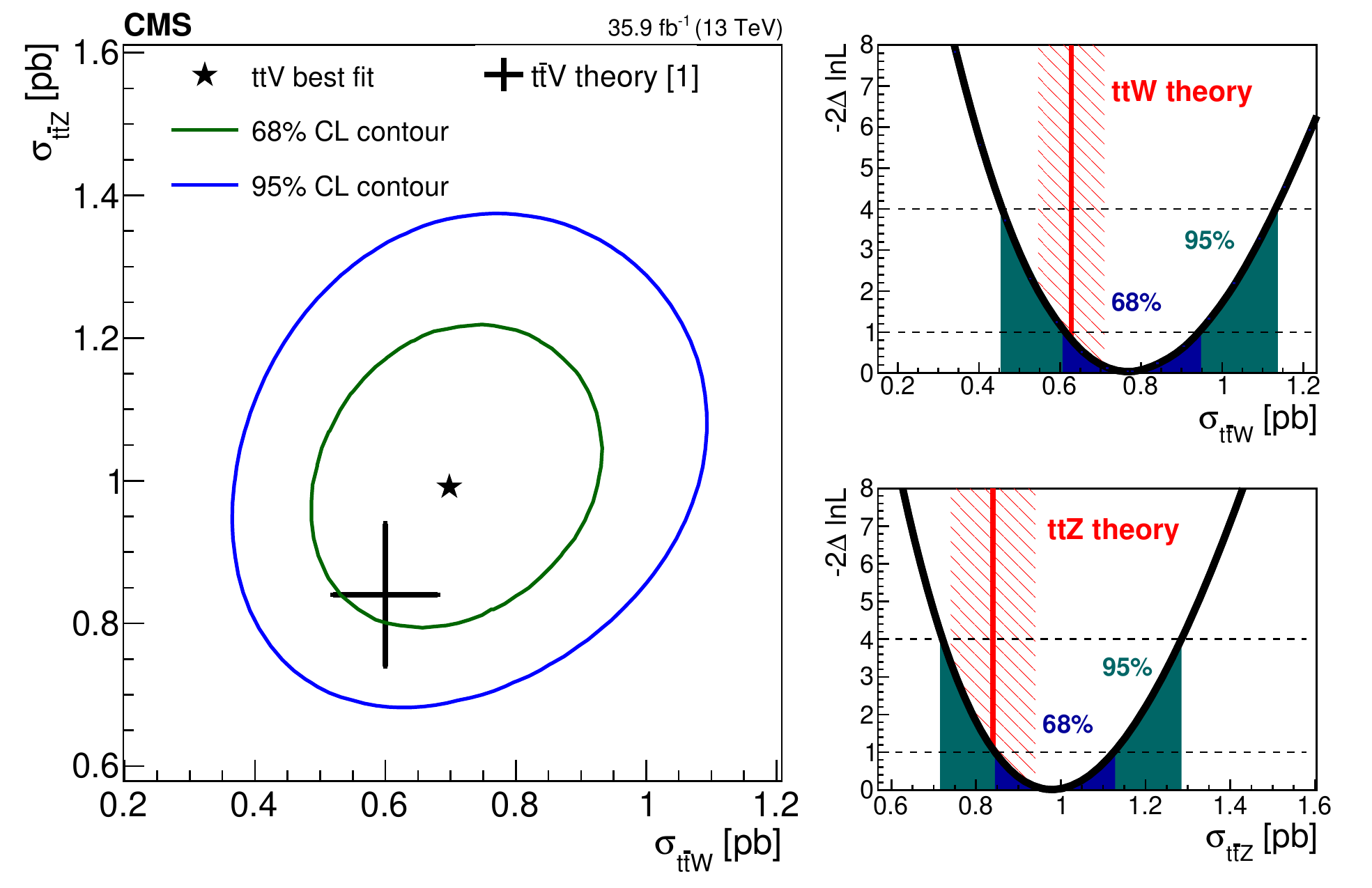}
    \caption{
	Result of the simultaneous fit for $\ttbar\rm{Z}$ and $\ttbar\rm{W}$ cross sections (star), along with its 68 and 95\% confidence level contours.
	The right panel presents the individual measured cross sections along with the 68 and 95\% confidence level intervals 
	and the theory prediction with their respective uncertainties for $\ttbar\rm{W}$ (upper right) and $\ttbar\rm{Z}$ (bottom right) ~\cite{ttV_run2}.
}
    \label{fig:ttV}
\end{figure}

Measurements of $\ttbar\rm{Z}$ and $\ttbar\rm{W}$ production have only become feasible with
the large LHC datasets~\cite{ttV_run1,ttV_run2}.  The processes feature very massive final states of $\mathcal{O}(0.5)$ TeV,
and the almost fourfold increase in production cross  sections  at  LHC  Run  2  compared  to  Run  1 
resulted in promptly re-observing both $\ttbar\rm{Z}$ and $\ttbar\rm{W}$.
The multilepton final states are studied in order to reduce the major background processes. 
The background in these events are generally low, though it is challenging to precisely estimate the residual number of wrongly
identified leptons. The fake lepton background is usually modeled from the data in control regions;
the probability for a loosely identified nonprompt lepton to fulfill the full set of tight requirements is estimated 
and validated using MC simulation.
While the requirement of two leptons compatible with the Z boson significantly improves the purity
of the sample for the measurement of $\ttbar\rm{Z}$ production, the selection of $\ttbar\rm{W}$ relies on a 
multivariate approach based on kinematic variables such as the missing missing, the leading lepton or jet $p_{\rm{T}}$.
A categorization of the selected events based on jet and b tagged jet multiplicity helps to better control the background.
The measurements~\cite{ttV_run1,ttV_run2}--with uncertainties equally shared between statistic and systematic uncertainties--are compatible 
with SM predictions (Fig.~\ref{fig:ttV}), and have been used to constrain BSM physics contributions to the Z$tt$, W$tt$, 
and H$tt$ couplings~\cite{ttV_run2}.

In view of the null results of the searches for new heavy particles decaying into top quarks, a comprehensive effective 
field theory (EFT) approach~\cite{eft} to study top quark couplings becomes attractive.  In such an approach the indirect effects
of BSM physics on the top quark couplings are treated in a consistent way, by
constructing a full set of effective operators that mediate top quark couplings with mass-dimension six (the single 
dimension-five operator violates lepton number conservation).  An extended Lagrangian is assumed based on a series expansion
in the inverse of the energy scale of the BSM physics, $1/\Lambda$, hence operators are suppressed as long as $\Lambda$
is large compared with the experimentally-accessible energy:
\begin{align*}
\mathcal{L_{\rm{eff}}} = \mathcal{L_{\rm{SM}}} + \sum_i\frac{c_i^{(6)} \cdot O_i^{(6)}}{\Lambda^2} + \mathcal{O} (\Lambda^{-4}) \ .
\end{align*}
In the above equation, the effective Lagrangian density $\mathcal{L_{\rm{eff}}}$ is given by the SM Lagrangian $\mathcal{L_{\rm{SM}}}$
and is complemented with a sum of dimension-six operators $O_i^{(6)}$, each weighted with a Wilson coefficient $c_i^{(6)}$,
calculable in LO QCD perturbation theory.  The EFT approach is a comprehensive description of the top couplings in a gauge invariant
and  renormalizable  way  that  respects  all  SM  symmetries, and the first NLO corrections in the top quark sector have been presented. 

\begin{figure}
\begin{minipage}{.5\linewidth}
\centering
\subfloat[]{\label{ttV_eft:a}\includegraphics[scale=0.85]{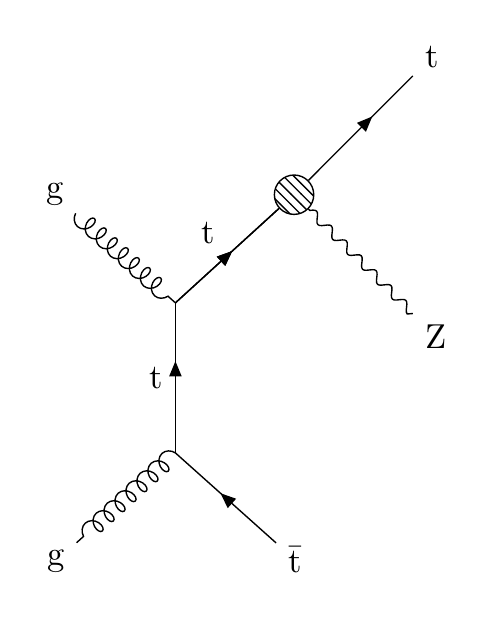}}
\end{minipage}
\begin{minipage}{.5\linewidth}
\centering
\subfloat[]{\label{ttV_eft:b}\includegraphics[scale=0.25]{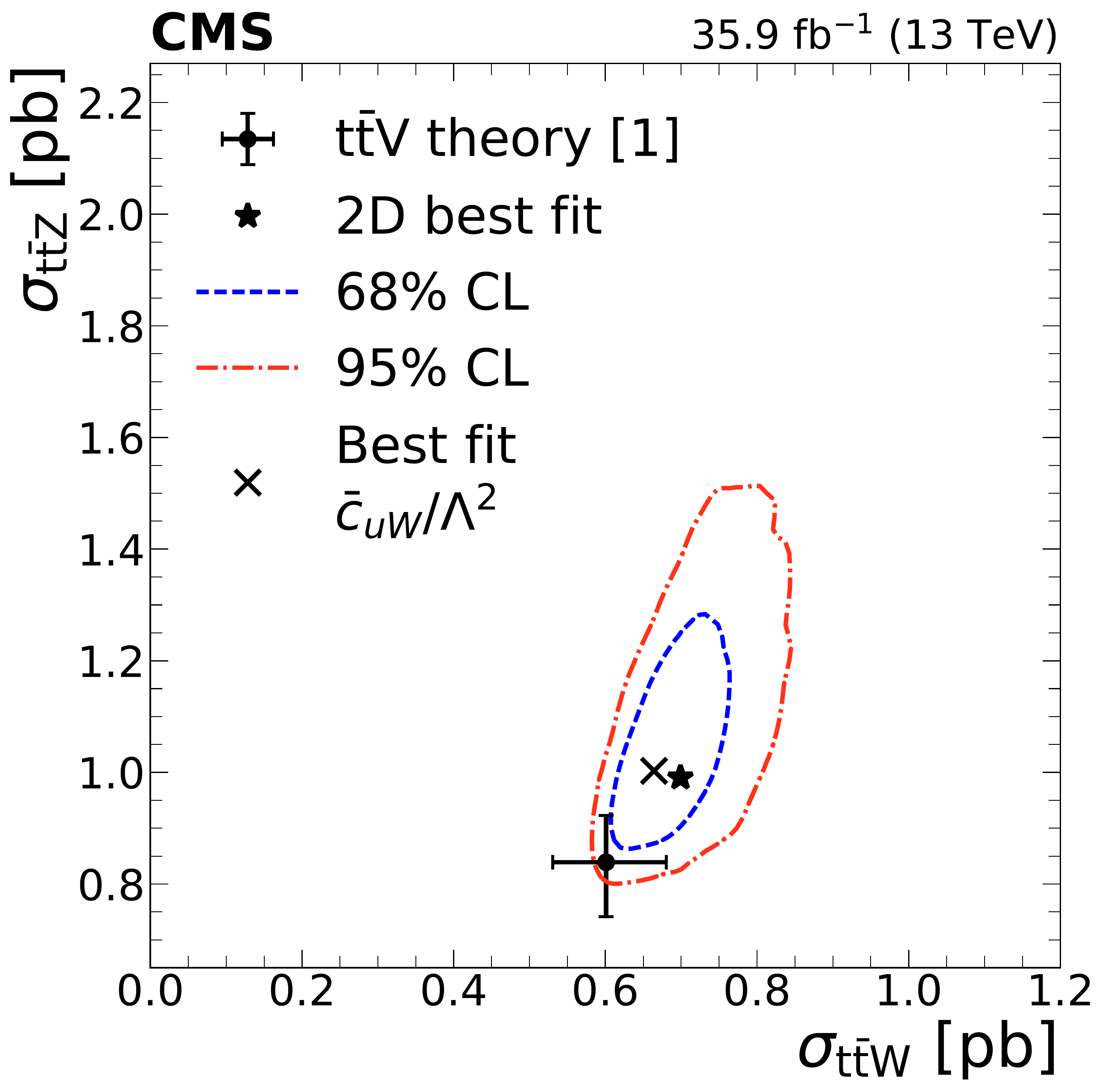}}
\end{minipage}%\par\medskip

\caption{
(a) Representative LO diagram with the most significant BSM effect for a given operator identified to affect
the contribution from $\ttbar\rm{Z}$ and $\ttbar\rm{W}$.
(b) The $\ttbar\rm{Z}$ and $\ttbar\rm{W}$ cross section corresponding to the best fit value for such an operator,
 along with the corresponding 68\% (dashed) and 95\% (dash-dotted) contours. 
 The theory predictions (dot) with their uncertainties (bars), and the two-dimensional best fit to the $\ttbar\rm{Z}$ and 
 $\ttbar\rm{W}$ cross sections (star) are superimposed~\cite{ttV_run2}.
}
\label{fig:ttV_eft}
\end{figure}

The $\ttbar\rm{Z}$ and $\ttbar\rm{W}$ results of Ref.~\cite{ttV_run2} have been then used to set constraints on the Wilson coefficients of dimension-six operators. 
Eight operators have been identified which are of particular interest because they change the expected cross sections of
$\ttbar\rm{Z}$, $\ttbar\rm{W}$, or $\ttbar\rm{H}$  without significantly impacting the expected background yields (Fig.~\ref{ttV_eft:a}). 
The constraints, obtained by considering one operator at a time (Fig.~\ref{ttV_eft:b}), are a useful first step towards more global approaches.

\clearpage
An additional rare process in the SM is the production of single top quarks in association with a Z boson,
that has been already searched for during Run 1~\cite{st_tZ_run1}.
Events containing three leptons, two or three jets, and significant missing momentum have been selected 
for the updated analysis using pp data corresponding to 36 fb$^{-1}$~\cite{st_tZ_run2}.
A higher separation of signal 
from background events, e.g., from $\ttbar\rm{Z}$ or WZ+jets, has been ensued by combining different types of multivariate 
discriminants that resulted in 20\% increase in the expected significance.
The search has exceeded the conventional $3\sigma$ threshold (``evidence'' for the existence of the signal process)
on the incompatibility with the background-only hypothesis, in contrast to the expected 1.8 standard deviations in Run 1.

\begin{figure}[!ht]
  \centering
  \includegraphics[scale=0.75]{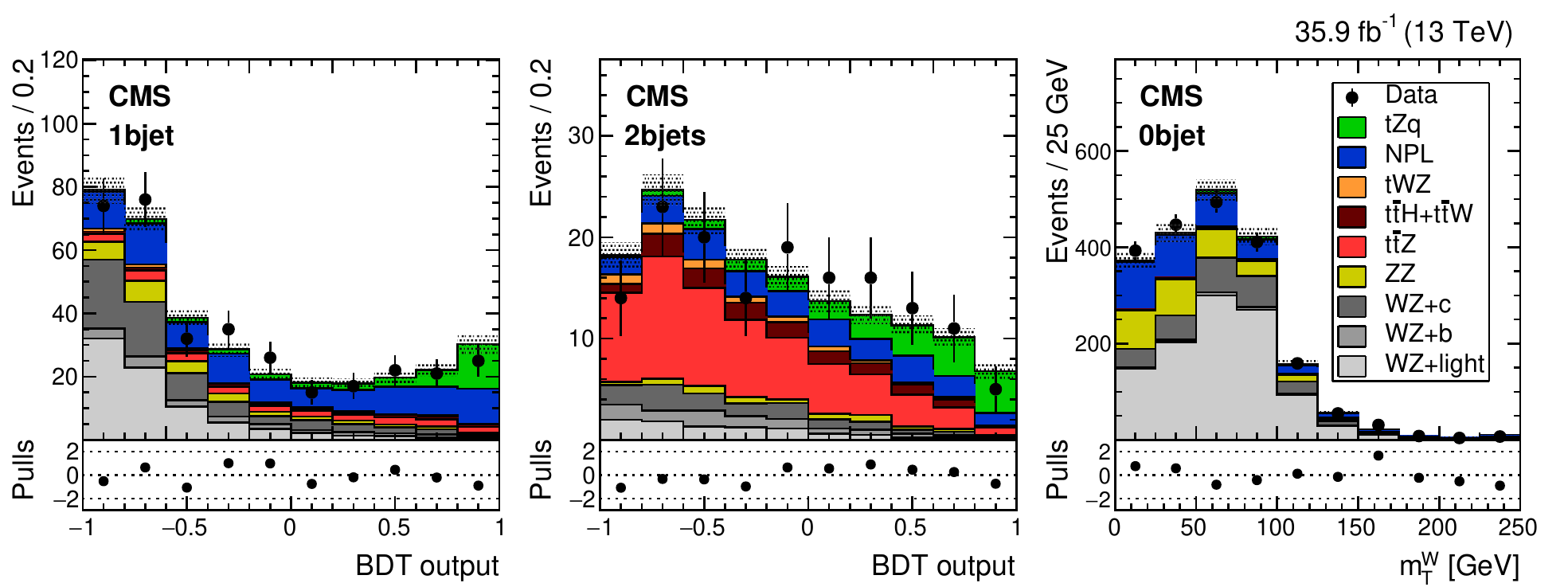}
    \caption{
       Post-fit data-to-prediction comparisons of the employed variables, i.e, the classifier output (a, b) 
       and transverse mass of the $\rm{W} \rightarrow \ell \nu$ system (c),  
       for events (a, signal enriched) with one b jet and one light flavor jet, 
       (b, $\ttbar\rm{Z}$ enriched) at least two jets with at least two of them b tagged,
       (c, WZ+jets enriched) and at least one jet without any b-tagged.
       The  hatched  bands  include  the  total  uncertainty  in  the
       background and signal contributions.  
       The pulls of the distributions, defined in each bin as the difference between the data and prediction, divided by the
       quadratic sum of total uncertainties in the predictions (systematic and statistical) and the data
       (statistical), are shown at the bottom of the plots~\cite{st_tZ_run2}.
}
    \label{fig:st_tZ}
\end{figure}

\begin{figure}
\begin{minipage}{.5\linewidth}
\centering
\subfloat[]{\label{tttt_prod:a}\includegraphics[scale=0.45]{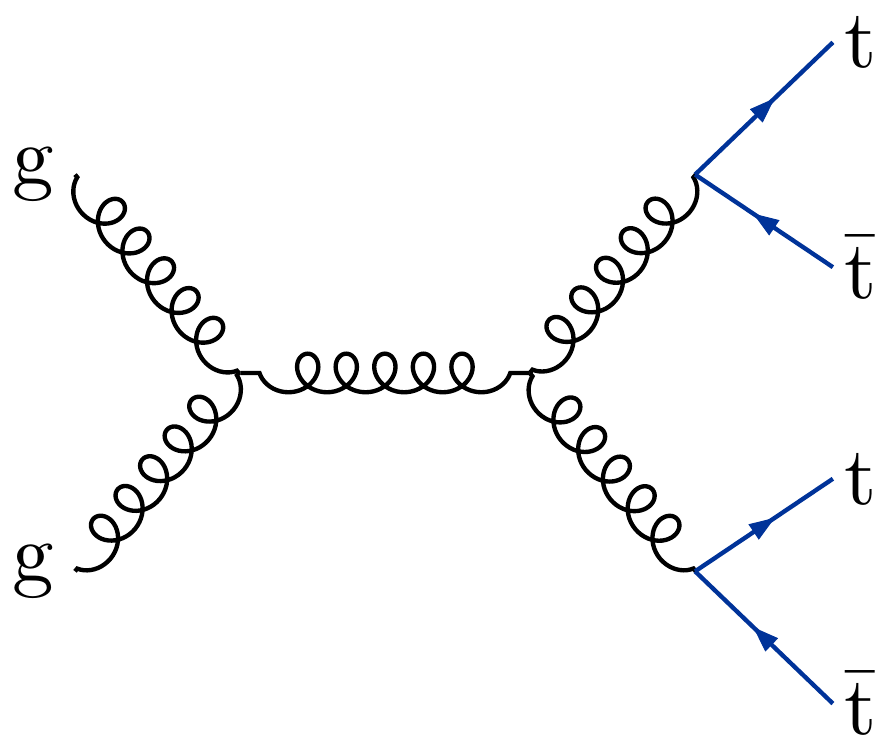}}
\end{minipage}
\begin{minipage}{.5\linewidth}
\centering
\subfloat[]{\label{tttt_prod:b}\includegraphics[scale=0.45]{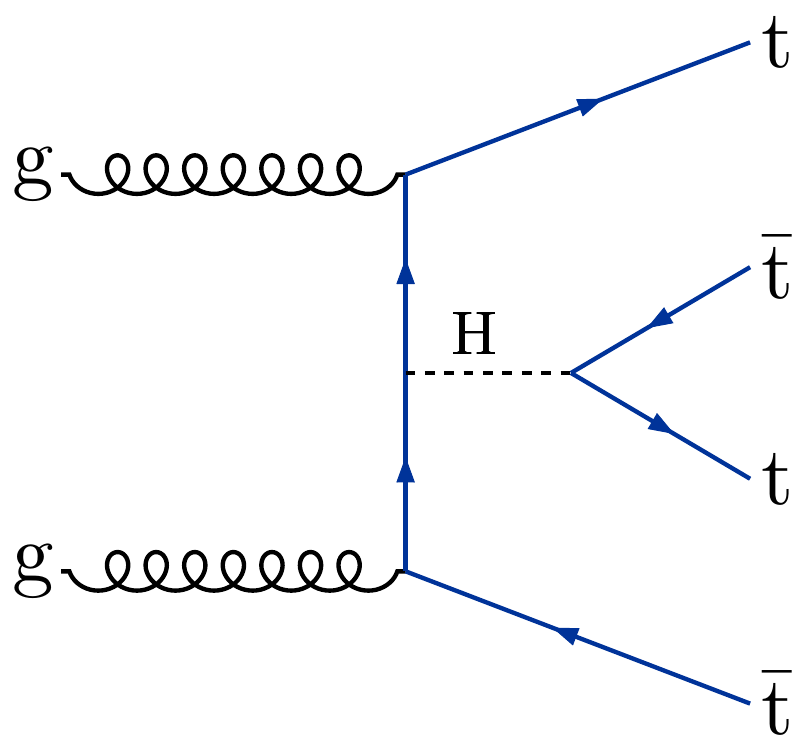}}
\end{minipage}%\par\medskip

\caption{
Representative diagrams for $\rm{t\bar{t}}$$\rm{t\bar{t}}$ production at LO in the SM through (a) gluon and (b) 
the Higgs boson mediation~\cite{tttt_13TeV}.
}
\label{fig:tttt_prod}
\end{figure}

Four top quark production ($\rm{t\bar{t}}$$\rm{t\bar{t}}$) is an even rarer process, 
and it is currently predicted at NLO precision~\cite{tttt_nlo} in the SM.
The representative Born level Feynman diagrams of $\rm{t\bar{t}}$$\rm{t\bar{t}}$ production, 
which occurs either through the gluon, the electroweak gauge boson, or  the  Higgs  boson  mediation, are shown in Fig.~\ref{fig:tttt_prod}.
The  Higgs-induced $\rm{t\bar{t}}$$\rm{t\bar{t}}$ production (Fig.~\ref{tttt_prod:b}) exhibits  no  dependence  on  the  Higgs  boson
width, with the cross section proportional to the top quark Yukawa coupling to the fourth power~\cite{tttt_lo}.
Although a high integrated luminosity is needed to reach a $5\sigma$ discovery of the rare production (outside the LHC Run 2 reach),
null  searching  results  in  the  low  luminosity  operation of  the  LHC  are  also  useful  because  they  can  be  used
to  constrain  the  top  Yukawa  coupling. 

\begin{figure}
\begin{minipage}{.5\linewidth}
\centering
\subfloat[]{\label{tttt_results:a}\includegraphics[scale=0.35]{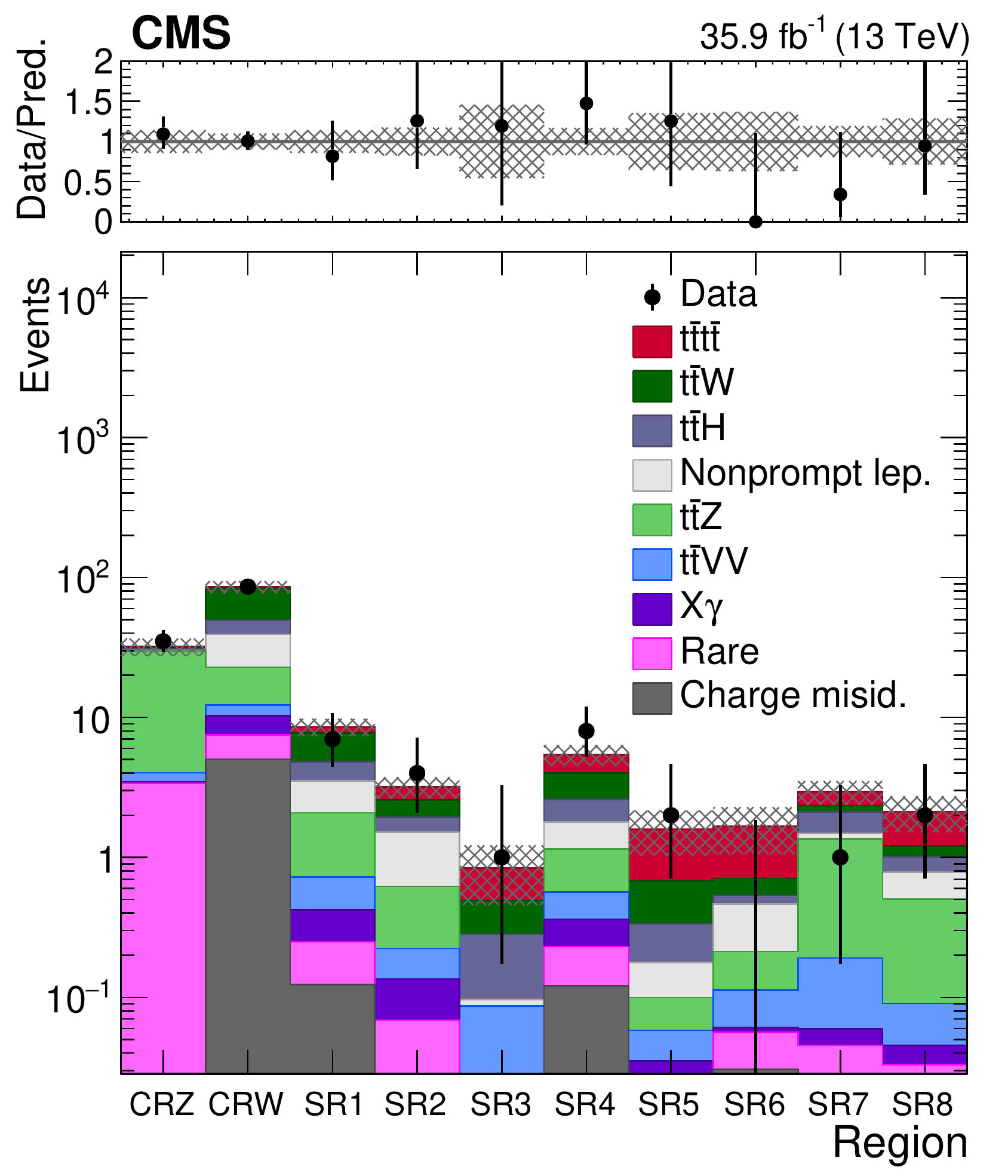}}
\end{minipage}
\begin{minipage}{.5\linewidth}
\centering
\subfloat[]{\label{tttt_results:b}\includegraphics[scale=0.35]{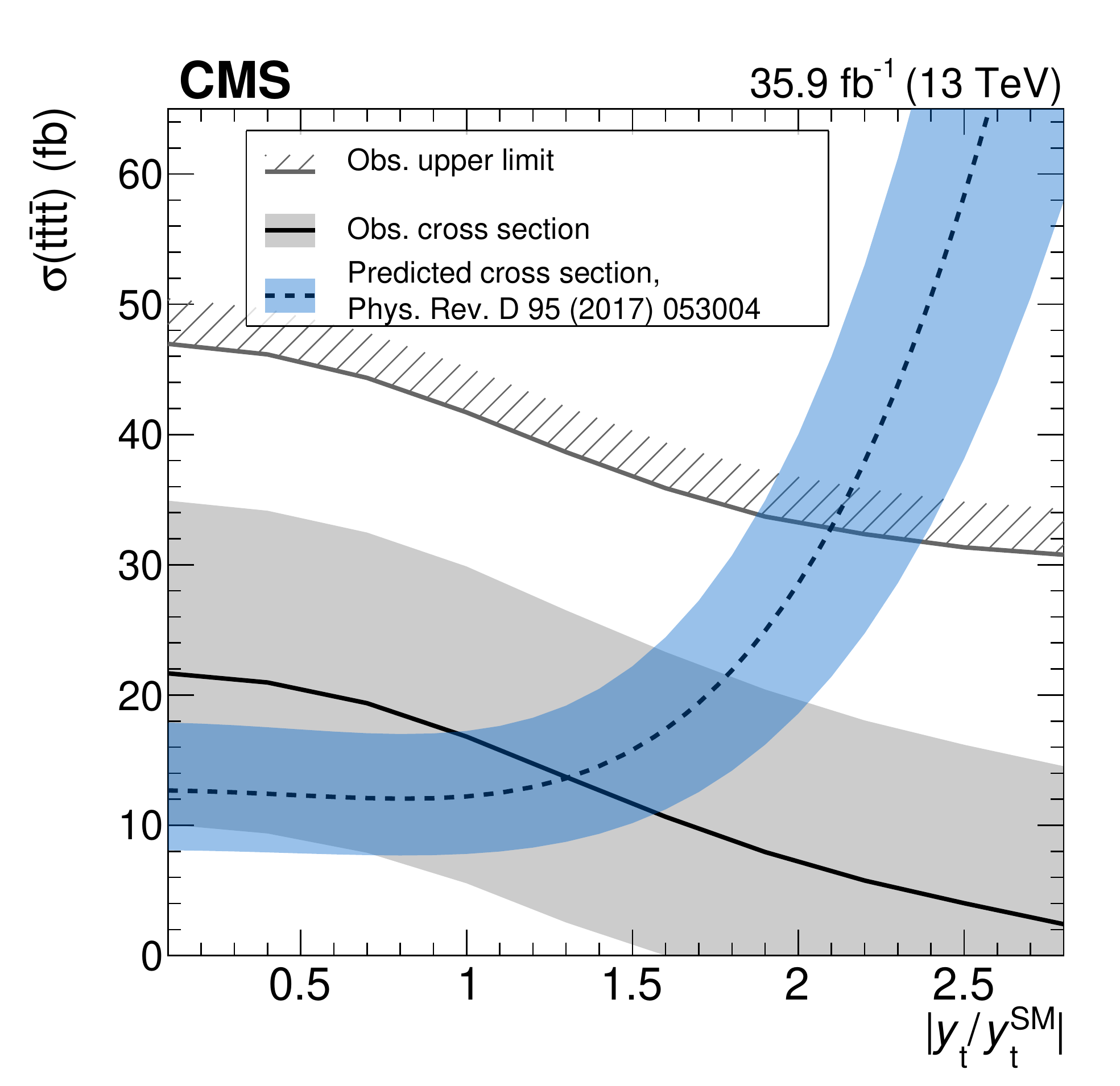}}
\end{minipage}%\par\medskip

\caption{
(a)
Observed yields in all (control and signal) regions, compared to the post-fit predictions 
for signal and background processes. The hatched areas represent the total uncertainties in the signal and background 
predictions. The upper panel shows the ratios of the observed event yield to the total prediction.
(b)
The predicted SM value for $\rm{t\bar{t}}$$\rm{t\bar{t}}$ production~\cite{tttt_lo} (taking into account a NLO/LO factor at $y_{\rm{t}}^{\rm{SM}}$~\cite{tttt_nlo})
as a function of the ratio of the top quark Yukawa coupling to 
its SM value (dashed line), $|y_{\rm{t}}/y_{\rm{t}}^{\rm{SM}}|$, compared with the observed value (solid line), 
and with the 95\% confidence level upper limit (hatched line). The measured cross section dependence on $|y_{\rm{t}}/y_{\rm{t}}^{\rm{SM}}|$
is induced due to the rescaling of the $\ttbar\rm{H}$ background by a factor of $|y_{\rm{t}}/y_{\rm{t}}^{\rm{SM}}|^2$~\cite{tttt_13TeV}.
}
\label{fig:tttt_results}
\end{figure}

The same-sign dilepton and the three- (or more) lepton final states, though represent a very low branching fraction,
are the most sensitive to $\rm{t\bar{t}}$$\rm{t\bar{t}}$ production in the regime with SM-like kinematic properties,
due to the low level of backgrounds.
By optimizing the signal selection for sensitivity to SM $\rm{t\bar{t}}$$\rm{t\bar{t}}$ production, by using an improved b jet 
identification algorithm, and by employing background estimation techniques that are adapted to take into
account the higher jet and b jet multiplicity requirements in the signal regions (Fig.~\ref{tttt_results:a}), 
the corresponding expected upper limits of have improved on previous (by 1.6 times relative to Ref.~\cite{tttt_8TeV}), 
and most recent (by 3.5 relative to Ref.~\cite{tttt_13TeV_part} and 1.4 times relative to Ref.~\cite{tttt_bsm_13TeV}) searches. 
Reinterpreting the results to constrain the ratio of the top quark Yukawa coupling to its SM value, 
a lower bound of 1.9 at 95\% confidence level (Fig.~\ref{tttt_results:b}) is attained~\cite{tttt_13TeV}. 

\clearpage

\section[``I like the dreams of the future better than the history of the past'']{``I like the dreams of the future better than the history of the past''\ \footnote{To J. Adams from T. Jefferson, 1 August 1816.}}

Top physics has come a long way from discovery and first measurements
to increasingly sophisticated analyses using the LHC data. 
All measurements so far are compatible with
the expectations of the SM and contribute significantly to constraining possible BSM extensions.  
As  data  accumulate,  uncertainties  in  signal  modeling  becomes  of a hindrance 
for further progress though. 
The rich variety of inclusive and differential cross sections
can be extensively used to constrain theoretical predictions based on MC simulation and analytical calculations.
More specifically, they serve as substantial input for MC tuning or performing
dedicated ancillary measurements, such as the studies of color flow in \ttbar events~\cite{cms_color_flow}.
The importance of higher-order corrections is also validated, albeit the gradual decrease in experimental uncertainties 
reveals discrepancies to be further scrutinized. 

One unique property of top quark events is the high fraction of b-flavored quarks in the partonic final state. 
This is routinely exploited through ``in-situ'' measurements of the b-tagging efficiency in challenging experimental conditions
characterized with several jets and charged leptons, resembling the signal region of many BSM physics  searches.
Meanwhile, the top quark properties~\cite{kuhl} are remarkably well understood, and their precise determination is essential 
for testing the overall consistency of the SM through precision electroweak fits. 
In parallel, top quark has an extraordinary impact on the Higgs sector and on the SM extrapolation to high-energies.
Precision measurements and ever increasing in accuracy calculations for top quark production 
can be used to include \ttbar and single top quark data into free and bound proton PDF fits. 
It is important though to test what is the bias in PDF-related uncertainties on the predictions given the potential circularity.

\begin{figure}
\begin{minipage}{.5\linewidth}
\centering
\subfloat[]{\label{project:a}\includegraphics[scale=0.4]{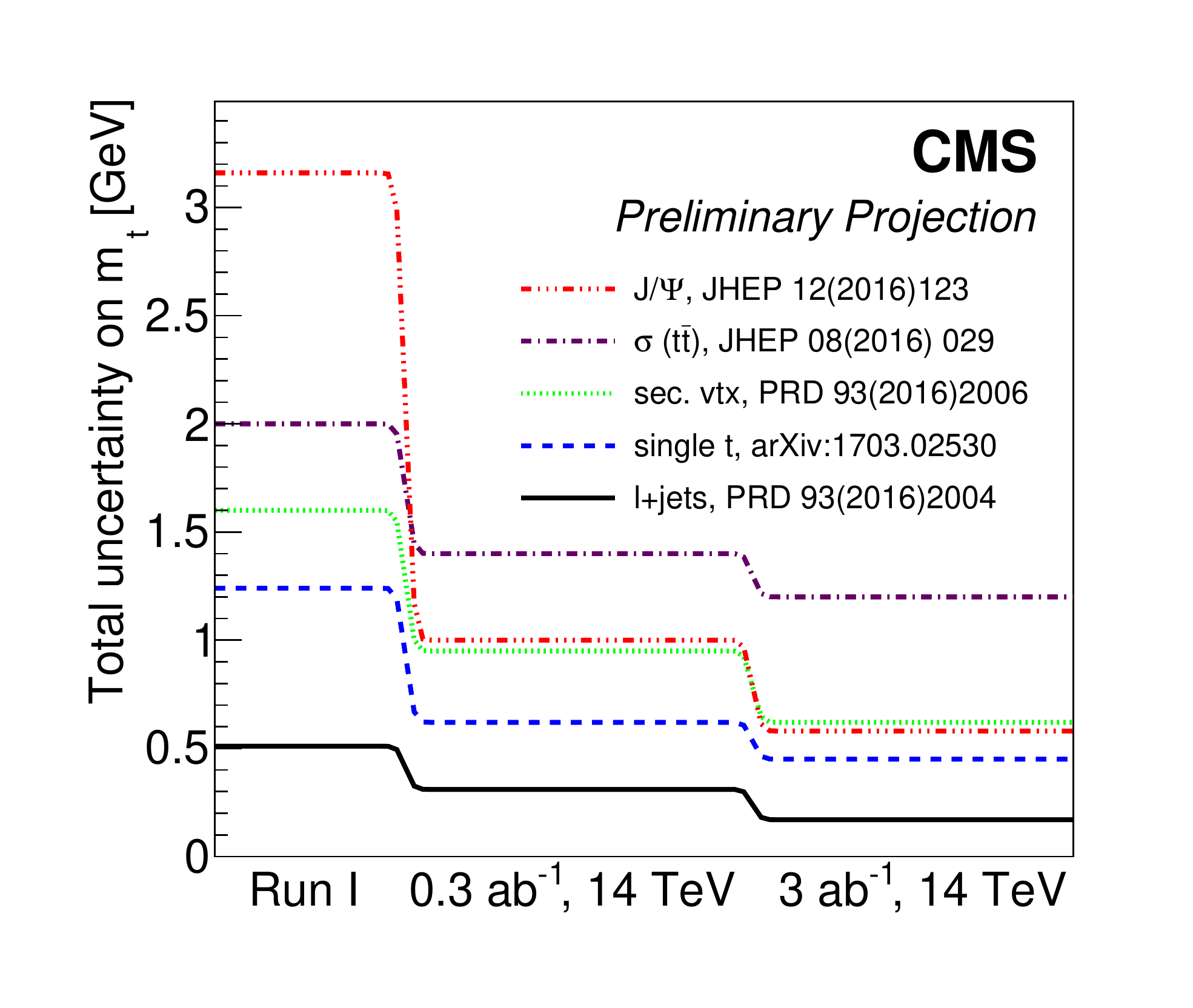}}
\end{minipage}
\begin{minipage}{.5\linewidth}
\centering
\subfloat[]{\label{project:b}\includegraphics[scale=0.22]{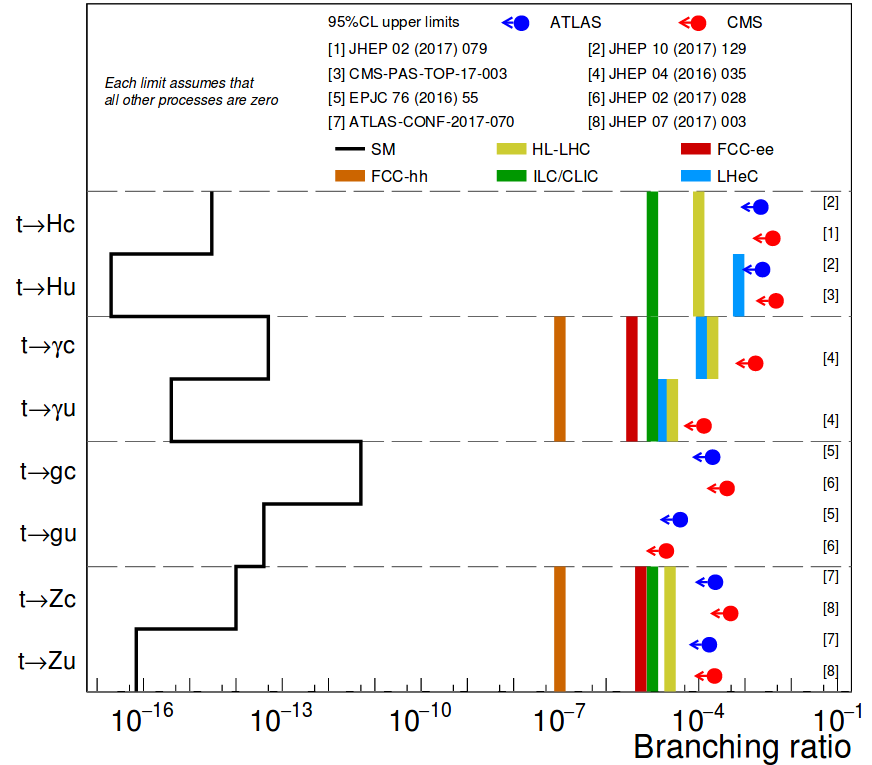}}
\end{minipage}%\par\medskip

\caption{
(a)
Total uncertainty on top quark mass obtained with different methods  and  their  projections  to  the  
HL-LHC  conditions upgrade. The projections for $\sqrt{s}=14$ TeV, with 0.3 ab$^{-1}$ or 3 ab$^{-1}$
of data, are based on measurements performed at Run 1, 
assuming that an upgraded detector will maintain the same physics performance despite of a severe increase in pileup~\cite{ecfa}.
(b)
Compilation~\cite{kirill} of branching ratio limits for top quark FCNC decays (points), compared to the SM predictions (black lines), 
as well as to various projections for HL-LHC conditions and future experiments (colored lines).
}
\label{fig:project}
\end{figure}

Top-quark physics will remain important after the scheduled upgrades of the LHC experiments~\cite{gouzevitch} (HL-LHC), and infrastructures  
at future colliders~\cite{list,boyko} may take the quest for the top to the next level.
Projections of the expected uncertainties from key results help to reveal the physics potential at the HL-LHC.
Most prominently, the uncertainty in the top quark mass could reach a rather optimistic estimate of 200 MeV~\cite{ecfa}, 
with its simultaneous determination with the production cross section grazing the 1 GeV threshold (Fig.~\ref{project:a}).  
The former method rivals a prominent and long-sought way of precision measurement of the top quark mass; 
a threshold scan of the toponium state~\cite{simon} to be performed by a linear collider whose initial energy is controlled. 
The couplings of the top  quark  as  well  as  rare  processes,  such  as  flavour-changing neutral currents (FCNCs), 
may  reveal  deviations from the SM.
Future projections~\cite{kirill} represent good prospects for pushing top-FCNC boundaries~\cite{zarneck} to even higher constraints (Fig.~\ref{project:b}).   

\end{document}